\documentclass[%
reprint,
superscriptaddress,
 amsmath,
 amssymb,
 aps,
 prl,
 showpacs, 
 floatfix,
]{revtex4-2}

\usepackage{graphicx}
\usepackage{dcolumn}
\usepackage{bm}
\usepackage{comment}

\begin{document}

\preprint{APS/123-QED}

\title{Antiscarring in Chaotic Quantum Wells}

\author{J.~Keski-Rahkonen}
\affiliation{Computational Physics Laboratory, Tampere University, P.O. Box 692, FI-33101 Tampere, Finland}
\affiliation{Department of Physics, Harvard University, Cambridge, Massachusetts 02138, USA}
\affiliation{Department of Chemistry and Chemical Biology, Harvard University, Cambridge, Massachusetts 02138, USA}

\author{A.M.~Graf}
\affiliation{Department of Chemistry and Chemical Biology, Harvard University, Cambridge, Massachusetts 02138, USA}

\author{E.J.~Heller}
\affiliation{Department of Physics, Harvard University, Cambridge, Massachusetts 02138, USA}
\affiliation{Department of Chemistry and Chemical Biology, Harvard University, Cambridge, Massachusetts 02138, USA}

\date{\today}

\begin{abstract}

Chaos plays a crucial role in numerous natural phenomena, but its quantum nature has remained large elusive. One intriguing quantum-chaotic phenomenon is the scarring of a single-particle wavefunction, where the quantum probability density is enhanced in the vicinity of a classical periodic orbit. These quantum scars illustrate the quantum suppression of classical chaos, offering a unique way to explore the classical-quantum relationship beyond conventional limits. In this study, we establish an ergodicity theorem for slacking a group of adjacent eigenstates, revealing the aspect of antiscarring -- the reduction of probability density along a periodic orbit generating the corresponding scars. We thereafter apply these two concepts to variational scars in a disordered quantum well, and finally discuss their broader implications, suggesting potential experimental approaches to observe this phenomenon.  

\end{abstract}

\maketitle

\section{Introduction}

Chaos lurks everywhere in our classical-perceived world~\cite{Strogatz_book}: For example, it is the main player behind the ever so infamous mission to forecast tomorrow's weather~\cite{lorentz_j.atmos.sci_20_130_1963}. Nevertheless, the quantum nature of chaos has remained ambiguous. This paradigmatic role of chaos was already mused over by none other than A. Einstein in the early days of quantum mechanics~\cite{einstein_verh.dtsch.phys.ges_19_82_1917, stone_phys.today_58_37_2005}. Even though ``genuine'' chaos in the spirit of classical mechanics does not exist, chaos does appear in the quantum realm, but in a different fashion~\cite{berry_phys.scripta_40_335_1989, heller_phys.today_46_38_2008, jensen_nature_355_311_1992, jensen_nature_355_591_1992}. Curiously, the investigation of quantum signatures of chaos differs distinctly from the study of chaos in classical mechanics: the latter considers the time-evolution of phase space points (see, e.g., Refs.~\cite{Goldstein_book, Arnold_book, Gutzwiller_book}), whereas the former conventionally leans on the statistical analysis of a sequence of the eigenvalues of the Hamiltonian, or on an examination of the probability density distribution of individual eigenstates (see, e.g., Refs.~\cite{Gutzwiller_book, Stockmann_book, nakamura_2004_QD, Haake_book, nakamura2010quantum, nakamura1994quantum, casati1995quantum}).

\begin{figure}[h!]
    \centering
    \includegraphics[width=0.95\linewidth]{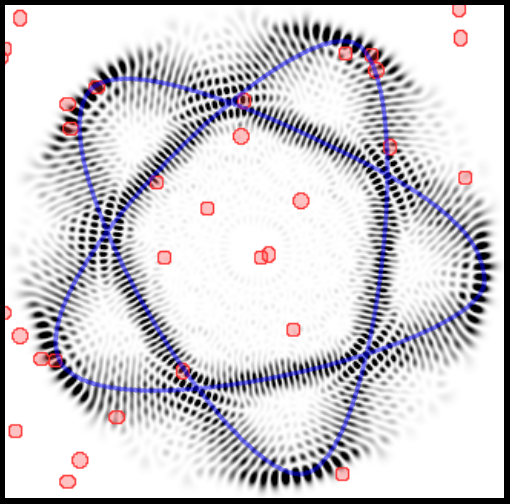}
    \caption{Probability density distribution of the eigenstate $n = 3223$ ($E \sim 549$) of a quantum well confined by a $r^5$ potential and disturbed by potential bumps (red dots denoting the locations and full widths at half maximum of the bumps). The shown eigenstate of the \emph{perturbed} system is strongly scarred by a pentagram-shape periodic orbit (blue solid line) of the \emph{unperturbed} potential. Note that several bumps are located on the scar path; in fact the scar is orientated in a such a way that it maximizes the overlap with the perturbation stemming from the bumps.}
    \label{fig:Pentagram_scar}
\end{figure}

Since a classically chaotic system is ergodic, i.e., almost all of its trajectories eventually explore the entire accessible phase space in a uniform manner, a simple solution would be to associate the corresponding high-energy quantum states with the Wigner functions constructed as homogeneous over the energy shell~\cite{Berry_rspa_413_182_1987, Berry_j.phys.a_10_2083_1977, oconnor_phys.rev.lett_61_2288_1988}. Furthermore, the Gutzwiller trace formula~\cite{Gutzwiller_book, Gutzwiller_j.math.phys_12_343_1971} revealing the energy eigenvalues in terms of classical periodic orbits (POs) is very egalitarian: there is no particular reason of why a given PO would have an outsized contribution to a given eigenstate. These two points of view are further supported by the quantum ergodicity theorems of Shnirelman, Colin de Verdiere, and Zelditch~\cite{Shnirelman_Uspekhi.Mat.Nauk_29_181_1974, Colindeverdiere_comm.math.phys_102_497_1985, zelditch_duke.math.j_55_919_1987} stating that the expectation value of an operator coincides with the microcanonical average of the classical function corresponding to the operator in the semiclassical limit $\hbar \rightarrow 0$. However, the theorem leaves the possibility of a subset of macroscopically nonergodic eigenstates with vanishing contribution in the limit $\hbar \rightarrow 0$.~\cite{Heller_book_2008} Therefore, in contrast to a tempting fallacy, \emph{not all} eigenstates of a classically chaotic system are doomed to be random and features. For instance, as consequence of quantum interference, the probability density of a quantum state can be enhanced in the vicinity of a short, moderately unstable periodic orbit PO of the chaotic classical counterpart, and hence the quantum state bears an imprint of the PO -- a \emph{quantum scar}~\cite{Heller_phys.rev.lett_53_1515_1984, Heller_book_2008,kaplan_ann.phys_264_171_1998, Kaplan_nonlinearity_12_R1_1999}.

The scarring of a single-particle wave function was first introduced by one of the present authors in Ref.~\cite{Heller_phys.rev.lett_53_1515_1984}, after which it has then evolved to be one of the prominent phenomena in the field of quantum chaos. Quantum scars bridge the gap between the classical and quantum worlds. Besides a vast amount of theoretical interest~\cite{bogomolny_physica.d_31_169_1988, berry_proc_r_soc_lond_a_423_219_1989, kaplan.phys.rev.e_59_6609_1999, kus_phys.rev.a_43_4244_1991, dariano_phys.rev.a_45_3646_1992, tomsovic_phys.rev.lett_70_1405_1993, revuelta_phys.rev.e_102_042210_2020, agam_j.phys.a.math_26_2113_1993,bohigas_phys.rep_223_43_1993, wisniacki_phys.rev.lett_97_094101_2006}, quantum scars are reported in a diverse range of experiments~\cite{Fromhol_phys.rev.lett_75_1142_1995, Wilkinson_nature_380_608_1996, narimanov_phys.rev.lett_80_49_1998, Honig_phys.rev.a_39_5642_1989,bogomolny_phys.rev.lett_97_254102_2006,kim_phys.rev.b_65_165317_2002, stockman_phys.rev.lett.64.2215_1990, dorr_phys.rev.lett_80_1030_1998, Sridhaar_phys.rev.lett_67_785_1991, Stein_phys.rev.lett_68_2867_1992, nockel_nature_385_45_1997, Lee_phys.rev.lett_88_033903_2002, Harayama_phys.rev.e_67_015207_2003, chinnery_phys.rev.e_53_272_1996} and simulations~\cite{Larson_phys.rev.a_87_013624_2013, Huang_phys.rev.lett_103_054101_2009, Xu_phys.rev.lett_110_064102_2013, xuan_phys.rev.E_86_016702_2012, song_phys_rev.research_1_033008_2019,bogomolny_phys.rev.lett_92_244102_2004, aberg_phys.rev.lett_100_204101_2008, muller_j.phys.b_27_2693_1994,kudrolli_phys.rev.e_63_026208_2001}.
While quantum scars typically emerge in a closed system, an analog of scars do exist in open systems~\cite{wisniacki_phys.rev.E_77_045201_2008, akis_phys.rev.lett_79_123_1997, burke_phys.rev.lett_104_176801_2010} which are connected to the concept of a pointer state~\cite{zurek_rev.mod.phys_75_715_2003, ferry_phys.rev.lett_93_026803_2004, brunner_phys.rev.lett_101_024102_2008}, and can have a significant effect on the transport properties~\cite{Huang_phys.rev.lett_103_054101_2009, kaplan_phys.rev.e_59_5325_1999, fromhold_phys.rev.lett_72_2608_1994, baranger_phys.rev.lett_70_3876_1993,marcus_phys.rev.lett_69_506_1992, fromhold_phys.rev.lett_75_1142_1995, jalabert_phys.rev.lett_65_2442_1990}. In fact, from the properties of scars in the closed system, one can infer the extent of conductance fluctuations in the corresponding open system. 

In addition to the original narrative above, two new chapters have recently been opened in the saga of quantum scarring. A new field of quantum many-body scaring has emerged from the experimental observation of weak ergodicity breaking in Rydberg atom quantum simulators~\cite{bernien_nature_551_579_2017, turner_nat.phys_14_745_2018, Ho_phys.rev.lett_122_040603_2019, serbyn_nat.phys_17_675_2021}, and in optical lattices~\cite{scherg_nat.commun_12_1_2021, zhao_phys.rev.lett_124_160604_2020} that has yielded a deluxe of theoretical knowledge upon the topic, including an ambition to connect many-body scars to the conventional scarring~\cite{Hummel_phys.rev.Lett_130_250402_2023, Evrard_phys.rev.lett_132_020401_2024}. These special states in a many-body Hilbert space evade thermalization at finite energy densities producing surprising persistent oscillations of local observables without relying on the aspects of (near-)integrability or the protection given by a global symmetry (see, e.g., Refs.~\cite{alhambra_phys.rev.b_101_205107_2020, zhao_phys.rev.lett_124_160604_2020, pai_phys.rev.lett_123_136401_2019, bull_phys.rev.lett_123_030601_2019, ok_phys.rev.research_1_033144_2019, mukherjee_phys.rev.b_101_245107_2020, mark_phys.rev.b_101_195131_2020, hudomal_commun.phys_3_1_2020, voorden_phys.rev.b_101_220305_2020, lee_phys.rev.b_101_241111_2020, shiraishi_phys.rev.lett_119_030601_2017, moudgalya_phys.rev.b_98_235156_2018, lin_phys.rev.lett_122_173401_2019, chattopadhyay_phys.rev.b_101_174308_2020, mizuta_phys.rev.research_2_033284_2020, moudgalya_phys.revb_102_085140_2020, moudgalya_phys.rev.b_102_085120_2020, iadecola_phys.rev.b_101_024306_2020, choi_phys.rev.lett_122_220603_2019, bull_phys.rev.b_101_165139_2020, odea_phys.rev.research_2_043305_2020, khemani_phys.rev.b_99_161101_2019, yao_phys.rev.b_105_125123_2022, turner_phys.rev.x_11_021021_2021, michailidis_phys.rev.x_10_011055_2020}). In general, this hybrid phenomenon of quantum scarring and many-body localization could have profound implications for fundamental and practical reason, such as in quantum-boosted metrology~\cite{desaules_phys.rev.lett_129_020601_2022}. In this regard, a key question has been the fate of many-body scarring under disorder present in realistic quantum devices~\cite{mondrago_PRXQuantum_2_030349_2021, michalidis_phys.rev.research_2_022065_2020, turner_phys.rev.b_98_155134_2018}.  

On the other hand, the deeper understanding about the role of disorder in two-dimensional nanostructures was the initial spark for the second new branch of quantum scarring~\cite{Luukko_sci.rep_6_37656_2016, keski-rahkonen_phys.rev.b_97_094204_2017, keski-rahkonen_j.phys.conden.matter_31_105301_2019, keski-rahkonen_phys.rev.lett_123_214101_2019, luukko_phys.rev.lett_119_203001_2017} that we consider here in this Letter. In this phenomenon, the scars in a \emph{perturbated quantum} system are however formed around POs of the corresponding \emph{classical unperturbed} counterpart (see Fig.~\ref{fig:Pentagram_scar}). This kind of scarring, coined the epithet variational, has fundamentally distinct mechanism compared to the conventional scars~\cite{keski-rahkonen_phys.rev.lett_123_214101_2019}: these perturbation-induced (PI) scars arise as a a unique corpus of a special near-degeneracy structure and the localized nature of a perturbation. Overall, PI scars are not a rare occurrence, instead they thrive in various disordered potential landscapes~\cite{Luukko_sci.rep_6_37656_2016, keski-rahkonen_phys.rev.b_97_094204_2017, keski-rahkonen_j.phys.conden.matter_31_105301_2019,keski-rahkonen_phys.rev.lett_123_214101_2019, luukko_phys.rev.lett_119_203001_2017}.  

From a more pragmatic viewpoint, if high-energy eigenstates of a generic QD were indeed featureless and random due to chaos, e.g., stemming from the effect of the boundary (see, e.g., Refs.~\cite{kroetz_phys.rev.E_94_022218_2016, kang_phys.rev.lett_83_4144_1999, marcus_chaos_3_643_1993, Ketzmerick_phys.rev.b_54_10841_1996, Sachrajda_phys.rev.lett_80_1948_1998}) or the magnetic field (see, e.g., Refs.~\cite{nakamura_phys.rev.lett_61_247_1988, magnusdottir_Phys.Rev.B.61.10229_2000, zhen_phys.rev.B_52_1745_1995}), or from different kind of impurities and defects (see, e.g. Refs.~\cite{rasanen_phys.rev.B_70_115308_2004, guclu_phys.rev.B_68_035304_2003, halonen_phys.rev.B_53_6971_1996}), controlled applications in this regime would be tedious to realize. However, PI scars provide a way to regulate this chaos by suppressing it locally in the shape of classical POs. It has been demonstrated~\cite{Luukko_sci.rep_6_37656_2016} that PI scars can be utilized to efficiently propagate quantum wave packets in a \emph{perturbed} system, counterintuitively even with higher fidelity than in the \emph{unperturbed} counterpart containing no scars. Besides being a common feature in nanostructures plagued by impurities, PI scars also have a salient property being highly controllable. First, the existence and geometry of the scars can be tuned by adjusting the confining potential~\cite{keski-rahkonen_phys.rev.lett_123_214101_2019} or with an external magnetic field~\cite{keski-rahkonen_phys.rev.b_97_094204_2017}. In addition to random scattered impurities, a distinct PI scar in a QD can arise from a single focused perturbative potential, generated in a controlled manner by, e.g., a conducting nanotip, which can pin the scar into the desired orientation.~\cite{keski-rahkonen_phys.rev.b_97_094204_2017} Combined, these two features may open a pathway into coherently modulating quantum transport in nanoscale devices by exploiting the scarring -- into scartronics, with significant potential as a resource in the quantum-enhanced nanoelectonics of tomorrow. 
 
In this paper, we however investigate the inevitable consequence of quantum scars, namely \emph{antiscarring}~\cite{Heller_book_2008}: a depression of the probability density in a quantum state along the path of the scar-generating PO. For instance, it has been shown~\cite{kaplan_phys.rev.e_59_5325_1999} that the existence of antiscarred states yields anomalously long escape times in some decay processes. Despite of the proliferation on the list of possible antiscaring effects (see, e.g., Refs.~\cite{Pilatowsky_New.J.Phys_23_033045_2021, Lee_ann.phys_307_392_2003, Bies_phys.rev.e_64_016204_2001, Wang_phys.rev.e_63_056208_2001}), antiscarring as a phenomenon has remained elusive, in particular being a lingering puzzle in the context of PI scarring.

Reflecting upon the quandary around the subject, our pursuit is therefore to settle the dilemma whether ergodicity is totally suppressed by scars in quantum-chaotic systems, and to constitute a solid foundation for further studies on antiscarring, and relating topics, such as eigenfunction thermalization~\cite{Deutsch_Phys.Rev.A_43_2046_1991, Srdnicki_Phys.Rev.E_50_888_1994, Rigol_Nature_452_854_2008, Polkovnikov_Rev.Mod.Phys_83_863_2011, Deutsch_rep.prog.phys_81_082001_2018}. With this motive as our guiding directive, the paper is organized as follows: in the following Sec. II, the investigated system of a quantum well affected by impurities is described in detail, and a brief theoretical outline for PI scars in the selected system is given. In Sec. III, we then report and characterize scarring-enhanced localization of the probability density in some high-energy eigenstates of the disordered quantum well. In particular, we observe strong PI scars with the same geometry and orientation in multiple eigenstates within moderately large energy windows; whereas its classical analog exhibits chaotic behavior. This ostensible quantum-classical contradiction segues us into the central theme of the manuscript, i.e., the enigmatic antiscarring that is presented in two parts in Sec. IV: First, we shall present our stacking theorem of quantum ergodicity, and then gauge the necessary of antiscarring on the account of the theorem in the presence of quantum scars. Next, besides being operative in all forms of scarring, we analyze and explicitly demonstrate the effect of antiscarring in the selected test framework of PI scars. Finally, in Sec. V, we draw some conclusions with future perspectives, illuminating some of widely held views on the quantum nature of ergodicity, and conclude with the brief summary in Sec. VI.

\section{Perturbed Quantum Well}
 
To elucidate and further investigate antiscarring, we consider a disordered quantum well (QW) described by the following generic single-electron Hamiltonian without spin given in atomic units:
\begin{equation}\label{Hamiltonian}
    \mathcal{H} = -\frac{1}{2m}\nabla^2  + V_{\textrm{ext}} + V_{\textrm{imp}}.
\end{equation}

The external confining potential $V_{\textrm{ext}}$ defines the geometry of a QW. In the absence of impurities, it also determines the energy scales and the integrability of the system. Among various confinements, the power-law potentials $V_{\textrm{ext}} \propto r^{d}$ with integer $d > 0$, are particularly convenient for investigating the classical POs of the underlining unperturbed system. In the following, we focus on the potential of
\begin{equation*}
    V_{\textrm{ext}} = \frac{1}{2} \lambda r^{5},
\end{equation*}
that has become an archetypical system for studying PI scars; it is in many ways a natural “sweet spot” for studying the phenomenon as elaborated below (see also Ref.~\cite{Luukko_sci.rep_6_37656_2016}). 

The local perturbations can be modeled by adding randomly located bumps to the otherwise smooth confining potential $V_{\textrm{ext}}\propto r^5$, thus the total perturbation to be 
\begin{equation*}
    V_{\textrm{imp}} = A \sum_i \exp\left({\frac{\vert \mathbf{r} - \mathbf{r}_i \vert^2}{2\sigma^2}}\right),
\end{equation*}
where the impurities are distributed over the confining potential in uniform manner of the average impurity density being two per unit square. The individual bumps are assumed to be Gaussian-like with amplitude $A$ and width $\sigma$ that are set to $A = 24$ and $\sigma = 0.1$ yielding strong PI scarring within the considered impurity density. A similar disorder model of randomly scattered bumps has been studied with density-functional theory~\cite{hirose_phys.rev.B_65_193305_2002, hirose_phys.Rev.B_63_075301_2001} and the diffusive quantum Monte-Carlo approach~\cite{guclu_phys.rev.B_68_035304_2003}. Furthermore, a Gaussian bump is a good approximation for a perturbation caused by a conducting nanotip~\cite{blasi_phys.rev.B_87_241303_2013, boyd_nanotechnology_22_185201_2011,bleszynski_nano.lett_7_2559_2007}. The role of such external impurities in QDs can be quantitatively identified through the measured differential magnetoconductance displaying the quantum eigenstates~\cite{rasanen_phys.rev.B_70_115308_2004}.

In general, our Hamiltonian above has direct experimental relevance as being a prototype model for semiconductor quantum wells influenced by impurities (see, e.g., Refs.~\cite{reimann_rev.mod.phys_74_1283_2002, rasanen_phys.rev.B_70_115308_2004, kouwenhoven_rep.prog.phys_64_701_2001, bruce_phys.rev.B_61_4718_2000, stopa_phys.rev.B_54_13767_1996, rogge_phys.rev.lett_105_046802_2010, rasanen_phys.rev.B_77_041302_2008}). It also offers an excellent platform as a quantum counterpart of classical billiard with realistic soft walls, to probe the nature of quantum chaos~\cite{Stockmann_book, nakamura_2004_QD}, e.g., with a statistical analysis of the energy levels in addition to investigate PI scarring and ergodicity, like in Ref.~\cite{keski-rahkonen_j.phys.conden.matter_31_105301_2019}. Besides indirectly observing conductance fluctuations caused by scarred states, open QWs are suitable for wave function imaging based on shifts in the energy of the single-particle resonances, induced by an AFM tip~\cite{mendoze_phys.rev.B_68_205302_2003, zozoulenko_phys.rev.B_58_10597_1998, bird_2013_QD}. In addition, the scarred eigenstates of an electron in a QW may be measured with quantum tomography~\cite{jullien_nature_514_603_2014}. Outside of QWs, another suggested route towards a proof-of-principle experiment is to realize an analog condition in an optical system~\cite{keski-rahkonen_phys.rev.lett_123_214101_2019}, a frequently employed avenue to observe conventional quantum scars (see, e.g., Refs.~\cite{Sridhaar_phys.rev.lett_67_785_1991, Stein_phys.rev.lett_68_2867_1992, Lee_phys.rev.lett_88_033903_2002, Harayama_phys.rev.e_67_015207_2003, nockel_nature_385_45_1997, dorr_phys.rev.lett_80_1030_1998, stockman_phys.rev.lett.64.2215_1990}).

While further experimental evidence on scarring will be accrued, our Hamiltonian also presents an interesting framework to study the phenomenon and its consequences from a theoretical perspective. For instance, the chosen confinement potential $V_{\textrm{ext}} \propto r^5$ belongs to the special class of functions exhibiting multiplicative scaling behaviour, often referred as homogeneous. In a homogeneous, circular potential, POs at different energies are similar up to a scaling in space and time. Thus, the geometry of a PO is independent of its energy. Furthermore, due to the circular symmetry, different POs can be easily enumerated with two integers $a/b$ (see, e.g., Refs.~\cite{Goldstein_book, reynolds2010closed, kotkin2020exploring}): after $a$ oscillations around the radial turning points, the particle has traveled around the origin $b$ times before returning to its original configuration. In our specific potential, a distinct non-trivial PO, i.e., excluding circular orbits and the bouncing balls corresponding to the zero angular momentum case, is a five-pointed star: the orbit closes on itself after two rounds around the origin ($a =2$) in the time of five radial oscillation ($b = 5$). In addition, there exist other POs, such as $3/7$ and $5/11$, but they are longer and their role will be less important in our later analysis. For comparison, the shortest non-trivial POs are longer in the wells of $d=1,3,4$; whereas the harmonic potential ($d=2$) is too special for drawing general conclusions, even though an interesting case for studying scarring and its possible applications~\cite{keski-rahkonen_phys.rev.lett_123_214101_2019,keski-rahkonen_phys.rev.b_97_094204_2017}. 

When introduced to the disorder $V_{\textrm{imp}}$ caused by local potential bumps, our classical system will become \emph{mixed} in the sense that it exhibits both chaotic and regular behavior. In general, chaos settles as the relative size of the non-integrable part of the classical Hamiltonian is increased according to the Kolmogorov-Arnold-Moser (KAM) theorem~\cite{Kolmogorov_dokl.akad.nauk_93_763_1953, Arnold_uspehi.mat.nauk_18_13_1963,Moser_nachr.akad.wiss_II_1_1962}. In our prototype set-up, the nature of chaos is characterized by two facts. First, the bumps are relatively small in size $(\sigma = 0.1)$, and scattered with the low-enough impurity density to preclude the bumps overlapping and thus forming ''superbumps''. In other words, the bumps can be thus seen as individual, spatially localized perturbations in the confining potential. However, in the energy range considered here, $E \sim 500$, tens of bumps exist in the classically allowed region. Second, the chosen amplitude of the bumps ($A = 24$) is small in comparison to the considered energy scale of $E \sim 500$, thus setting each bump as a small, local perturbation in the confinement potential. In the end, together the bumps nevertheless form sufficient perturbation enough to destroy classical long-time stability. In particular, any stable structures  (KAM islands) present in the otherwise chaotic Poincaré surface of section are small or only comparable to $\hbar = 1$. Thus, our perturbed QW is by far too chaotic to support distinct scars in the conventional fashion~\cite{Heller_phys.rev.lett_53_1515_1984, Heller_book_2008,kaplan_ann.phys_264_171_1998, Kaplan_nonlinearity_12_R1_1999} where a scarred state corresponds to a moderately unstable PO in the classical counterpart.

On the quantum side, the eigenstates $\vert r, m\rangle$ of the unperturbed, circularly symmetric system are labeled by two quantum numbers $(r, m)$, corresponding to radial and angular motion, respectively. In the absence of a magnetic field, the states $\vert r, \pm m \rangle$ are exactly degenerate. Moreover, there are also near-degeneracies, or quasi-degeneracies, intimately connected to classical POs: According to the Bohr-Sommerfeld quantization, if a state defined by the quantum numbers $(r, m)$ is nearby in action to a classical PO with a ratio $a/b$ in the radial and angular oscillation frequencies, the relative states $\vert r + ka, m - kb\rangle$, where $k \in \mathbf{N}$, will consequently be nearby in energy. In general, the smaller values of $a$ and $b$ yield more accurate near-degeneracy, which will also result in stronger scarring. These groups of nearly-degenerate states are colloquially referred as a resonant set, which is the first ingredient on the scar recipe.

\begin{figure}[h!]
    \centering
    \includegraphics[width = 1.0\linewidth]{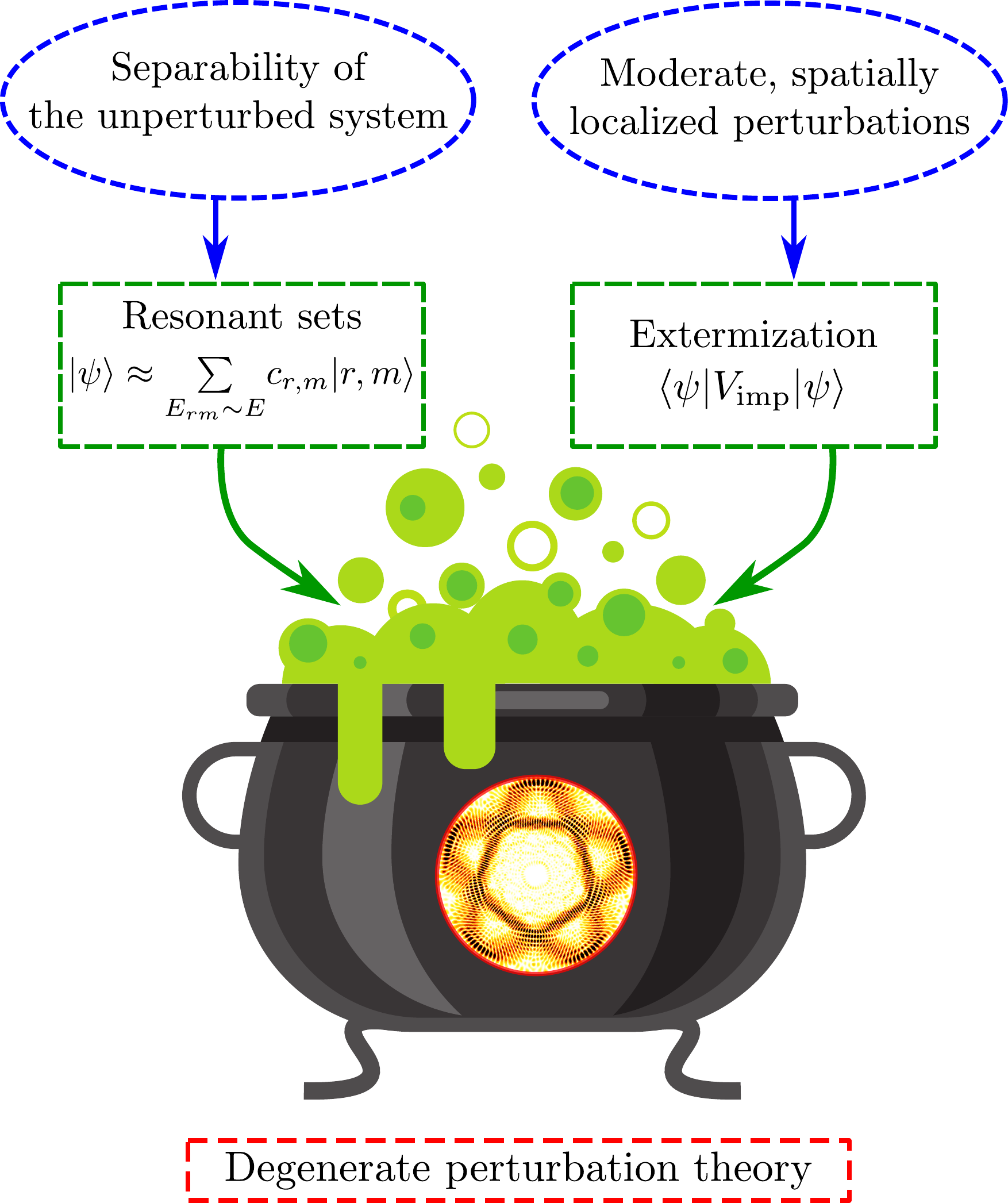}
    \caption{Variational scarring: To bake up perturbation-induced scars, such as shown in Fig.~\ref{fig:Pentagram_scar}, the degenerate-perturbation-theory recipe consists of two general ingredients: the separibility of the unperturbed system, and moderate, localized perturbations. First, due the separability, there are special near-degeneracies, so-called resonant sets, in the eigenstates $\vert r, m \rangle$ of the unperturbed system that are connected to the periodic orbits of its classical counterpart. In particular, a sufficiently minor perturbation will generate linear combinations $\vert \psi \rangle$ that are mostly contained in the near-degenerate part of some resonant set ($E_{rm} \sim E$). Second, scarred eigenstates are preferred since the total perturbation $V_{\textrm{imp}}$ is composed of individual perturbations narrow in space, e.g., separated into distinct bumps scattered through the confinement potential. Second, driven by the variational principle, an effective way to extremize the expectation value of the perturbation $\langle \psi \vert V_{\textrm{imp}} \vert \psi \rangle$ is achieved by selecting a scarred state, and tuning its orientation coinciding with as many or as few spatially localized perturbations as possible. This kind of heuristic argumentation founded in the perturbation theory explains well the existence of scarred states, and their properties, such as the preferred orientations discussed later.} 
    \label{fig:Scarring_mechanism}
\end{figure}

Like illustrated in Fig.~\ref{fig:Scarring_mechanism}, the second ingredient is a set of impurities localized in space that is mixed with the presence of resonant sets in the cauldron of (degenerate) perturbation theory to cook up scarred eigenstates. In a more precise manner, a superposition of the states from a common resonant set will exhibit beating in both the radial and angular directions. Because the beat frequencies of the states are in sync, the interference pattern will trace out the shape of the classical PO of the classical resonant $a/b$. Subsequently, an adequately perturbation generate eigenstates that are linear combinations of a single resonant set. Based on the variational theorem, the states corresponding to extremal eigenvalues extremize the perturbed Hamiltonian. Because the states in a resonant set are (nearly) degenerate, this basically means extremizing the perturbation. In the extremization, the system prefers the scarred states since the bumps causing the perturbation are localized. Thus, scarred states can effectively maximize (minimize) the perturbation by selecting paths coinciding with as many (few) bumps as possible. As a result, the extremal eigenstates arising from each resonant set often contain scars of the corresponding PO.

As a general remark, we want to point out the ubiquitous nature of PI scarring. First, our heuristic line of reasoning above goes beyond the circular QW considered here, for instance, substantiated in Refs.~\cite{luukko_phys.rev.lett_119_203001_2017, keski-rahkonen_phys.rev.lett_123_214101_2019}. Second, our argumentation does not assert if the bumps are repulsive or attractive. Although we focus on the repulsive bumps, PI scars persist also in perturbed potential landscapes with attractive dips. On the other hand, our argumentation does not quantify the localization of the bumps for PI scarring to occur; the specific details depend on the given system, e.g., on the density, width, and amplitude of the bumps along with the level of the near degeneracy in the unperturbed quantum system. Finally, we stress the robustness of PI scars against the changes in system features, such as a deviation from the optimal bump parameters~\cite{Luukko_sci.rep_6_37656_2016, keski-rahkonen_j.phys.conden.matter_31_105301_2019}, or the inclusion of an external magnetic field~\cite{keski-rahkonen_phys.rev.b_97_094204_2017, keski-rahkonen_j.phys.conden.matter_31_105301_2019}.

\section{Scar Observations}

As expected from the scar phenomenology above, when perturbed by randomly positioned Gaussian-like bumps, some of the high-energy eigenstates of the QW are strongly scarred by POs of the unperturbed system. Figure~\ref{fig:Pentagram_scar} shows an example of a scar found in the eigenstates of the prototype system of our interest. To be specific, we solve the four thousand lowest eigenstates and energies of the Hamiltonian of Eq.~\ref{Hamiltonian} by employing the \texttt{itp2d} software~\cite{luukko_comput.phys.commun_184_769_2013} based on the imaginary time propagation method~\cite{aichinger_comput.phys.commun_171_197_2005}. 

\begin{figure}[h!]
    \centering
    \includegraphics[width=0.95\linewidth]{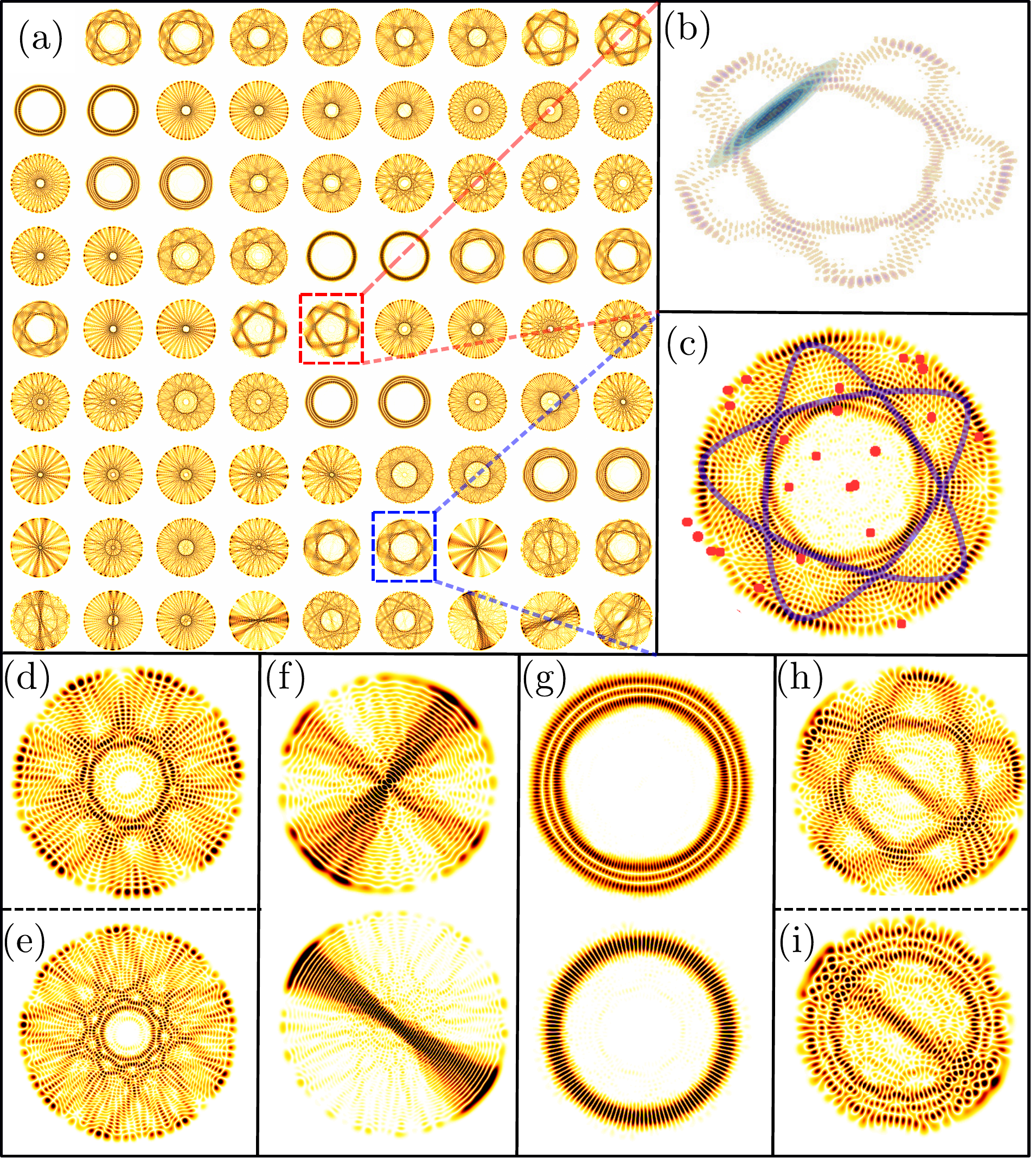}
    \caption{Kaleidoscope of high-energy eigenstates in the prototype potential well perturbed by Gaussian bumps. Subplot (a) shows the probability densities of 80 eigenstates lying in the neighborhood of the scar presented in Fig.~\ref{fig:Pentagram_scar}, indicated by a red box. As illustrated in (b), this state is taken a reference scar to prepare a Gaussian wavepacket for analyzing the stability and existence of preferred scar orientations (see Fig.~\ref{fig:Preferred_orientations}), and for the scarmometer (see Fig.~\ref{fig:Scarmometer}). In addition to pentagram scars, such as in (b) oriented to overlap with the bumps as much as possible, there are counter scars which have instead aligned to minimize the perturbation. An example of this kind of antipodal state is tagged with a blue box in (a), and enlarged in (b) along with the corresponding PO of the unperturbed potential drawn as a solid blue line and red markers denoting the locations of the bumps. The eigenstate carpet (a) reveals also other interesting features than the pentagrams emphasized in (b) and (c). For instance, there are states scarred longer POs than associated with the $5/2$ resonance: Subplots (d) and (e) portray two scars connected to classical resonances $7/3$ and $9/4$ appearing in the form of a heptagram and enneagram that are the second and third shortest nontrivial POs of the unperturbed system, respectively. Besides various scars, there are several bouncing-ball states, such as shown in (f), and ring-like states, called remnants, such as shown in (g). In addition to these kind of single geometrical shapes, some states are mixtures of the defined echelons. For example, subplots (h) and (i) display a crossbreed of bouncing ball with a pentagram scar and remnant, respectively.    
    }
    \label{fig:Scar_examples}
\end{figure}

Like present in the menagerie of Fig.~\ref{fig:Scar_examples}a), some of the numerically solved states are very strongly scarred. For instance, in our paragon shown in Fig.~\ref{fig:Pentagram_scar} almost $80\%$ of the probability density resides in the vicinity of the pentagon path. In Fig.~\ref{fig:Scar_examples}b), the same state is re-illustrated with a Gaussian wavepacket which is designed to align with the pentagram for evaluating the level of scarring below.
As visible in Fig.~\ref{fig:Pentagram_scar}, the pentagram in question is oriented so that it maximizes the overlap with the bump, but these type of pinning scars are partnered by ``counter-scarred'' states that instead minimize the perturbation by avoiding bumps, an example presented in Fig.~\ref{fig:Scar_examples}c). Moreover, as also palpable in Fig.~\ref{fig:Pentagram_scar}, in the energy zone of order$\sim 500$, the local wavelength of the scarred eigenstates is comparable to the full width at half maximum of the Gaussian bumps, which is $2\sigma\sqrt{2 \textrm{ln}(2)} \approx 0.235$. For clarity, we want to point out that scars are not a rare occurrence. Although the exact proportion and strength of scars alter between different random realizations of bumpy potential landscapes, we note that scarred states among all the first 4000 eigenstates seem to vary from $10\, \%$ up to $60\, \%$ at the considered impurity parameters of $A = 24$ and $\sigma = 0.1$. In this work, we have not attempted to specify quantitatively how common the scars or preferred orientations are among all random realizations of the perturbation locations.

Besides the most evident five-pointed-star geometry by the PO of resonance $2/5$, we as well observe scarred states connected to higher-order classical resonances, for example, Figs.~\ref{fig:Scar_examples}d) and e) show two scars reflecting the PO of $7/13$ and $9/4$ resonance, respectively. Furthermore, excluded from the scar taxonomy into a separative family, some states contain features of classical ``bouncing-ball-like'' motion, as seen in Fig.~\ref{fig:Scar_examples}f). Another respective category of eigenstates encompasses remnants, or lingering phantoms of the destroyed circular symmetry of the unperturbed quantum system, as illustrated in Fig.~\ref{fig:Scar_examples}g). These remnant states are related to the eigenstates of the unperturbed system that are weakly affected by the applied perturbation $V_{\textrm{imp}}$. Blurring the boundaries in this kind of crude classes, there also exist mixes states, such as a combination of a scar and a bouncing-ball state (see Fig.~\ref{fig:Scar_examples}h), or a merge of a bouncing-ball and a remnant state (see Fig.~\ref{fig:Scar_examples}i). There is also a possibility that a state contains a trace of two scars.

Prior to venturing back to the observed classical beings living in the quantum-chaotic realm, we want to emphasize two points. First, the observed scars cannot be explained by dynamical localization~\cite{casati_dynamical_localization, germpel_phys.rev.A_29_1639, shepelyansky_physica.d_28_103_1987, izrailev_phys.rep_196__299_1990}: it corresponds to localization in angular momentum space, whereas the scars are localized in position space. Second, a similar reconstruction of classical-like states from (nearly) degenerate basis states has also been considered previously in Refs.~\cite{liu_phys.rev.E_74_046214_2006, chen_phys.rev.E_66_046215_2002, li_phys.rev.E_65_056220_2002, kumar_j.phys.A_41_075306_2008, chen._phys.rev.A_83_032124_2011, chen_phys.rev.E_72_056210_2005, chen_j.phys.A_36_7751_2003, lee_j.phys.A_41_275102_2008}, but this kind of pseudo-scarring is substantially weaker to the PI scarring reported here; such scars being rare with no indication on the role of local perturbations. In the same vein, either dynamical localization nor pseudo-scarring are able to explain that PI scars generally orient to coincide with as many bumps as possible, and have preferred orientations.

The indicated existence of these special scar orientations can be demonstrated by identifying scarred states with wavepackets. Inspired by the studies of conventional scarring~\cite{Heller_book_2008}, we can take advantage of a test wavepacket initialized on a PO to pin point eigenstates scarred by a given PO. In extension, the same strategy can be further utilized to systemically analyze preferred scar orientations. Here, we have selected a squeezed Gaussian wavepacket $\vert \phi(\alpha) \rangle$ engineered according to the chosen reference scar in Fig.~\ref{fig:Pentagram_scar}. The orientation of the pentagram orbit employed to devise the probe wavepacket is described by the coordinate $\alpha$: it is determined in a such way that the wavepacket aligns with the positive y-axis $\alpha = 0$, rotating clockwise with increasing angle. For example, in Fig.~\ref{fig:Scar_examples}(b), we present our test Gaussian superimposed over the reference scar of Fig.~\ref{fig:Pentagram_scar} that has been employed to design the wavepacket. In this particular case, the angle $\alpha$ is $\tilde{\alpha} = 13\pi/50 \approx 0.80$.  

\begin{figure}[h!]
    \centering
    \includegraphics[width=1.0\linewidth]{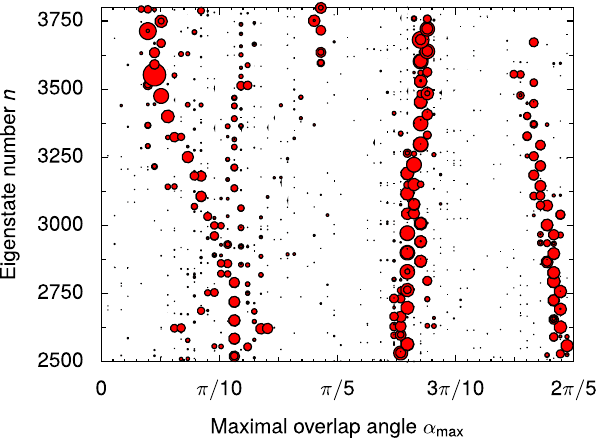}
    \caption{Scatter plot demonstrating the existence and stability of the preferred scar orientations for $n = 2500 \cdots 3800$ which converts into the energy range of $458 \cdots 618$. For each eigenstate, a circle is marked on the orientation angle $\alpha$ determined by the highest overlap between the eigenstate and the reference Gaussian wavepacket presented in Fig.~\ref{fig:Scar_examples}. The radius of the circle depicts the squared magnitude of the overlap. In the figure, three branches of high overlaps are visible that account to the preferred scar orientations. Whereas the leftmost and rightmost branches evolve relatively slowly in angle implying rotation in the corresponding scar orientations, the middle branch is instead approximately upright within the whole energy window of interest. The upright branch of $\tilde{\alpha} \approx 13\pi/50$ stands for the reference state shown in Fig.~\ref{fig:Scar_examples} that is further employed to pick up states scarred in the similar fashion within the framework of scarmometer, embellished in Fig.~\ref{fig:Scarmometer}.} 
    \label{fig:Preferred_orientations}
\end{figure}

First, we consider the maximal orientation angle $\alpha_{\textrm{max}}$ yielding the highest overlap between a target eigenstate and the probe Gaussian wavepacket $\vert \phi(\alpha) \rangle$ with the given orientation. Figure~\ref{fig:Preferred_orientations} shows the overlap of the wavepacket with the target eigenstates of $n = 2500 \dots 3800$ as function of the angle $\alpha_{\textrm{max}}$ within an angular window of $2\Pi/5$ after which the PO is the same. The considered target states cover an energy range of $458 \cdot 618$. For each eigenstate, a circle is marked on the maximal angle with the radius determined by the squared magnitude of its overlap. For this particular realization of the random bumps, three orientations are clearly preferred. The rightmost and leftmost branches undergo angle drifting as a function of energy: the later in a less dramatic degree; whereas the former also seem to spill out two weaker sub-branches. On the other hand, the middle branch is stable and close to vertical, evincing that the corresponding scars share the same orientation. 

The robustness of the preferred orientations revealed in Fig.~\ref{fig:Preferred_orientations} can be understood within the framework of PI scarring outlined above, and illustrated in Fig.~\ref{fig:Scarring_mechanism}). By the previous argumentation, the scar orientations are mostly selected by the positions of the perturbations. Since the inner and outer radii of the POs change slowly with energy, the orientations that extremize the total perturbation $V_{\textrm{imp}}$ will be determined largely by the same impurities for many different resonant sets. In the particular case of Fig.~\ref{fig:Preferred_orientations}, the increase of energy results in the average radius of the PO swelling by roughly $0.24$ units, which is commensurate to the characteristic length scale of the individual bumps ($\sigma=0.1$).

\begin{figure}[h!]
\includegraphics[width=1\linewidth]{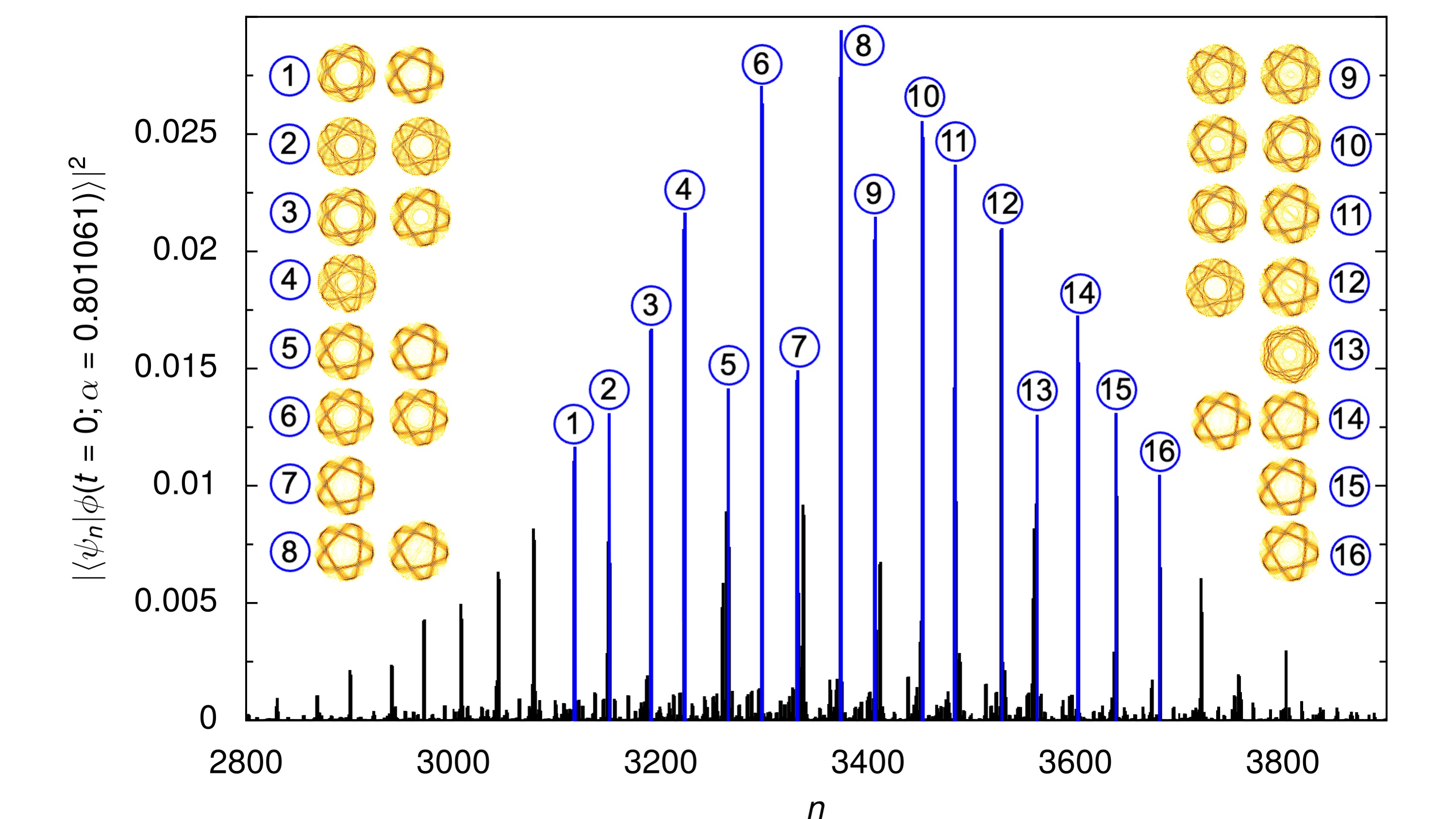}
\caption{Extracting scars with a specific geometry and orientation with a wavepacket ``scarmometer''. Figure shows the overlap of the eigenstates $\psi_n$ with a Gaussian wavepacket $\phi$ that is designed in respect to the scarred state $n= 3223$, as illustrated in Fig.~\ref{fig:Scar_examples}(b). For the strongest peaks with $\mathcal{F}(n) > 0.1$, the associated probability density of the eigenstate is also included, underlining that the shown states are all scarred to a varying degree by the five-pointed PO, but the orientation among the scars is same, as perused in Fig.~\ref{fig:Preferred_orientations}. The states situated between the scar peaks showcase a diverse array of patterns, like heptagram-shape scars, remnants and bouncing balls as illustrated in Fig.~\ref{fig:Scar_examples}.}
\label{fig:Scarmometer}
\end{figure}

Since the scars appear in the same preferred orientations across a wide energy range, a single wavepacket can cover many scarred eigenstates in its spectrum. In subsequent, a wavepacket initialized on a specific PO of the unperturbed system can be employed to pinpoint scars with a particular orientation in the perturbed system. This kind of method to isolate scars is illustrated in Fig.~\ref{fig:Scarmometer}. In particular, the average energy of the chosen test Gaussian $\phi$ as well its orientation, parametrized by a rotation angle $\alpha$, were set to matched with the sample scarred eigenstate of $n=3223$, shown in Fig.~\ref{fig:Pentagram_scar}. Further, we can also squeeze the wavepacket so that its full width at half maximum profile roughly corresponds to the width of the scar, thus yielding a sharper detection tool. 

Figure~\ref{fig:Scarmometer} shows this scar detector where the states scarred similarly to the target state $n=3223$ appearing as large overlaps with the test wavepacket illustrated in Fig.~\ref{fig:Scar_examples}(b):
\begin{equation}
    \mathcal{F}(n) = \vert \langle E_n \vert \phi(t_0 = 0; \alpha) \vert^2.
\end{equation}
In Fig.~~\ref{fig:Scarmometer}, we show the corresponding eigenstates respect to the largest peaks satisfying the criteria $\mathcal{F}(n) > 0.01$ (blue lines) that not only share the similar geometry but also the orientation. To clarify, there are this type of scars outside the considered energy window, but these states are not recognized by the given scarmometer for two main reasons: the size of the scars scales as the classical that is a function of energy, and at same time their orientation can change, thus yielding diminishing overlap with the test wavepacket. The states in-between the scar peaks show a rich collections of probability density patterns, as depicted in Fig.\ref{fig:Scar_examples}.

\onecolumngrid\
\twocolumngrid\

\section{Antiscarring}

In this section, we shall address a shadow cast by the observed PI scars -- \emph{antiscarring}. This is carried out in two steps: first providing an ergodicity theorem for a group of eigenstates, subsequently implies the existence of antiscarring that we exemplify in case of the studied disordered QW.    

\subsection{Stacking theorem}

To amalgamate the conceptual idea behind the existence of antiscarring, we first argue that the collective, average probability density of the eigenstates must eventually become ergodic in the following sense:

\begin{quote}
      \textbf{(Eigenstate) Stacking Theorem}: A sum of eigenstates of a Hamiltonian is uniform in phase space over a stacking window of $\Delta \varepsilon \geq \hbar/T_{\textrm{min}}$, where $T_{\textrm{min}}$ is the fundamental period of the shortest PO in the corresponding classical system.
\end{quote}

To prove the stacking theorem above, we can, without losing generality, start by selecting an eigenstate $\vert E_0 \rangle$ from the spectrum of the Hamiltonian describing the given quantum system. This state $\vert E_0 \rangle$ serves as a reference point to our stacking neighborhood $\Delta  E = \left[E_0 - \Delta \varepsilon/2, E_0 +\Delta \varepsilon/2 \right]$ where the energy window $\Delta \epsilon$ is chosen to be larger than set by the shortest PO  in the corresponding classical system ($\Delta \epsilon \gtrsim \hbar/T_{\textrm{min}}$). 

To proceed, we need to decide a method to measure the uniformity of a phase space. Adopting an insight presented in Ref.~\cite{Heller_book_2008}, a very convenient litmus test for the uniformity is to project all the eigenstates onto sufficiently spatially narrow Gaussians $\vert \Phi(t_0) \rangle$ over phase space (but by no means necessary option). These Gaussians are prepared so that they all share the energy expectation value of $E_0$ corresponding to our reference state $\vert E_0 \rangle$. Furthermore, 
we require that all the test states $\vert \Phi(t_0) \rangle$ have the same width of $\Delta \varepsilon$, thus preventing the time-evolved states to return and to overlap with themselves within the window of
\begin{equation}\label{eq:stacking_window}
    \Delta t = \frac{\hbar}{\Delta \varepsilon} \le T_{\textrm{min}}.
\end{equation}
In addition, the assumption of the equal dispersion for all test Gaussians ensures an eligible probe for the population of the phase space. 

Next we stack together eigenstates in our energy interval $\Delta E$ measuring them in respect to our test Gaussian as
 \begin{equation*}
 \begin{split}
     &\sum_{E_n \in \Delta E} \left \vert \langle \Phi(t_0) \vert E_n \rangle \right \vert^2\\
     &= \int_{\Delta E} \int_{-\infty}^{\infty} e^{i E t/\hbar} \langle \Phi(t_0) \vert \Phi(t) \rangle\, \textrm{d}t \, \textrm{d} E\\
     &= \int_{-\infty}^{\infty} \Omega_{\Delta t}(t) e^{i E_0 t/\hbar} P(t) \, \textrm{d} t,
     \end{split}
 \end{equation*}
 where $P(t) = \langle \Phi(t_0) \vert \Phi(t) \rangle$ is the fidelity of the test Gaussian, and variable energy window $\Delta E$ is implied by the finite time window $\Delta t$ in the transform limited cut-off function $\Omega_{\Delta t}(t)$. Notably, this integral carries only information about behavior the before any critical or interesting dynamics occurs, gleaned from the long-term system-specific details. 
 
By construction, the integral and hence the sum is the same for every test state $\vert \Psi(t_0) \rangle$. In other words, their population summing over the stack of states within $\Delta E$ is uniform. Furthermore, these test states are situated everywhere in phase space centered on the energy shell $E_0$. Therefore an sum over a sufficiently large range of eigenstates is uniform in phase space, as claimed by the theorem.

\subsection{Antiscars}

More interestingly, since there can be strongly scarred states among the eigenstates as illustrated in disordered QW above, the stacking theorem warrants the existence of antiscarred states, with low instead of high probability in the region of strong scars. However, this effect can be obscure at an single eigenstate level due to the multitude of various antiscarred states that support the scars. Therefore, in order to make this phenomenon more tangible, we construct the cumulative scar density
\begin{equation*}
    \rho^{\textrm{scar}}(\mathbf{r}) = \sum_{\substack{E_n \in F \\ \\ \mathcal{F}(n) > \mathcal{F}_c}} \vert \langle \mathbf{r} \vert E_n \rangle \vert^2
\end{equation*}
and in a complementary to it, the scar-reduced probability density
\begin{equation*}
    \rho^{\textrm{antiscar}}(\mathbf{r}) = \sum_{\substack{E_n \in F \\ \\ \mathcal{F}(n) < \mathcal{F}_c}} \vert \langle \mathbf{r} \vert E_n \rangle \vert^2
\end{equation*}
where $\Delta E$ is the stacking window and $\mathcal{F}_c$ is a criteria for (strongly) scarred stated according to the scarmometer. Notable, if the stacking window is large enough in the sense of Eq.~\ref{eq:stacking_window}, the sum of these densities $\rho^{\textrm{scar}}$ and $\rho^{\textrm{antiscar}}$ is uniform as dictated by the theorem proven above. 

\begin{figure}[h!]
\includegraphics[width=1\linewidth]{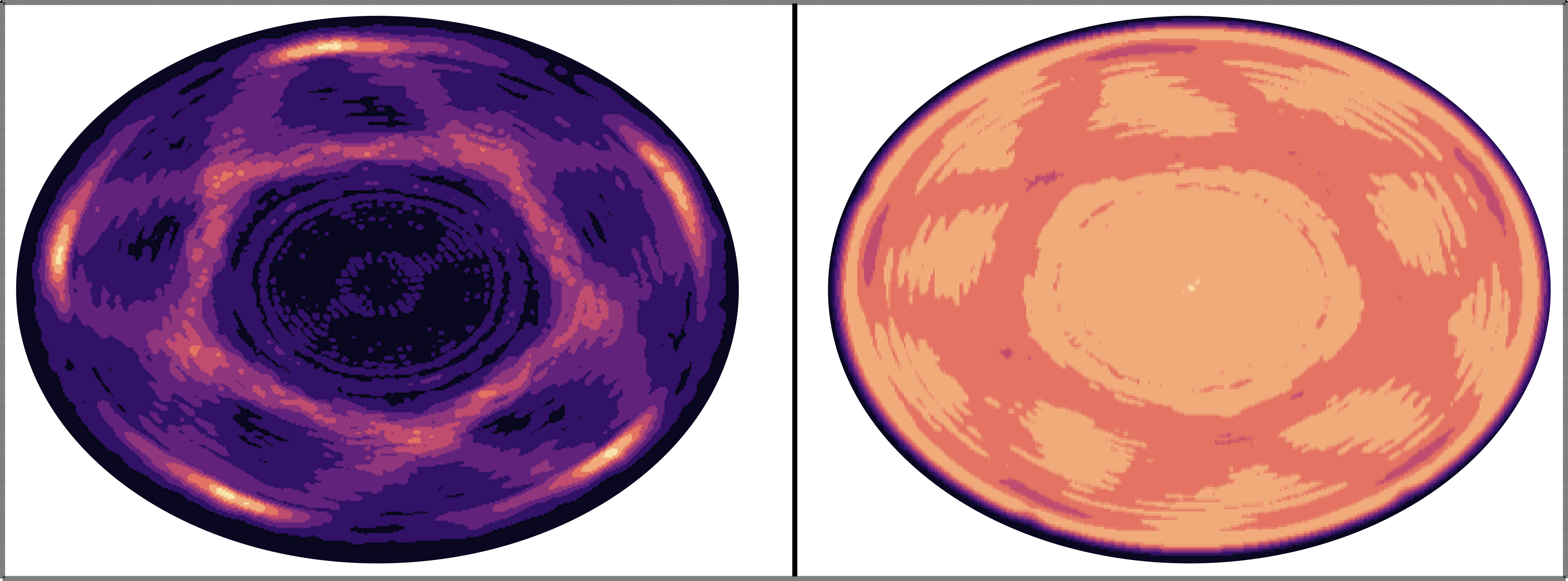}
\caption{Antiscarring: The left panel displays a cumulative density plot for a set of scarred eigenstates with a scarmometer value $\mathcal{F}(n)$ greater than $0.005$. These scars are strong and exhibit similar spatial orientations, resulting in a concentrated collective density along the five-point periodic orbit. In contrast, the right panel shows the remaining non-scarred states within the range of $n = 3000, \cdots, 3700$. While these non-scarred states do not exhibit pentagram-shaped patterns, their cumulative density closely resembles that of the scarred states down to minute details. This construction reveals the subtle impact of antiscarring on the non-scarred states.}
\label{fig:Antiscar}
\end{figure}

Figure~\ref{fig:Antiscar} illustrates these densities calculated for a specific portion of the system spectrum as shown in Fig.~\ref{fig:Scarmometer}. The scar criterion is defined as $\mathcal{F}_c = 0.005$ , represented by the red dashed line in Fig.~\ref{fig:Scarmometer}. This criterion effectively filters out pentagram-shaped scars from other types of states. The left panel of Fig.~\ref{fig:Antiscar} displays the cumulative density of scarred states $\rho^{\textrm{scar}}$, while the right panel showcases the scar-reduced density $\rho^{\textrm{antiscar}}$, which represents the cumulative density of states not identified by the scarmometer within the energy range of $n=3000, \cdots, 3700$. Given that the included scars exhibit a shared orientation, their collective density pattern reflects the periodic orbit responsible for the scarring. Notably, there are significant probability density accumulations at the tips and self-crossing point of the pentagram, where the classical particle spends more time on the orbit and is therefore more likely to be located. The blurriness observed at the tips is attributed to the slight rotational drift of the scars as a function of energy (see Fig.~\ref{fig:Preferred_orientations}). On the other hand, despite the diverse appearance of non-scarred states depicted in Fig.~\ref{fig:Scar_examples}, their cumulative density reveals a depression emerging alongside the scarring PO. Even more notably, the intricate details of this antiscar construction mirror precisely with the cumulative density composed of the identified scars. This observation validates the assertion of the stacking theorem that these two densities will precisely combine to form a uniform density.

\onecolumngrid\
\twocolumngrid\

In this works, we have only illustrated antiscarring specifically in a disordered quantum well, where the strength and similar orientation of perturbation-induced scars accentuates the effect. However, conventional scarring mechanisms also remain effective in this scenario. An important avenue of future research is to assessment the impact of antiscarring in the context of many-body scars. 

From an experimental standpoint, the presence of antiscarred states due to the conventional scars leads to unusually extended escape times in certain decay processes~\cite{kaplan_phys.rev.e_59_5325_1999}, making them possibly detectable in transport measurements. Additionally, open QWs offer a suitable platform for imaging wave functions by observing shifts in the energy of single-particle resonances induced by an atomic force microscopy tip~\cite{mendoza_phys.rev.b_68_205302_2003, zozoulenko_phys.rev.B_58_10597_1998, bird_2013_QD}, opening a possible pathway to detect antiscarring signatures. Moreover, the scarred eigenstates of an electron in a quantum well can be measured using quantum tomography~\cite{jullien_nature_514_603_2014} to study the fingerprints of antiscarring. On the other hand, we see that this phenomenon could be directly visualized through scanning tunneling microscopy in graphene-based quantum dots, in a manner discussed in Refs.~\cite{Ge_nano.lett_20_8682_2020, Ge_nano.lett_21_8993_2021}. Conversely, the antiscarring signatures could also be detected through classical wave experiments commonly used to visualize conventional scarring, such as those involving microwave cavities~\cite{Sridhaar_phys.rev.lett_67_785_1991, Stein_phys.rev.lett_68_2867_1992}, acoustic cavities~\cite{chinnery_phys.rev.e_53_272_1996}, fluid surface waves~\cite{kudrolli_phys.rev.e_63_026208_2001}, or optical systems~\cite{nockel_nature_385_45_1997, dorr_phys.rev.lett_80_1030_1998, stockman_phys.rev.lett.64.2215_1990}. Finally, we want to point out the potential presence of already observed antiscarring effects in simulations and experiments, such as Refs.~\cite{Pilatowsky_New.J.Phys_23_033045_2021, Lee_ann.phys_307_392_2003, Bies_phys.rev.e_64_016204_2001, Wang_phys.rev.e_63_056208_2001}

\section{Summary}

In conclusion, our results confirm a foundational quantum feature of ergodicity, which has been an quintessential concept in the edifice of classical chaos. Our main qualitative conclusion is the fact that an enhancement of a quantum scar is always accompanied by a compensation of depressed probability density in the neighborhood of the scar-tracing orbit in other eigenstates, colloquially known as antiscarring. Besides theoretically justifying the onset of the phenomenon, we explicitly demonstrate the antiscarring effect in a disordered quantum well that hosts multiple strong, pentagram-like scars sharing a common orientation within a large-enough energy window to support quantum ergodic behaviour.   

This work benchmarks and thereby solidifies the dual phenomenon of ubiquous quantum scarring. In the current epoch of single and many-body scarring, straightforward applications offer the possibility to enter unexplored regimes that give further insight into some of the most subtle and strange aspects of the quantum nature of chaos. At a more concrete level, the results may pave a way to enhance QW performance by utilizing scars in a controlled manner. In a bigger picture, we expect that our work will inspire future investigations on quantum scarring and on the quantum nature of ergodicity intimately connected to the eigenstate thermalization hypothesis, not to exclude the new window of shedding further light upon the conundrum called quantum chaos.

\begin{acknowledgments}
We thank J. Cotler, E. Räsänen, and L. Kaplan for useful discussions. Furthermore, J. K.-R. thanks the Emil Aaltonen Foundation and the Oskar Huttunen Foundation for financial support.
\end{acknowledgments}


\bibliography{references}

\begin{thebibliography}{171}%
\makeatletter
\providecommand \@ifxundefined [1]{%
 \@ifx{#1\undefined}
}%
\providecommand \@ifnum [1]{%
 \ifnum #1\expandafter \@firstoftwo
 \else \expandafter \@secondoftwo
 \fi
}%
\providecommand \@ifx [1]{%
 \ifx #1\expandafter \@firstoftwo
 \else \expandafter \@secondoftwo
 \fi
}%
\providecommand \natexlab [1]{#1}%
\providecommand \enquote  [1]{``#1''}%
\providecommand \bibnamefont  [1]{#1}%
\providecommand \bibfnamefont [1]{#1}%
\providecommand \citenamefont [1]{#1}%
\providecommand \href@noop [0]{\@secondoftwo}%
\providecommand \href [0]{\begingroup \@sanitize@url \@href}%
\providecommand \@href[1]{\@@startlink{#1}\@@href}%
\providecommand \@@href[1]{\endgroup#1\@@endlink}%
\providecommand \@sanitize@url [0]{\catcode `\\12\catcode `\$12\catcode
  `\&12\catcode `\#12\catcode `\^12\catcode `\_12\catcode `\%12\relax}%
\providecommand \@@startlink[1]{}%
\providecommand \@@endlink[0]{}%
\providecommand \url  [0]{\begingroup\@sanitize@url \@url }%
\providecommand \@url [1]{\endgroup\@href {#1}{\urlprefix }}%
\providecommand \urlprefix  [0]{URL }%
\providecommand \Eprint [0]{\href }%
\providecommand \doibase [0]{https://doi.org/}%
\providecommand \selectlanguage [0]{\@gobble}%
\providecommand \bibinfo  [0]{\@secondoftwo}%
\providecommand \bibfield  [0]{\@secondoftwo}%
\providecommand \translation [1]{[#1]}%
\providecommand \BibitemOpen [0]{}%
\providecommand \bibitemStop [0]{}%
\providecommand \bibitemNoStop [0]{.\EOS\space}%
\providecommand \EOS [0]{\spacefactor3000\relax}%
\providecommand \BibitemShut  [1]{\csname bibitem#1\endcsname}%
\let\auto@bib@innerbib\@empty
\bibitem [{\citenamefont {Strogatz}(2007)}]{Strogatz_book}%
  \BibitemOpen
  \bibfield  {author} {\bibinfo {author} {\bibfnamefont {S.}~\bibnamefont
  {Strogatz}},\ }\href@noop {} {\emph {\bibinfo {title} {Nonlinear Dynamics And
  Chaos}}},\ Studies in nonlinearity\ (\bibinfo  {publisher} {Sarat Book
  House},\ \bibinfo {year} {2007})\BibitemShut {NoStop}%
\bibitem [{\citenamefont {Lorenz}(1963)}]{lorentz_j.atmos.sci_20_130_1963}%
  \BibitemOpen
  \bibfield  {author} {\bibinfo {author} {\bibfnamefont {E.~N.}\ \bibnamefont
  {Lorenz}},\ }\bibfield  {title} {\bibinfo {title} {Deterministic nonperiodic
  flow},\ }\href {https://doi.org/10.1175/1520-0469(1963)020<0130:DNF>2.0.CO;2}
  {\bibfield  {journal} {\bibinfo  {journal} {J. Atmos. Sci.}\ }\textbf
  {\bibinfo {volume} {20}},\ \bibinfo {pages} {130} (\bibinfo {year}
  {1963})}\BibitemShut {NoStop}%
\bibitem [{\citenamefont
  {Einstein}(1917)}]{einstein_verh.dtsch.phys.ges_19_82_1917}%
  \BibitemOpen
  \bibfield  {author} {\bibinfo {author} {\bibfnamefont {A.}~\bibnamefont
  {Einstein}},\ }\bibfield  {title} {\bibinfo {title} {Zum quantensatz von
  sommerfeld und epstein},\ }\href@noop {} {\bibfield  {journal} {\bibinfo
  {journal} {Verh. Dtsch. Phys. Ges.}\ }\textbf {\bibinfo {volume} {19}},\
  \bibinfo {pages} {82} (\bibinfo {year} {1917})}\BibitemShut {NoStop}%
\bibitem [{\citenamefont {Stone}(2005)}]{stone_phys.today_58_37_2005}%
  \BibitemOpen
  \bibfield  {author} {\bibinfo {author} {\bibfnamefont {A.~D.}\ \bibnamefont
  {Stone}},\ }\bibfield  {title} {\bibinfo {title} {Einstein's unknown insight
  and the problem of quantizing chaos},\ }\href
  {https://doi.org/10.1063/1.2062917} {\bibfield  {journal} {\bibinfo
  {journal} {Phys. Today}\ }\textbf {\bibinfo {volume} {58}},\ \bibinfo {pages}
  {37} (\bibinfo {year} {2005})}\BibitemShut {NoStop}%
\bibitem [{\citenamefont
  {Berry}(1989{\natexlab{a}})}]{berry_phys.scripta_40_335_1989}%
  \BibitemOpen
  \bibfield  {author} {\bibinfo {author} {\bibfnamefont {M.}~\bibnamefont
  {Berry}},\ }\bibfield  {title} {\bibinfo {title} {Quantum chaology, not
  quantum chaos},\ }\href {http://stacks.iop.org/1402-4896/40/i=3/a=013}
  {\bibfield  {journal} {\bibinfo  {journal} {Phys. Scripta}\ }\textbf
  {\bibinfo {volume} {40}},\ \bibinfo {pages} {335} (\bibinfo {year}
  {1989}{\natexlab{a}})}\BibitemShut {NoStop}%
\bibitem [{\citenamefont {Heller}\ and\ \citenamefont
  {Tomsovic}(2008)}]{heller_phys.today_46_38_2008}%
  \BibitemOpen
  \bibfield  {author} {\bibinfo {author} {\bibfnamefont {E.~J.}\ \bibnamefont
  {Heller}}\ and\ \bibinfo {author} {\bibfnamefont {S.}~\bibnamefont
  {Tomsovic}},\ }\bibfield  {title} {\bibinfo {title} {Postmodern quantum
  mechanics},\ }\href {https://doi.org/10.1063/1.881358} {\bibfield  {journal}
  {\bibinfo  {journal} {Phys. Today}\ }\textbf {\bibinfo {volume} {46}},\
  \bibinfo {pages} {38} (\bibinfo {year} {2008})}\BibitemShut {NoStop}%
\bibitem [{\citenamefont
  {Jensen}(1992{\natexlab{a}})}]{jensen_nature_355_311_1992}%
  \BibitemOpen
  \bibfield  {author} {\bibinfo {author} {\bibfnamefont {R.}~\bibnamefont
  {Jensen}},\ }\bibfield  {title} {\bibinfo {title} {Quantum chaos},\ }\href
  {https://doi.org/10.1038/355311a0} {\bibfield  {journal} {\bibinfo  {journal}
  {Nature (London)}\ }\textbf {\bibinfo {volume} {355}},\ \bibinfo {pages}
  {311–318} (\bibinfo {year} {1992}{\natexlab{a}})}\BibitemShut {NoStop}%
\bibitem [{\citenamefont
  {Jensen}(1992{\natexlab{b}})}]{jensen_nature_355_591_1992}%
  \BibitemOpen
  \bibfield  {author} {\bibinfo {author} {\bibfnamefont {R.}~\bibnamefont
  {Jensen}},\ }\bibfield  {title} {\bibinfo {title} {Bringing order out of
  chaos},\ }\href {https://doi.org/10.1038/355591a0} {\bibfield  {journal}
  {\bibinfo  {journal} {Nature (London)}\ }\textbf {\bibinfo {volume} {355}},\
  \bibinfo {pages} {591–592} (\bibinfo {year}
  {1992}{\natexlab{b}})}\BibitemShut {NoStop}%
\bibitem [{\citenamefont {Goldstein}\ \emph {et~al.}(2014)\citenamefont
  {Goldstein}, \citenamefont {Poole},\ and\ \citenamefont
  {Safko}}]{Goldstein_book}%
  \BibitemOpen
  \bibfield  {author} {\bibinfo {author} {\bibfnamefont {H.}~\bibnamefont
  {Goldstein}}, \bibinfo {author} {\bibfnamefont {C.}~\bibnamefont {Poole}},\
  and\ \bibinfo {author} {\bibfnamefont {J.}~\bibnamefont {Safko}},\
  }\href@noop {} {\emph {\bibinfo {title} {Classical Mechanics: Pearson New
  International Edition}}}\ (\bibinfo  {publisher} {Pearson Education
  Limited},\ \bibinfo {year} {2014})\BibitemShut {NoStop}%
\bibitem [{\citenamefont {Vogtmann}\ \emph {et~al.}(1997)\citenamefont
  {Vogtmann}, \citenamefont {Weinstein},\ and\ \citenamefont
  {Arnold}}]{Arnold_book}%
  \BibitemOpen
  \bibfield  {author} {\bibinfo {author} {\bibfnamefont {K.}~\bibnamefont
  {Vogtmann}}, \bibinfo {author} {\bibfnamefont {A.}~\bibnamefont
  {Weinstein}},\ and\ \bibinfo {author} {\bibfnamefont {V.}~\bibnamefont
  {Arnold}},\ }\href@noop {} {\emph {\bibinfo {title} {Mathematical Methods of
  Classical Mechanics}}},\ Graduate Texts in Mathematics\ (\bibinfo
  {publisher} {Springer New York},\ \bibinfo {year} {1997})\BibitemShut
  {NoStop}%
\bibitem [{\citenamefont {Gutzwiller}(1991)}]{Gutzwiller_book}%
  \BibitemOpen
  \bibfield  {author} {\bibinfo {author} {\bibfnamefont {M.}~\bibnamefont
  {Gutzwiller}},\ }\href@noop {} {\emph {\bibinfo {title} {Chaos in Classical
  and Quantum Mechanics}}},\ Interdisciplinary Applied Mathematics\ (\bibinfo
  {publisher} {Springer New York},\ \bibinfo {year} {1991})\BibitemShut
  {NoStop}%
\bibitem [{\citenamefont {St{\"o}ckmann}(1999)}]{Stockmann_book}%
  \BibitemOpen
  \bibfield  {author} {\bibinfo {author} {\bibfnamefont {H.-J.}\ \bibnamefont
  {St{\"o}ckmann}},\ }\href {https://doi.org/10.1017/CBO9780511524622} {\emph
  {\bibinfo {title} {Quantum Chaos: An Introduction}}}\ (\bibinfo  {publisher}
  {Cambridge University Press},\ \bibinfo {year} {1999})\BibitemShut {NoStop}%
\bibitem [{\citenamefont {Nakamura}\ and\ \citenamefont
  {Harayama}(2004)}]{nakamura_2004_QD}%
  \BibitemOpen
  \bibfield  {author} {\bibinfo {author} {\bibfnamefont {K.}~\bibnamefont
  {Nakamura}}\ and\ \bibinfo {author} {\bibfnamefont {T.}~\bibnamefont
  {Harayama}},\ }\href@noop {} {\emph {\bibinfo {title} {Quantum Chaos and
  Quantum Dots}}},\ Mesoscopic physics and nanotechnology\ (\bibinfo
  {publisher} {Oxford University Press},\ \bibinfo {year} {2004})\BibitemShut
  {NoStop}%
\bibitem [{\citenamefont {Haake}(1991)}]{Haake_book}%
  \BibitemOpen
  \bibfield  {author} {\bibinfo {author} {\bibfnamefont {F.}~\bibnamefont
  {Haake}},\ }\href@noop {} {\emph {\bibinfo {title} {Quantum Signatures of
  Chaos}}},\ Springer series in synergetics\ (\bibinfo  {publisher}
  {Springer-Verlag},\ \bibinfo {year} {1991})\BibitemShut {NoStop}%
\bibitem [{\citenamefont {Nakamura}(2010)}]{nakamura2010quantum}%
  \BibitemOpen
  \bibfield  {author} {\bibinfo {author} {\bibfnamefont {K.}~\bibnamefont
  {Nakamura}},\ }\href@noop {} {\emph {\bibinfo {title} {Quantum versus Chaos:
  Questions Emerging from Mesoscopic Cosmos}}},\ Fundamental Theories of
  Physics\ (\bibinfo  {publisher} {Springer Netherlands},\ \bibinfo {year}
  {2010})\BibitemShut {NoStop}%
\bibitem [{\citenamefont {Nakamura}(1994)}]{nakamura1994quantum}%
  \BibitemOpen
  \bibfield  {author} {\bibinfo {author} {\bibfnamefont {K.}~\bibnamefont
  {Nakamura}},\ }\href@noop {} {\emph {\bibinfo {title} {Quantum Chaos: A New
  Paradigm of Nonlinear Dynamics}}},\ Cambridge Nonlinear Science Series\
  (\bibinfo  {publisher} {Cambridge University Press},\ \bibinfo {year}
  {1994})\BibitemShut {NoStop}%
\bibitem [{\citenamefont {Casati}\ and\ \citenamefont
  {Chirikov}(1995)}]{casati1995quantum}%
  \BibitemOpen
  \bibfield  {author} {\bibinfo {author} {\bibfnamefont {G.}~\bibnamefont
  {Casati}}\ and\ \bibinfo {author} {\bibfnamefont {B.}~\bibnamefont
  {Chirikov}},\ }\href@noop {} {\emph {\bibinfo {title} {Quantum Chaos: Between
  Order and Disorder}}}\ (\bibinfo  {publisher} {Cambridge University Press},\
  \bibinfo {year} {1995})\BibitemShut {NoStop}%
\bibitem [{\citenamefont {Berry}\ \emph {et~al.}(1987)\citenamefont {Berry},
  \citenamefont {Percival},\ and\ \citenamefont
  {Weiss}}]{Berry_rspa_413_182_1987}%
  \BibitemOpen
  \bibfield  {author} {\bibinfo {author} {\bibfnamefont {M.~V.}\ \bibnamefont
  {Berry}}, \bibinfo {author} {\bibfnamefont {I.~C.}\ \bibnamefont
  {Percival}},\ and\ \bibinfo {author} {\bibfnamefont {N.~O.}\ \bibnamefont
  {Weiss}},\ }\bibfield  {title} {\bibinfo {title} {The bakerian lecture, 1987.
  quantum chaology},\ }\href {https://doi.org/10.1098/rspa.1987.0109}
  {\bibfield  {journal} {\bibinfo  {journal} {P. Roy. Soc. A-Math. Phys.}\
  }\textbf {\bibinfo {volume} {413}},\ \bibinfo {pages} {183} (\bibinfo {year}
  {1987})}\BibitemShut {NoStop}%
\bibitem [{\citenamefont {Berry}(1977)}]{Berry_j.phys.a_10_2083_1977}%
  \BibitemOpen
  \bibfield  {author} {\bibinfo {author} {\bibfnamefont {M.~V.}\ \bibnamefont
  {Berry}},\ }\bibfield  {title} {\bibinfo {title} {Regular and irregular
  semiclassical wavefunctions},\ }\href
  {https://doi.org/10.1088/0305-4470/10/12/016} {\bibfield  {journal} {\bibinfo
   {journal} {J. Phys. A}\ }\textbf {\bibinfo {volume} {10}},\ \bibinfo {pages}
  {2083} (\bibinfo {year} {1977})}\BibitemShut {NoStop}%
\bibitem [{\citenamefont {O'Connor}\ and\ \citenamefont
  {Heller}(1988)}]{oconnor_phys.rev.lett_61_2288_1988}%
  \BibitemOpen
  \bibfield  {author} {\bibinfo {author} {\bibfnamefont {P.~W.}\ \bibnamefont
  {O'Connor}}\ and\ \bibinfo {author} {\bibfnamefont {E.~J.}\ \bibnamefont
  {Heller}},\ }\bibfield  {title} {\bibinfo {title} {Quantum localization for a
  strongly classically chaotic system},\ }\href
  {https://doi.org/10.1103/PhysRevLett.61.2288} {\bibfield  {journal} {\bibinfo
   {journal} {Phys. Rev. Lett.}\ }\textbf {\bibinfo {volume} {61}},\ \bibinfo
  {pages} {2288} (\bibinfo {year} {1988})}\BibitemShut {NoStop}%
\bibitem [{\citenamefont
  {Gutzwiller}(1971)}]{Gutzwiller_j.math.phys_12_343_1971}%
  \BibitemOpen
  \bibfield  {author} {\bibinfo {author} {\bibfnamefont {M.~C.}\ \bibnamefont
  {Gutzwiller}},\ }\bibfield  {title} {\bibinfo {title} {Periodic orbits and
  classical quantization conditions},\ }\href
  {https://doi.org/10.1063/1.1665596} {\bibfield  {journal} {\bibinfo
  {journal} {J. Math. Phys}\ }\textbf {\bibinfo {volume} {12}},\ \bibinfo
  {pages} {343} (\bibinfo {year} {1971})}\BibitemShut {NoStop}%
\bibitem [{\citenamefont
  {Shnirel'man}(1974)}]{Shnirelman_Uspekhi.Mat.Nauk_29_181_1974}%
  \BibitemOpen
  \bibfield  {author} {\bibinfo {author} {\bibfnamefont {A.~I.}\ \bibnamefont
  {Shnirel'man}},\ }\bibfield  {title} {\bibinfo {title} {Ergodic properties of
  eigenfunctions},\ }\href {http://www.ams.org/mathscinet-getitem?mr=402834}
  {\bibfield  {journal} {\bibinfo  {journal} {Uspekhi Mat. Nauk}\ }\textbf
  {\bibinfo {volume} {29}},\ \bibinfo {pages} {181} (\bibinfo {year}
  {1974})}\BibitemShut {NoStop}%
\bibitem [{\citenamefont {Colin~de
  Verdi{\'e}re}(1985)}]{Colindeverdiere_comm.math.phys_102_497_1985}%
  \BibitemOpen
  \bibfield  {author} {\bibinfo {author} {\bibfnamefont {Y.}~\bibnamefont
  {Colin~de Verdi{\'e}re}},\ }\bibfield  {title} {\bibinfo {title}
  {Ergodicit{\'e} et fonctions propres du laplacien},\ }\href
  {https://projecteuclid.org:443/euclid.cmp/1104114465} {\bibfield  {journal}
  {\bibinfo  {journal} {Comm. Math. Phys.}\ }\textbf {\bibinfo {volume}
  {102}},\ \bibinfo {pages} {497} (\bibinfo {year} {1985})}\BibitemShut
  {NoStop}%
\bibitem [{\citenamefont {Zelditch}(1987)}]{zelditch_duke.math.j_55_919_1987}%
  \BibitemOpen
  \bibfield  {author} {\bibinfo {author} {\bibfnamefont {S.}~\bibnamefont
  {Zelditch}},\ }\bibfield  {title} {\bibinfo {title} {Uniform distribution of
  eigenfunctions on compact hyperbolic surfaces},\ }\href
  {https://doi.org/10.1215/S0012-7094-87-05546-3} {\bibfield  {journal}
  {\bibinfo  {journal} {Duke Math. J.}\ }\textbf {\bibinfo {volume} {55}},\
  \bibinfo {pages} {919} (\bibinfo {year} {1987})}\BibitemShut {NoStop}%
\bibitem [{\citenamefont {Heller}(2018)}]{Heller_book_2008}%
  \BibitemOpen
  \bibfield  {author} {\bibinfo {author} {\bibfnamefont {E.~J.}\ \bibnamefont
  {Heller}},\ }\href@noop {} {\emph {\bibinfo {title} {The Semiclassical Way to
  Dynamics and Spectroscopy}}}\ (\bibinfo  {publisher} {Princeton University
  Press},\ \bibinfo {year} {2018})\BibitemShut {NoStop}%
\bibitem [{\citenamefont {Heller}(1984)}]{Heller_phys.rev.lett_53_1515_1984}%
  \BibitemOpen
  \bibfield  {author} {\bibinfo {author} {\bibfnamefont {E.~J.}\ \bibnamefont
  {Heller}},\ }\bibfield  {title} {\bibinfo {title} {Bound-state eigenfunctions
  of classically chaotic hamiltonian systems: Scars of periodic orbits},\
  }\href {https://doi.org/10.1103/PhysRevLett.53.1515} {\bibfield  {journal}
  {\bibinfo  {journal} {Phys. Rev. Lett.}\ }\textbf {\bibinfo {volume} {53}},\
  \bibinfo {pages} {1515} (\bibinfo {year} {1984})}\BibitemShut {NoStop}%
\bibitem [{\citenamefont {Kaplan}\ and\ \citenamefont
  {Heller}(1998)}]{kaplan_ann.phys_264_171_1998}%
  \BibitemOpen
  \bibfield  {author} {\bibinfo {author} {\bibfnamefont {L.}~\bibnamefont
  {Kaplan}}\ and\ \bibinfo {author} {\bibfnamefont {E.~J.}\ \bibnamefont
  {Heller}},\ }\bibfield  {title} {\bibinfo {title} {Linear and nonlinear
  theory of eigenfunction scars},\ }\href
  {https://doi.org/https://doi.org/10.1006/aphy.1997.5773} {\bibfield
  {journal} {\bibinfo  {journal} {Ann. Phys. (N. Y.)}\ }\textbf {\bibinfo
  {volume} {264}},\ \bibinfo {pages} {171 } (\bibinfo {year}
  {1998})}\BibitemShut {NoStop}%
\bibitem [{\citenamefont
  {Kaplan}(1999{\natexlab{a}})}]{Kaplan_nonlinearity_12_R1_1999}%
  \BibitemOpen
  \bibfield  {author} {\bibinfo {author} {\bibfnamefont {L.}~\bibnamefont
  {Kaplan}},\ }\bibfield  {title} {\bibinfo {title} {Scars in quantum chaotic
  wavefunctions},\ }\href {https://doi.org/10.1088/0951-7715/12/2/009}
  {\bibfield  {journal} {\bibinfo  {journal} {Nonlinearity}\ }\textbf {\bibinfo
  {volume} {12}},\ \bibinfo {pages} {R1} (\bibinfo {year}
  {1999}{\natexlab{a}})}\BibitemShut {NoStop}%
\bibitem [{\citenamefont {Bogomolny"}(1988)}]{bogomolny_physica.d_31_169_1988}%
  \BibitemOpen
  \bibfield  {author} {\bibinfo {author} {\bibfnamefont {E.~B.}\ \bibnamefont
  {Bogomolny"}},\ }\bibfield  {title} {\bibinfo {title} {Smoothed wave
  functions of chaotic quantum systems},\ }\href
  {https://doi.org/https://doi.org/10.1016/0167-2789(88)90075-9} {\bibfield
  {journal} {\bibinfo  {journal} {Physica D: Nonlinear Phenomena}\ }\textbf
  {\bibinfo {volume} {31}},\ \bibinfo {pages} {169 } (\bibinfo {year}
  {1988})}\BibitemShut {NoStop}%
\bibitem [{\citenamefont
  {Berry}(1989{\natexlab{b}})}]{berry_proc_r_soc_lond_a_423_219_1989}%
  \BibitemOpen
  \bibfield  {author} {\bibinfo {author} {\bibfnamefont {M.~V.}\ \bibnamefont
  {Berry}},\ }\bibfield  {title} {\bibinfo {title} {Quantum scars of classical
  closed orbits in phase space},\ }\href
  {https://doi.org/10.1098/rspa.1989.0052} {\bibfield  {journal} {\bibinfo
  {journal} {Proc. R. Soc. Lond. A}\ }\textbf {\bibinfo {volume} {423}},\
  \bibinfo {pages} {219} (\bibinfo {year} {1989}{\natexlab{b}})}\BibitemShut
  {NoStop}%
\bibitem [{\citenamefont {Kaplan}\ and\ \citenamefont
  {Heller}(1999)}]{kaplan.phys.rev.e_59_6609_1999}%
  \BibitemOpen
  \bibfield  {author} {\bibinfo {author} {\bibfnamefont {L.}~\bibnamefont
  {Kaplan}}\ and\ \bibinfo {author} {\bibfnamefont {E.~J.}\ \bibnamefont
  {Heller}},\ }\bibfield  {title} {\bibinfo {title} {Measuring scars of
  periodic orbits},\ }\href {https://doi.org/10.1103/PhysRevE.59.6609}
  {\bibfield  {journal} {\bibinfo  {journal} {Phys. Rev. E}\ }\textbf {\bibinfo
  {volume} {59}},\ \bibinfo {pages} {6609} (\bibinfo {year}
  {1999})}\BibitemShut {NoStop}%
\bibitem [{\citenamefont {Ku\ifmmode~\acute{s}\else \'{s}\fi{}}\ \emph
  {et~al.}(1991)\citenamefont {Ku\ifmmode~\acute{s}\else \'{s}\fi{}},
  \citenamefont {Zakrzewski},\ and\ \citenamefont {\ifmmode~\dot{Z}\else
  \.{Z}\fi{}yczkowski}}]{kus_phys.rev.a_43_4244_1991}%
  \BibitemOpen
  \bibfield  {author} {\bibinfo {author} {\bibfnamefont {M.}~\bibnamefont
  {Ku\ifmmode~\acute{s}\else \'{s}\fi{}}}, \bibinfo {author} {\bibfnamefont
  {J.}~\bibnamefont {Zakrzewski}},\ and\ \bibinfo {author} {\bibfnamefont
  {K.}~\bibnamefont {\ifmmode~\dot{Z}\else \.{Z}\fi{}yczkowski}},\ }\bibfield
  {title} {\bibinfo {title} {Quantum scars on a sphere},\ }\href
  {https://doi.org/10.1103/PhysRevA.43.4244} {\bibfield  {journal} {\bibinfo
  {journal} {Phys. Rev. A}\ }\textbf {\bibinfo {volume} {43}},\ \bibinfo
  {pages} {4244} (\bibinfo {year} {1991})}\BibitemShut {NoStop}%
\bibitem [{\citenamefont {D'Ariano}\ \emph {et~al.}(1992)\citenamefont
  {D'Ariano}, \citenamefont {Evangelista},\ and\ \citenamefont
  {Saraceno}}]{dariano_phys.rev.a_45_3646_1992}%
  \BibitemOpen
  \bibfield  {author} {\bibinfo {author} {\bibfnamefont {G.~M.}\ \bibnamefont
  {D'Ariano}}, \bibinfo {author} {\bibfnamefont {L.~R.}\ \bibnamefont
  {Evangelista}},\ and\ \bibinfo {author} {\bibfnamefont {M.}~\bibnamefont
  {Saraceno}},\ }\bibfield  {title} {\bibinfo {title} {Classical and quantum
  structures in the kicked-top model},\ }\href
  {https://doi.org/10.1103/PhysRevA.45.3646} {\bibfield  {journal} {\bibinfo
  {journal} {Phys. Rev. A}\ }\textbf {\bibinfo {volume} {45}},\ \bibinfo
  {pages} {3646} (\bibinfo {year} {1992})}\BibitemShut {NoStop}%
\bibitem [{\citenamefont {Tomsovic}\ and\ \citenamefont
  {Heller}(1993)}]{tomsovic_phys.rev.lett_70_1405_1993}%
  \BibitemOpen
  \bibfield  {author} {\bibinfo {author} {\bibfnamefont {S.}~\bibnamefont
  {Tomsovic}}\ and\ \bibinfo {author} {\bibfnamefont {E.~J.}\ \bibnamefont
  {Heller}},\ }\bibfield  {title} {\bibinfo {title} {Semiclassical construction
  of chaotic eigenstates},\ }\href
  {https://doi.org/10.1103/PhysRevLett.70.1405} {\bibfield  {journal} {\bibinfo
   {journal} {Phys. Rev. Lett.}\ }\textbf {\bibinfo {volume} {70}},\ \bibinfo
  {pages} {1405} (\bibinfo {year} {1993})}\BibitemShut {NoStop}%
\bibitem [{\citenamefont {Revuelta}\ \emph {et~al.}(2020)\citenamefont
  {Revuelta}, \citenamefont {Vergini}, \citenamefont {Benito},\ and\
  \citenamefont {Borondo}}]{revuelta_phys.rev.e_102_042210_2020}%
  \BibitemOpen
  \bibfield  {author} {\bibinfo {author} {\bibfnamefont {F.}~\bibnamefont
  {Revuelta}}, \bibinfo {author} {\bibfnamefont {E.}~\bibnamefont {Vergini}},
  \bibinfo {author} {\bibfnamefont {R.~M.}\ \bibnamefont {Benito}},\ and\
  \bibinfo {author} {\bibfnamefont {F.}~\bibnamefont {Borondo}},\ }\bibfield
  {title} {\bibinfo {title} {Short-periodic-orbit method for excited chaotic
  eigenfunctions},\ }\href {https://doi.org/10.1103/PhysRevE.102.042210}
  {\bibfield  {journal} {\bibinfo  {journal} {Phys. Rev. E}\ }\textbf {\bibinfo
  {volume} {102}},\ \bibinfo {pages} {042210} (\bibinfo {year}
  {2020})}\BibitemShut {NoStop}%
\bibitem [{\citenamefont {Agam}\ and\ \citenamefont
  {Fishman}(1993)}]{agam_j.phys.a.math_26_2113_1993}%
  \BibitemOpen
  \bibfield  {author} {\bibinfo {author} {\bibfnamefont {O.}~\bibnamefont
  {Agam}}\ and\ \bibinfo {author} {\bibfnamefont {S.}~\bibnamefont {Fishman}},\
  }\bibfield  {title} {\bibinfo {title} {Quantum eigenfunctions in terms of
  periodic orbits of chaotic systems},\ }\href
  {https://doi.org/10.1088/0305-4470/26/9/010} {\bibfield  {journal} {\bibinfo
  {journal} {J. Phys. A Math.}\ }\textbf {\bibinfo {volume} {26}},\ \bibinfo
  {pages} {2113} (\bibinfo {year} {1993})}\BibitemShut {NoStop}%
\bibitem [{\citenamefont {Bohigas}\ \emph {et~al.}(1993)\citenamefont
  {Bohigas}, \citenamefont {Tomsovic},\ and\ \citenamefont
  {Ullmo}}]{bohigas_phys.rep_223_43_1993}%
  \BibitemOpen
  \bibfield  {author} {\bibinfo {author} {\bibfnamefont {O.}~\bibnamefont
  {Bohigas}}, \bibinfo {author} {\bibfnamefont {S.}~\bibnamefont {Tomsovic}},\
  and\ \bibinfo {author} {\bibfnamefont {D.}~\bibnamefont {Ullmo}},\ }\bibfield
   {title} {\bibinfo {title} {Manifestations of classical phase space
  structures in quantum mechanics},\ }\href
  {https://doi.org/https://doi.org/10.1016/0370-1573(93)90109-Q} {\bibfield
  {journal} {\bibinfo  {journal} {Phys. rep.}\ }\textbf {\bibinfo {volume}
  {223}},\ \bibinfo {pages} {43} (\bibinfo {year} {1993})}\BibitemShut
  {NoStop}%
\bibitem [{\citenamefont {Wisniacki}\ \emph {et~al.}(2006)\citenamefont
  {Wisniacki}, \citenamefont {Vergini}, \citenamefont {Benito},\ and\
  \citenamefont {Borondo}}]{wisniacki_phys.rev.lett_97_094101_2006}%
  \BibitemOpen
  \bibfield  {author} {\bibinfo {author} {\bibfnamefont {D.~A.}\ \bibnamefont
  {Wisniacki}}, \bibinfo {author} {\bibfnamefont {E.}~\bibnamefont {Vergini}},
  \bibinfo {author} {\bibfnamefont {R.~M.}\ \bibnamefont {Benito}},\ and\
  \bibinfo {author} {\bibfnamefont {F.}~\bibnamefont {Borondo}},\ }\bibfield
  {title} {\bibinfo {title} {Scarring by homoclinic and heteroclinic orbits},\
  }\href {https://doi.org/10.1103/PhysRevLett.97.094101} {\bibfield  {journal}
  {\bibinfo  {journal} {Phys. Rev. Lett.}\ }\textbf {\bibinfo {volume} {97}},\
  \bibinfo {pages} {094101} (\bibinfo {year} {2006})}\BibitemShut {NoStop}%
\bibitem [{\citenamefont {Fromhold}\ \emph
  {et~al.}(1995{\natexlab{a}})\citenamefont {Fromhold}, \citenamefont
  {Wilkinson}, \citenamefont {Sheard}, \citenamefont {Eaves}, \citenamefont
  {Miao},\ and\ \citenamefont {Edwards}}]{Fromhol_phys.rev.lett_75_1142_1995}%
  \BibitemOpen
  \bibfield  {author} {\bibinfo {author} {\bibfnamefont {T.~M.}\ \bibnamefont
  {Fromhold}}, \bibinfo {author} {\bibfnamefont {P.~B.}\ \bibnamefont
  {Wilkinson}}, \bibinfo {author} {\bibfnamefont {F.~W.}\ \bibnamefont
  {Sheard}}, \bibinfo {author} {\bibfnamefont {L.}~\bibnamefont {Eaves}},
  \bibinfo {author} {\bibfnamefont {J.}~\bibnamefont {Miao}},\ and\ \bibinfo
  {author} {\bibfnamefont {G.}~\bibnamefont {Edwards}},\ }\bibfield  {title}
  {\bibinfo {title} {Manifestations of classical chaos in the energy level
  spectrum of a quantum well},\ }\href
  {https://doi.org/10.1103/PhysRevLett.75.1142} {\bibfield  {journal} {\bibinfo
   {journal} {Phys. Rev. Lett.}\ }\textbf {\bibinfo {volume} {75}},\ \bibinfo
  {pages} {1142} (\bibinfo {year} {1995}{\natexlab{a}})}\BibitemShut {NoStop}%
\bibitem [{\citenamefont {Wilkinson}\ \emph {et~al.}(1996)\citenamefont
  {Wilkinson}, \citenamefont {Fromhold}, \citenamefont {Eaves}, \citenamefont
  {Sheard}, \citenamefont {Miura},\ and\ \citenamefont
  {Takamasu}}]{Wilkinson_nature_380_608_1996}%
  \BibitemOpen
  \bibfield  {author} {\bibinfo {author} {\bibfnamefont {P.~B.}\ \bibnamefont
  {Wilkinson}}, \bibinfo {author} {\bibfnamefont {T.~M.}\ \bibnamefont
  {Fromhold}}, \bibinfo {author} {\bibfnamefont {L.}~\bibnamefont {Eaves}},
  \bibinfo {author} {\bibfnamefont {F.~W.}\ \bibnamefont {Sheard}}, \bibinfo
  {author} {\bibfnamefont {N.}~\bibnamefont {Miura}},\ and\ \bibinfo {author}
  {\bibfnamefont {T.}~\bibnamefont {Takamasu}},\ }\bibfield  {title} {\bibinfo
  {title} {Observation of 'scarred' wavefunctions in a quantum well with
  chaotic electron dynamics},\ }\href {https://doi.org/10.1038/380608a0}
  {\bibfield  {journal} {\bibinfo  {journal} {Nature (London)}\ }\textbf
  {\bibinfo {volume} {380}},\ \bibinfo {pages} {608} (\bibinfo {year}
  {1996})}\BibitemShut {NoStop}%
\bibitem [{\citenamefont {Narimanov}\ and\ \citenamefont
  {Stone}(1998)}]{narimanov_phys.rev.lett_80_49_1998}%
  \BibitemOpen
  \bibfield  {author} {\bibinfo {author} {\bibfnamefont {E.~E.}\ \bibnamefont
  {Narimanov}}\ and\ \bibinfo {author} {\bibfnamefont {A.~D.}\ \bibnamefont
  {Stone}},\ }\bibfield  {title} {\bibinfo {title} {Origin of strong scarring
  of wave functions in quantum wells in a tilted magnetic field},\ }\href
  {https://doi.org/10.1103/PhysRevLett.80.49} {\bibfield  {journal} {\bibinfo
  {journal} {Phys. Rev. Lett.}\ }\textbf {\bibinfo {volume} {80}},\ \bibinfo
  {pages} {49} (\bibinfo {year} {1998})}\BibitemShut {NoStop}%
\bibitem [{\citenamefont {H{\"o}nig}\ and\ \citenamefont
  {Wintgen}(1989)}]{Honig_phys.rev.a_39_5642_1989}%
  \BibitemOpen
  \bibfield  {author} {\bibinfo {author} {\bibfnamefont {A.}~\bibnamefont
  {H{\"o}nig}}\ and\ \bibinfo {author} {\bibfnamefont {D.}~\bibnamefont
  {Wintgen}},\ }\bibfield  {title} {\bibinfo {title} {Spectral properties of
  strongly perturbed coulomb systems: Fluctuation properties},\ }\href
  {https://doi.org/10.1103/PhysRevA.39.5642} {\bibfield  {journal} {\bibinfo
  {journal} {Phys. Rev. A}\ }\textbf {\bibinfo {volume} {39}},\ \bibinfo
  {pages} {5642} (\bibinfo {year} {1989})}\BibitemShut {NoStop}%
\bibitem [{\citenamefont {Bogomolny}\ \emph {et~al.}(2006)\citenamefont
  {Bogomolny}, \citenamefont {Dietz}, \citenamefont {Friedrich}, \citenamefont
  {Miski-Oglu}, \citenamefont {Richter}, \citenamefont {Sch{\"a}fer},\ and\
  \citenamefont {Schmit}}]{bogomolny_phys.rev.lett_97_254102_2006}%
  \BibitemOpen
  \bibfield  {author} {\bibinfo {author} {\bibfnamefont {E.}~\bibnamefont
  {Bogomolny}}, \bibinfo {author} {\bibfnamefont {B.}~\bibnamefont {Dietz}},
  \bibinfo {author} {\bibfnamefont {T.}~\bibnamefont {Friedrich}}, \bibinfo
  {author} {\bibfnamefont {M.}~\bibnamefont {Miski-Oglu}}, \bibinfo {author}
  {\bibfnamefont {A.}~\bibnamefont {Richter}}, \bibinfo {author} {\bibfnamefont
  {F.}~\bibnamefont {Sch{\"a}fer}},\ and\ \bibinfo {author} {\bibfnamefont
  {C.}~\bibnamefont {Schmit}},\ }\bibfield  {title} {\bibinfo {title} {First
  experimental observation of superscars in a pseudointegrable barrier
  billiard},\ }\href {https://doi.org/10.1103/PhysRevLett.97.254102} {\bibfield
   {journal} {\bibinfo  {journal} {Phys. Rev. Lett.}\ }\textbf {\bibinfo
  {volume} {97}},\ \bibinfo {pages} {254102} (\bibinfo {year}
  {2006})}\BibitemShut {NoStop}%
\bibitem [{\citenamefont {Kim}\ \emph {et~al.}(2002)\citenamefont {Kim},
  \citenamefont {Barth}, \citenamefont {St\"ockmann},\ and\ \citenamefont
  {Bird}}]{kim_phys.rev.b_65_165317_2002}%
  \BibitemOpen
  \bibfield  {author} {\bibinfo {author} {\bibfnamefont {Y.-H.}\ \bibnamefont
  {Kim}}, \bibinfo {author} {\bibfnamefont {M.}~\bibnamefont {Barth}}, \bibinfo
  {author} {\bibfnamefont {H.-J.}\ \bibnamefont {St\"ockmann}},\ and\ \bibinfo
  {author} {\bibfnamefont {J.~P.}\ \bibnamefont {Bird}},\ }\bibfield  {title}
  {\bibinfo {title} {Wave function scarring in open quantum dots: A
  microwave-billiard analog study},\ }\href
  {https://doi.org/10.1103/PhysRevB.65.165317} {\bibfield  {journal} {\bibinfo
  {journal} {Phys. Rev. B}\ }\textbf {\bibinfo {volume} {65}},\ \bibinfo
  {pages} {165317} (\bibinfo {year} {2002})}\BibitemShut {NoStop}%
\bibitem [{\citenamefont {St{\"o}ckmann}\ and\ \citenamefont
  {Stein}(1990)}]{stockman_phys.rev.lett.64.2215_1990}%
  \BibitemOpen
  \bibfield  {author} {\bibinfo {author} {\bibfnamefont {H.-J.}\ \bibnamefont
  {St{\"o}ckmann}}\ and\ \bibinfo {author} {\bibfnamefont {J.}~\bibnamefont
  {Stein}},\ }\bibfield  {title} {\bibinfo {title} {``quantum'' chaos in
  billiards studied by microwave absorption},\ }\href
  {https://doi.org/10.1103/PhysRevLett.64.2215} {\bibfield  {journal} {\bibinfo
   {journal} {Phys. Rev. Lett.}\ }\textbf {\bibinfo {volume} {64}},\ \bibinfo
  {pages} {2215} (\bibinfo {year} {1990})}\BibitemShut {NoStop}%
\bibitem [{\citenamefont {D{\"o}rr}\ \emph {et~al.}(1998)\citenamefont
  {D{\"o}rr}, \citenamefont {St{\"o}ckmann}, \citenamefont {Barth},\ and\
  \citenamefont {Kuhl}}]{dorr_phys.rev.lett_80_1030_1998}%
  \BibitemOpen
  \bibfield  {author} {\bibinfo {author} {\bibfnamefont {U.}~\bibnamefont
  {D{\"o}rr}}, \bibinfo {author} {\bibfnamefont {H.-J.}\ \bibnamefont
  {St{\"o}ckmann}}, \bibinfo {author} {\bibfnamefont {M.}~\bibnamefont
  {Barth}},\ and\ \bibinfo {author} {\bibfnamefont {U.}~\bibnamefont {Kuhl}},\
  }\bibfield  {title} {\bibinfo {title} {Scarred and chaotic field
  distributions in a three-dimensional {S}inai-microwave resonator},\ }\href
  {https://doi.org/10.1103/PhysRevLett.80.1030} {\bibfield  {journal} {\bibinfo
   {journal} {Phys. Rev. Lett.}\ }\textbf {\bibinfo {volume} {80}},\ \bibinfo
  {pages} {1030} (\bibinfo {year} {1998})}\BibitemShut {NoStop}%
\bibitem [{\citenamefont {Sridhar}(1991)}]{Sridhaar_phys.rev.lett_67_785_1991}%
  \BibitemOpen
  \bibfield  {author} {\bibinfo {author} {\bibfnamefont {S.}~\bibnamefont
  {Sridhar}},\ }\bibfield  {title} {\bibinfo {title} {Experimental observation
  of scarred eigenfunctions of chaotic microwave cavities},\ }\href
  {https://doi.org/10.1103/PhysRevLett.67.785} {\bibfield  {journal} {\bibinfo
  {journal} {Phys. Rev. Lett.}\ }\textbf {\bibinfo {volume} {67}},\ \bibinfo
  {pages} {785} (\bibinfo {year} {1991})}\BibitemShut {NoStop}%
\bibitem [{\citenamefont {Stein}\ and\ \citenamefont
  {St{\"o}ckmann}(1992)}]{Stein_phys.rev.lett_68_2867_1992}%
  \BibitemOpen
  \bibfield  {author} {\bibinfo {author} {\bibfnamefont {J.}~\bibnamefont
  {Stein}}\ and\ \bibinfo {author} {\bibfnamefont {H.-J.}\ \bibnamefont
  {St{\"o}ckmann}},\ }\bibfield  {title} {\bibinfo {title} {Experimental
  determination of billiard wave functions},\ }\href
  {https://doi.org/10.1103/PhysRevLett.68.2867} {\bibfield  {journal} {\bibinfo
   {journal} {Phys. Rev. Lett.}\ }\textbf {\bibinfo {volume} {68}},\ \bibinfo
  {pages} {2867} (\bibinfo {year} {1992})}\BibitemShut {NoStop}%
\bibitem [{\citenamefont {N{\"o}ckel}\ and\ \citenamefont
  {Stone}(1997)}]{nockel_nature_385_45_1997}%
  \BibitemOpen
  \bibfield  {author} {\bibinfo {author} {\bibfnamefont {J.}~\bibnamefont
  {N{\"o}ckel}}\ and\ \bibinfo {author} {\bibfnamefont {A.}~\bibnamefont
  {Stone}},\ }\bibfield  {title} {\bibinfo {title} {Ray and wave chaos in
  asymmetric resonant optical cavities},\ }\href
  {https://doi.org/10.1038/385045a0} {\bibfield  {journal} {\bibinfo  {journal}
  {Nature}\ }\textbf {\bibinfo {volume} {385}},\ \bibinfo {pages} {45}
  (\bibinfo {year} {1997})}\BibitemShut {NoStop}%
\bibitem [{\citenamefont {Lee}\ \emph {et~al.}(2002)\citenamefont {Lee},
  \citenamefont {Lee}, \citenamefont {Chang}, \citenamefont {Moon},
  \citenamefont {Kim},\ and\ \citenamefont
  {An}}]{Lee_phys.rev.lett_88_033903_2002}%
  \BibitemOpen
  \bibfield  {author} {\bibinfo {author} {\bibfnamefont {S.-B.}\ \bibnamefont
  {Lee}}, \bibinfo {author} {\bibfnamefont {J.-H.}\ \bibnamefont {Lee}},
  \bibinfo {author} {\bibfnamefont {J.-S.}\ \bibnamefont {Chang}}, \bibinfo
  {author} {\bibfnamefont {H.-J.}\ \bibnamefont {Moon}}, \bibinfo {author}
  {\bibfnamefont {S.~W.}\ \bibnamefont {Kim}},\ and\ \bibinfo {author}
  {\bibfnamefont {K.}~\bibnamefont {An}},\ }\bibfield  {title} {\bibinfo
  {title} {Observation of scarred modes in asymmetrically deformed
  microcylinder lasers},\ }\href
  {https://doi.org/10.1103/PhysRevLett.88.033903} {\bibfield  {journal}
  {\bibinfo  {journal} {Phys. Rev. Lett.}\ }\textbf {\bibinfo {volume} {88}},\
  \bibinfo {pages} {033903} (\bibinfo {year} {2002})}\BibitemShut {NoStop}%
\bibitem [{\citenamefont {Harayama}\ \emph {et~al.}(2003)\citenamefont
  {Harayama}, \citenamefont {Fukushima}, \citenamefont {Davis}, \citenamefont
  {Vaccaro}, \citenamefont {Miyasaka}, \citenamefont {Nishimura},\ and\
  \citenamefont {Aida}}]{Harayama_phys.rev.e_67_015207_2003}%
  \BibitemOpen
  \bibfield  {author} {\bibinfo {author} {\bibfnamefont {T.}~\bibnamefont
  {Harayama}}, \bibinfo {author} {\bibfnamefont {T.}~\bibnamefont {Fukushima}},
  \bibinfo {author} {\bibfnamefont {P.}~\bibnamefont {Davis}}, \bibinfo
  {author} {\bibfnamefont {P.~O.}\ \bibnamefont {Vaccaro}}, \bibinfo {author}
  {\bibfnamefont {T.}~\bibnamefont {Miyasaka}}, \bibinfo {author}
  {\bibfnamefont {T.}~\bibnamefont {Nishimura}},\ and\ \bibinfo {author}
  {\bibfnamefont {T.}~\bibnamefont {Aida}},\ }\bibfield  {title} {\bibinfo
  {title} {Lasing on scar modes in fully chaotic microcavities},\ }\href
  {https://doi.org/10.1103/PhysRevE.67.015207} {\bibfield  {journal} {\bibinfo
  {journal} {Phys. Rev. E}\ }\textbf {\bibinfo {volume} {67}},\ \bibinfo
  {pages} {015207} (\bibinfo {year} {2003})}\BibitemShut {NoStop}%
\bibitem [{\citenamefont {Chinnery}\ and\ \citenamefont
  {Humphrey}(1996)}]{chinnery_phys.rev.e_53_272_1996}%
  \BibitemOpen
  \bibfield  {author} {\bibinfo {author} {\bibfnamefont {P.~A.}\ \bibnamefont
  {Chinnery}}\ and\ \bibinfo {author} {\bibfnamefont {V.~F.}\ \bibnamefont
  {Humphrey}},\ }\bibfield  {title} {\bibinfo {title} {Experimental
  visualization of acoustic resonances within a stadium-shaped cavity},\
  }\href@noop {} {\bibfield  {journal} {\bibinfo  {journal} {Phys. Rev. E}\
  }\textbf {\bibinfo {volume} {53}},\ \bibinfo {pages} {272} (\bibinfo {year}
  {1996})}\BibitemShut {NoStop}%
\bibitem [{\citenamefont {Larson}\ \emph {et~al.}(2013)\citenamefont {Larson},
  \citenamefont {Anderson},\ and\ \citenamefont
  {Altland}}]{Larson_phys.rev.a_87_013624_2013}%
  \BibitemOpen
  \bibfield  {author} {\bibinfo {author} {\bibfnamefont {J.}~\bibnamefont
  {Larson}}, \bibinfo {author} {\bibfnamefont {B.~M.}\ \bibnamefont
  {Anderson}},\ and\ \bibinfo {author} {\bibfnamefont {A.}~\bibnamefont
  {Altland}},\ }\bibfield  {title} {\bibinfo {title} {Chaos-driven dynamics in
  spin-orbit-coupled atomic gases},\ }\href
  {https://doi.org/10.1103/PhysRevA.87.013624} {\bibfield  {journal} {\bibinfo
  {journal} {Phys. Rev. A}\ }\textbf {\bibinfo {volume} {87}},\ \bibinfo
  {pages} {013624} (\bibinfo {year} {2013})}\BibitemShut {NoStop}%
\bibitem [{\citenamefont {Huang}\ \emph {et~al.}(2009)\citenamefont {Huang},
  \citenamefont {Lai}, \citenamefont {Ferry}, \citenamefont {Goodnick},\ and\
  \citenamefont {Akis}}]{Huang_phys.rev.lett_103_054101_2009}%
  \BibitemOpen
  \bibfield  {author} {\bibinfo {author} {\bibfnamefont {L.}~\bibnamefont
  {Huang}}, \bibinfo {author} {\bibfnamefont {Y.-C.}\ \bibnamefont {Lai}},
  \bibinfo {author} {\bibfnamefont {D.~K.}\ \bibnamefont {Ferry}}, \bibinfo
  {author} {\bibfnamefont {S.~M.}\ \bibnamefont {Goodnick}},\ and\ \bibinfo
  {author} {\bibfnamefont {R.}~\bibnamefont {Akis}},\ }\bibfield  {title}
  {\bibinfo {title} {Relativistic quantum scars},\ }\href
  {https://doi.org/10.1103/PhysRevLett.103.054101} {\bibfield  {journal}
  {\bibinfo  {journal} {Phys. Rev. Lett.}\ }\textbf {\bibinfo {volume} {103}},\
  \bibinfo {pages} {054101} (\bibinfo {year} {2009})}\BibitemShut {NoStop}%
\bibitem [{\citenamefont {Xu}\ \emph {et~al.}(2013)\citenamefont {Xu},
  \citenamefont {Huang}, \citenamefont {Lai},\ and\ \citenamefont
  {Grebogi}}]{Xu_phys.rev.lett_110_064102_2013}%
  \BibitemOpen
  \bibfield  {author} {\bibinfo {author} {\bibfnamefont {H.}~\bibnamefont
  {Xu}}, \bibinfo {author} {\bibfnamefont {L.}~\bibnamefont {Huang}}, \bibinfo
  {author} {\bibfnamefont {Y.-C.}\ \bibnamefont {Lai}},\ and\ \bibinfo {author}
  {\bibfnamefont {C.}~\bibnamefont {Grebogi}},\ }\bibfield  {title} {\bibinfo
  {title} {Chiral scars in chaotic dirac fermion systems},\ }\href
  {https://doi.org/10.1103/PhysRevLett.110.064102} {\bibfield  {journal}
  {\bibinfo  {journal} {Phys. Rev. Lett.}\ }\textbf {\bibinfo {volume} {110}},\
  \bibinfo {pages} {064102} (\bibinfo {year} {2013})}\BibitemShut {NoStop}%
\bibitem [{\citenamefont {Ni}\ \emph {et~al.}(2012)\citenamefont {Ni},
  \citenamefont {Huang}, \citenamefont {Lai},\ and\ \citenamefont
  {Grebogi}}]{xuan_phys.rev.E_86_016702_2012}%
  \BibitemOpen
  \bibfield  {author} {\bibinfo {author} {\bibfnamefont {X.}~\bibnamefont
  {Ni}}, \bibinfo {author} {\bibfnamefont {L.}~\bibnamefont {Huang}}, \bibinfo
  {author} {\bibfnamefont {Y.-C.}\ \bibnamefont {Lai}},\ and\ \bibinfo {author}
  {\bibfnamefont {C.}~\bibnamefont {Grebogi}},\ }\bibfield  {title} {\bibinfo
  {title} {Scarring of dirac fermions in chaotic billiards},\ }\href
  {https://doi.org/10.1103/PhysRevE.86.016702} {\bibfield  {journal} {\bibinfo
  {journal} {Phys. Rev. E}\ }\textbf {\bibinfo {volume} {86}},\ \bibinfo
  {pages} {016702} (\bibinfo {year} {2012})}\BibitemShut {NoStop}%
\bibitem [{\citenamefont {Song}\ \emph {et~al.}(2019)\citenamefont {Song},
  \citenamefont {Li}, \citenamefont {Xu}, \citenamefont {Huang},\ and\
  \citenamefont {Lai}}]{song_phys_rev.research_1_033008_2019}%
  \BibitemOpen
  \bibfield  {author} {\bibinfo {author} {\bibfnamefont {M.-Y.}\ \bibnamefont
  {Song}}, \bibinfo {author} {\bibfnamefont {Z.-Y.}\ \bibnamefont {Li}},
  \bibinfo {author} {\bibfnamefont {H.-Y.}\ \bibnamefont {Xu}}, \bibinfo
  {author} {\bibfnamefont {L.}~\bibnamefont {Huang}},\ and\ \bibinfo {author}
  {\bibfnamefont {Y.-C.}\ \bibnamefont {Lai}},\ }\bibfield  {title} {\bibinfo
  {title} {Quantization of massive dirac billiards and unification of
  nonrelativistic and relativistic chiral quantum scars},\ }\href
  {https://doi.org/10.1103/PhysRevResearch.1.033008} {\bibfield  {journal}
  {\bibinfo  {journal} {Phys. Rev. Research}\ }\textbf {\bibinfo {volume}
  {1}},\ \bibinfo {pages} {033008} (\bibinfo {year} {2019})}\BibitemShut
  {NoStop}%
\bibitem [{\citenamefont {Bogomolny}\ and\ \citenamefont
  {Schmit}(2004)}]{bogomolny_phys.rev.lett_92_244102_2004}%
  \BibitemOpen
  \bibfield  {author} {\bibinfo {author} {\bibfnamefont {E.}~\bibnamefont
  {Bogomolny}}\ and\ \bibinfo {author} {\bibfnamefont {C.}~\bibnamefont
  {Schmit}},\ }\bibfield  {title} {\bibinfo {title} {Structure of wave
  functions of pseudointegrable billiards},\ }\href
  {https://doi.org/10.1103/PhysRevLett.92.244102} {\bibfield  {journal}
  {\bibinfo  {journal} {Phys. Rev. Lett.}\ }\textbf {\bibinfo {volume} {92}},\
  \bibinfo {pages} {244102} (\bibinfo {year} {2004})}\BibitemShut {NoStop}%
\bibitem [{\citenamefont {\AA{}berg}\ \emph {et~al.}(2008)\citenamefont
  {\AA{}berg}, \citenamefont {Guhr}, \citenamefont {Miski-Oglu},\ and\
  \citenamefont {Richter}}]{aberg_phys.rev.lett_100_204101_2008}%
  \BibitemOpen
  \bibfield  {author} {\bibinfo {author} {\bibfnamefont {S.}~\bibnamefont
  {\AA{}berg}}, \bibinfo {author} {\bibfnamefont {T.}~\bibnamefont {Guhr}},
  \bibinfo {author} {\bibfnamefont {M.}~\bibnamefont {Miski-Oglu}},\ and\
  \bibinfo {author} {\bibfnamefont {A.}~\bibnamefont {Richter}},\ }\bibfield
  {title} {\bibinfo {title} {Superscars in billiards: A model for doorway
  states in quantum spectra},\ }\href
  {https://doi.org/10.1103/PhysRevLett.100.204101} {\bibfield  {journal}
  {\bibinfo  {journal} {Phys. Rev. Lett.}\ }\textbf {\bibinfo {volume} {100}},\
  \bibinfo {pages} {204101} (\bibinfo {year} {2008})}\BibitemShut {NoStop}%
\bibitem [{\citenamefont {Muller}\ and\ \citenamefont
  {Wintgen}(1994)}]{muller_j.phys.b_27_2693_1994}%
  \BibitemOpen
  \bibfield  {author} {\bibinfo {author} {\bibfnamefont {K.}~\bibnamefont
  {Muller}}\ and\ \bibinfo {author} {\bibfnamefont {D.}~\bibnamefont
  {Wintgen}},\ }\bibfield  {title} {\bibinfo {title} {Scars in wavefunctions of
  the diamagnetic kepler problem},\ }\href
  {https://doi.org/10.1088/0953-4075/27/13/003} {\bibfield  {journal} {\bibinfo
   {journal} {J. Phys. B}\ }\textbf {\bibinfo {volume} {27}},\ \bibinfo {pages}
  {2693} (\bibinfo {year} {1994})}\BibitemShut {NoStop}%
\bibitem [{\citenamefont {Kudrolli}\ \emph {et~al.}(2001)\citenamefont
  {Kudrolli}, \citenamefont {Abraham},\ and\ \citenamefont
  {Gollub}}]{kudrolli_phys.rev.e_63_026208_2001}%
  \BibitemOpen
  \bibfield  {author} {\bibinfo {author} {\bibfnamefont {A.}~\bibnamefont
  {Kudrolli}}, \bibinfo {author} {\bibfnamefont {M.~C.}\ \bibnamefont
  {Abraham}},\ and\ \bibinfo {author} {\bibfnamefont {J.~P.}\ \bibnamefont
  {Gollub}},\ }\bibfield  {title} {\bibinfo {title} {Scarred patterns in
  surface waves},\ }\href {https://doi.org/10.1103/PhysRevE.63.026208}
  {\bibfield  {journal} {\bibinfo  {journal} {Phys. Rev. E}\ }\textbf {\bibinfo
  {volume} {63}},\ \bibinfo {pages} {026208} (\bibinfo {year}
  {2001})}\BibitemShut {NoStop}%
\bibitem [{\citenamefont {Wisniacki}\ and\ \citenamefont
  {Carlo}(2008)}]{wisniacki_phys.rev.E_77_045201_2008}%
  \BibitemOpen
  \bibfield  {author} {\bibinfo {author} {\bibfnamefont {D.}~\bibnamefont
  {Wisniacki}}\ and\ \bibinfo {author} {\bibfnamefont {G.~G.}\ \bibnamefont
  {Carlo}},\ }\bibfield  {title} {\bibinfo {title} {Scarring in open quantum
  systems},\ }\href {https://doi.org/10.1103/PhysRevE.77.045201} {\bibfield
  {journal} {\bibinfo  {journal} {Phys. Rev. E}\ }\textbf {\bibinfo {volume}
  {77}},\ \bibinfo {pages} {045201} (\bibinfo {year} {2008})}\BibitemShut
  {NoStop}%
\bibitem [{\citenamefont {Akis}\ \emph {et~al.}(1997)\citenamefont {Akis},
  \citenamefont {Ferry},\ and\ \citenamefont
  {Bird}}]{akis_phys.rev.lett_79_123_1997}%
  \BibitemOpen
  \bibfield  {author} {\bibinfo {author} {\bibfnamefont {R.}~\bibnamefont
  {Akis}}, \bibinfo {author} {\bibfnamefont {D.~K.}\ \bibnamefont {Ferry}},\
  and\ \bibinfo {author} {\bibfnamefont {J.~P.}\ \bibnamefont {Bird}},\
  }\bibfield  {title} {\bibinfo {title} {Wave function scarring effects in open
  stadium shaped quantum dots},\ }\href
  {https://doi.org/10.1103/PhysRevLett.79.123} {\bibfield  {journal} {\bibinfo
  {journal} {Phys. Rev. Lett.}\ }\textbf {\bibinfo {volume} {79}},\ \bibinfo
  {pages} {123} (\bibinfo {year} {1997})}\BibitemShut {NoStop}%
\bibitem [{\citenamefont {Burke}\ \emph {et~al.}(2010)\citenamefont {Burke},
  \citenamefont {Akis}, \citenamefont {Day}, \citenamefont {Speyer},
  \citenamefont {Ferry},\ and\ \citenamefont
  {Bennett}}]{burke_phys.rev.lett_104_176801_2010}%
  \BibitemOpen
  \bibfield  {author} {\bibinfo {author} {\bibfnamefont {A.~M.}\ \bibnamefont
  {Burke}}, \bibinfo {author} {\bibfnamefont {R.}~\bibnamefont {Akis}},
  \bibinfo {author} {\bibfnamefont {T.~E.}\ \bibnamefont {Day}}, \bibinfo
  {author} {\bibfnamefont {G.}~\bibnamefont {Speyer}}, \bibinfo {author}
  {\bibfnamefont {D.~K.}\ \bibnamefont {Ferry}},\ and\ \bibinfo {author}
  {\bibfnamefont {B.~R.}\ \bibnamefont {Bennett}},\ }\bibfield  {title}
  {\bibinfo {title} {Periodic scarred states in open quantum dots as evidence
  of quantum darwinism},\ }\href
  {https://doi.org/10.1103/PhysRevLett.104.176801} {\bibfield  {journal}
  {\bibinfo  {journal} {Phys. Rev. Lett.}\ }\textbf {\bibinfo {volume} {104}},\
  \bibinfo {pages} {176801} (\bibinfo {year} {2010})}\BibitemShut {NoStop}%
\bibitem [{\citenamefont {Zurek}(2003)}]{zurek_rev.mod.phys_75_715_2003}%
  \BibitemOpen
  \bibfield  {author} {\bibinfo {author} {\bibfnamefont {W.~H.}\ \bibnamefont
  {Zurek}},\ }\bibfield  {title} {\bibinfo {title} {Decoherence, einselection,
  and the quantum origins of the classical},\ }\href
  {https://doi.org/10.1103/RevModPhys.75.715} {\bibfield  {journal} {\bibinfo
  {journal} {Rev. Mod. Phys.}\ }\textbf {\bibinfo {volume} {75}},\ \bibinfo
  {pages} {715} (\bibinfo {year} {2003})}\BibitemShut {NoStop}%
\bibitem [{\citenamefont {Ferry}\ \emph {et~al.}(2004)\citenamefont {Ferry},
  \citenamefont {Akis},\ and\ \citenamefont
  {Bird}}]{ferry_phys.rev.lett_93_026803_2004}%
  \BibitemOpen
  \bibfield  {author} {\bibinfo {author} {\bibfnamefont {D.~K.}\ \bibnamefont
  {Ferry}}, \bibinfo {author} {\bibfnamefont {R.}~\bibnamefont {Akis}},\ and\
  \bibinfo {author} {\bibfnamefont {J.~P.}\ \bibnamefont {Bird}},\ }\bibfield
  {title} {\bibinfo {title} {Einselection in action: Decoherence and pointer
  states in open quantum dots},\ }\href
  {https://doi.org/10.1103/PhysRevLett.93.026803} {\bibfield  {journal}
  {\bibinfo  {journal} {Phys. Rev. Lett.}\ }\textbf {\bibinfo {volume} {93}},\
  \bibinfo {pages} {026803} (\bibinfo {year} {2004})}\BibitemShut {NoStop}%
\bibitem [{\citenamefont {Brunner}\ \emph {et~al.}(2008)\citenamefont
  {Brunner}, \citenamefont {Akis}, \citenamefont {Ferry}, \citenamefont
  {Kuchar},\ and\ \citenamefont
  {Meisels}}]{brunner_phys.rev.lett_101_024102_2008}%
  \BibitemOpen
  \bibfield  {author} {\bibinfo {author} {\bibfnamefont {R.}~\bibnamefont
  {Brunner}}, \bibinfo {author} {\bibfnamefont {R.}~\bibnamefont {Akis}},
  \bibinfo {author} {\bibfnamefont {D.~K.}\ \bibnamefont {Ferry}}, \bibinfo
  {author} {\bibfnamefont {F.}~\bibnamefont {Kuchar}},\ and\ \bibinfo {author}
  {\bibfnamefont {R.}~\bibnamefont {Meisels}},\ }\bibfield  {title} {\bibinfo
  {title} {Coupling-induced bipartite pointer states in arrays of electron
  billiards: Quantum darwinism in action?},\ }\href
  {https://doi.org/10.1103/PhysRevLett.101.024102} {\bibfield  {journal}
  {\bibinfo  {journal} {Phys. Rev. Lett.}\ }\textbf {\bibinfo {volume} {101}},\
  \bibinfo {pages} {024102} (\bibinfo {year} {2008})}\BibitemShut {NoStop}%
\bibitem [{\citenamefont
  {Kaplan}(1999{\natexlab{b}})}]{kaplan_phys.rev.e_59_5325_1999}%
  \BibitemOpen
  \bibfield  {author} {\bibinfo {author} {\bibfnamefont {L.}~\bibnamefont
  {Kaplan}},\ }\bibfield  {title} {\bibinfo {title} {Scar and antiscar quantum
  effects in open chaotic systems},\ }\href
  {https://doi.org/10.1103/PhysRevE.59.5325} {\bibfield  {journal} {\bibinfo
  {journal} {Phys. Rev. E}\ }\textbf {\bibinfo {volume} {59}},\ \bibinfo
  {pages} {5325} (\bibinfo {year} {1999}{\natexlab{b}})}\BibitemShut {NoStop}%
\bibitem [{\citenamefont {Fromhold}\ \emph {et~al.}(1994)\citenamefont
  {Fromhold}, \citenamefont {Eaves}, \citenamefont {Sheard}, \citenamefont
  {Leadbeater}, \citenamefont {Foster},\ and\ \citenamefont
  {Main}}]{fromhold_phys.rev.lett_72_2608_1994}%
  \BibitemOpen
  \bibfield  {author} {\bibinfo {author} {\bibfnamefont {T.~M.}\ \bibnamefont
  {Fromhold}}, \bibinfo {author} {\bibfnamefont {L.}~\bibnamefont {Eaves}},
  \bibinfo {author} {\bibfnamefont {F.~W.}\ \bibnamefont {Sheard}}, \bibinfo
  {author} {\bibfnamefont {M.~L.}\ \bibnamefont {Leadbeater}}, \bibinfo
  {author} {\bibfnamefont {T.~J.}\ \bibnamefont {Foster}},\ and\ \bibinfo
  {author} {\bibfnamefont {P.~C.}\ \bibnamefont {Main}},\ }\bibfield  {title}
  {\bibinfo {title} {Magnetotunneling spectroscopy of a quantum well in the
  regime of classical chaos},\ }\href
  {https://doi.org/10.1103/PhysRevLett.72.2608} {\bibfield  {journal} {\bibinfo
   {journal} {Phys. Rev. Lett.}\ }\textbf {\bibinfo {volume} {72}},\ \bibinfo
  {pages} {2608} (\bibinfo {year} {1994})}\BibitemShut {NoStop}%
\bibitem [{\citenamefont {Baranger}\ \emph {et~al.}(1993)\citenamefont
  {Baranger}, \citenamefont {Jalabert},\ and\ \citenamefont
  {Stone}}]{baranger_phys.rev.lett_70_3876_1993}%
  \BibitemOpen
  \bibfield  {author} {\bibinfo {author} {\bibfnamefont {H.~U.}\ \bibnamefont
  {Baranger}}, \bibinfo {author} {\bibfnamefont {R.~A.}\ \bibnamefont
  {Jalabert}},\ and\ \bibinfo {author} {\bibfnamefont {A.~D.}\ \bibnamefont
  {Stone}},\ }\bibfield  {title} {\bibinfo {title} {Weak localization and
  integrability in ballistic cavities},\ }\href
  {https://doi.org/10.1103/PhysRevLett.70.3876} {\bibfield  {journal} {\bibinfo
   {journal} {Phys. Rev. Lett.}\ }\textbf {\bibinfo {volume} {70}},\ \bibinfo
  {pages} {3876} (\bibinfo {year} {1993})}\BibitemShut {NoStop}%
\bibitem [{\citenamefont {Marcus}\ \emph {et~al.}(1992)\citenamefont {Marcus},
  \citenamefont {Rimberg}, \citenamefont {Westervelt}, \citenamefont
  {Hopkins},\ and\ \citenamefont {Gossard}}]{marcus_phys.rev.lett_69_506_1992}%
  \BibitemOpen
  \bibfield  {author} {\bibinfo {author} {\bibfnamefont {C.~M.}\ \bibnamefont
  {Marcus}}, \bibinfo {author} {\bibfnamefont {A.~J.}\ \bibnamefont {Rimberg}},
  \bibinfo {author} {\bibfnamefont {R.~M.}\ \bibnamefont {Westervelt}},
  \bibinfo {author} {\bibfnamefont {P.~F.}\ \bibnamefont {Hopkins}},\ and\
  \bibinfo {author} {\bibfnamefont {A.~C.}\ \bibnamefont {Gossard}},\
  }\bibfield  {title} {\bibinfo {title} {Conductance fluctuations and chaotic
  scattering in ballistic microstructures},\ }\href
  {https://doi.org/10.1103/PhysRevLett.69.506} {\bibfield  {journal} {\bibinfo
  {journal} {Phys. Rev. Lett.}\ }\textbf {\bibinfo {volume} {69}},\ \bibinfo
  {pages} {506} (\bibinfo {year} {1992})}\BibitemShut {NoStop}%
\bibitem [{\citenamefont {Fromhold}\ \emph
  {et~al.}(1995{\natexlab{b}})\citenamefont {Fromhold}, \citenamefont
  {Wilkinson}, \citenamefont {Sheard}, \citenamefont {Eaves}, \citenamefont
  {Miao},\ and\ \citenamefont {Edwards}}]{fromhold_phys.rev.lett_75_1142_1995}%
  \BibitemOpen
  \bibfield  {author} {\bibinfo {author} {\bibfnamefont {T.~M.}\ \bibnamefont
  {Fromhold}}, \bibinfo {author} {\bibfnamefont {P.~B.}\ \bibnamefont
  {Wilkinson}}, \bibinfo {author} {\bibfnamefont {F.~W.}\ \bibnamefont
  {Sheard}}, \bibinfo {author} {\bibfnamefont {L.}~\bibnamefont {Eaves}},
  \bibinfo {author} {\bibfnamefont {J.}~\bibnamefont {Miao}},\ and\ \bibinfo
  {author} {\bibfnamefont {G.}~\bibnamefont {Edwards}},\ }\bibfield  {title}
  {\bibinfo {title} {Manifestations of classical chaos in the energy level
  spectrum of a quantum well},\ }\href
  {https://doi.org/10.1103/PhysRevLett.75.1142} {\bibfield  {journal} {\bibinfo
   {journal} {Phys. Rev. Lett.}\ }\textbf {\bibinfo {volume} {75}},\ \bibinfo
  {pages} {1142} (\bibinfo {year} {1995}{\natexlab{b}})}\BibitemShut {NoStop}%
\bibitem [{\citenamefont {Jalabert}\ \emph {et~al.}(1990)\citenamefont
  {Jalabert}, \citenamefont {Baranger},\ and\ \citenamefont
  {Stone}}]{jalabert_phys.rev.lett_65_2442_1990}%
  \BibitemOpen
  \bibfield  {author} {\bibinfo {author} {\bibfnamefont {R.~A.}\ \bibnamefont
  {Jalabert}}, \bibinfo {author} {\bibfnamefont {H.~U.}\ \bibnamefont
  {Baranger}},\ and\ \bibinfo {author} {\bibfnamefont {A.~D.}\ \bibnamefont
  {Stone}},\ }\bibfield  {title} {\bibinfo {title} {Conductance fluctuations in
  the ballistic regime: A probe of quantum chaos?},\ }\href
  {https://doi.org/10.1103/PhysRevLett.65.2442} {\bibfield  {journal} {\bibinfo
   {journal} {Phys. Rev. Lett.}\ }\textbf {\bibinfo {volume} {65}},\ \bibinfo
  {pages} {2442} (\bibinfo {year} {1990})}\BibitemShut {NoStop}%
\bibitem [{\citenamefont {Bernien}\ \emph {et~al.}(2017)\citenamefont
  {Bernien}, \citenamefont {Schwartz}, \citenamefont {Keesling}, \citenamefont
  {Levine}, \citenamefont {Omran}, \citenamefont {Pichler}, \citenamefont
  {Choi}, \citenamefont {Zibrov}, \citenamefont {Endres}, \citenamefont
  {Greiner} \emph {et~al.}}]{bernien_nature_551_579_2017}%
  \BibitemOpen
  \bibfield  {author} {\bibinfo {author} {\bibfnamefont {H.}~\bibnamefont
  {Bernien}}, \bibinfo {author} {\bibfnamefont {S.}~\bibnamefont {Schwartz}},
  \bibinfo {author} {\bibfnamefont {A.}~\bibnamefont {Keesling}}, \bibinfo
  {author} {\bibfnamefont {H.}~\bibnamefont {Levine}}, \bibinfo {author}
  {\bibfnamefont {A.}~\bibnamefont {Omran}}, \bibinfo {author} {\bibfnamefont
  {H.}~\bibnamefont {Pichler}}, \bibinfo {author} {\bibfnamefont
  {S.}~\bibnamefont {Choi}}, \bibinfo {author} {\bibfnamefont {A.~S.}\
  \bibnamefont {Zibrov}}, \bibinfo {author} {\bibfnamefont {M.}~\bibnamefont
  {Endres}}, \bibinfo {author} {\bibfnamefont {M.}~\bibnamefont {Greiner}},
  \emph {et~al.},\ }\bibfield  {title} {\bibinfo {title} {Probing many-body
  dynamics on a 51-atom quantum simulator},\ }\href@noop {} {\bibfield
  {journal} {\bibinfo  {journal} {Nature}\ }\textbf {\bibinfo {volume} {551}},\
  \bibinfo {pages} {579} (\bibinfo {year} {2017})}\BibitemShut {NoStop}%
\bibitem [{\citenamefont {Turner}\ \emph
  {et~al.}(2018{\natexlab{a}})\citenamefont {Turner}, \citenamefont
  {Michailidis}, \citenamefont {Abanin}, \citenamefont {Serbyn},\ and\
  \citenamefont {Papi{\'c}}}]{turner_nat.phys_14_745_2018}%
  \BibitemOpen
  \bibfield  {author} {\bibinfo {author} {\bibfnamefont {C.~J.}\ \bibnamefont
  {Turner}}, \bibinfo {author} {\bibfnamefont {A.~A.}\ \bibnamefont
  {Michailidis}}, \bibinfo {author} {\bibfnamefont {D.~A.}\ \bibnamefont
  {Abanin}}, \bibinfo {author} {\bibfnamefont {M.}~\bibnamefont {Serbyn}},\
  and\ \bibinfo {author} {\bibfnamefont {Z.}~\bibnamefont {Papi{\'c}}},\
  }\bibfield  {title} {\bibinfo {title} {Weak ergodicity breaking from quantum
  many-body scars},\ }\href@noop {} {\bibfield  {journal} {\bibinfo  {journal}
  {Nat. Phys.}\ }\textbf {\bibinfo {volume} {14}},\ \bibinfo {pages} {745}
  (\bibinfo {year} {2018}{\natexlab{a}})}\BibitemShut {NoStop}%
\bibitem [{\citenamefont {Ho}\ \emph {et~al.}(2019)\citenamefont {Ho},
  \citenamefont {Choi}, \citenamefont {Pichler},\ and\ \citenamefont
  {Lukin}}]{Ho_phys.rev.lett_122_040603_2019}%
  \BibitemOpen
  \bibfield  {author} {\bibinfo {author} {\bibfnamefont {W.~W.}\ \bibnamefont
  {Ho}}, \bibinfo {author} {\bibfnamefont {S.}~\bibnamefont {Choi}}, \bibinfo
  {author} {\bibfnamefont {H.}~\bibnamefont {Pichler}},\ and\ \bibinfo {author}
  {\bibfnamefont {M.~D.}\ \bibnamefont {Lukin}},\ }\bibfield  {title} {\bibinfo
  {title} {Periodic orbits, entanglement, and quantum many-body scars in
  constrained models: Matrix product state approach},\ }\href
  {https://doi.org/10.1103/PhysRevLett.122.040603} {\bibfield  {journal}
  {\bibinfo  {journal} {Phys. Rev. Lett.}\ }\textbf {\bibinfo {volume} {122}},\
  \bibinfo {pages} {040603} (\bibinfo {year} {2019})}\BibitemShut {NoStop}%
\bibitem [{\citenamefont {Serbyn}\ \emph {et~al.}(2021)\citenamefont {Serbyn},
  \citenamefont {Abanin},\ and\ \citenamefont
  {Papi{\'c}}}]{serbyn_nat.phys_17_675_2021}%
  \BibitemOpen
  \bibfield  {author} {\bibinfo {author} {\bibfnamefont {M.}~\bibnamefont
  {Serbyn}}, \bibinfo {author} {\bibfnamefont {D.~A.}\ \bibnamefont {Abanin}},\
  and\ \bibinfo {author} {\bibfnamefont {Z.}~\bibnamefont {Papi{\'c}}},\
  }\bibfield  {title} {\bibinfo {title} {Quantum many-body scars and weak
  breaking of ergodicity},\ }\href@noop {} {\bibfield  {journal} {\bibinfo
  {journal} {Nat. Phys.}\ }\textbf {\bibinfo {volume} {17}},\ \bibinfo {pages}
  {675} (\bibinfo {year} {2021})}\BibitemShut {NoStop}%
\bibitem [{\citenamefont {Scherg}\ \emph {et~al.}(2021)\citenamefont {Scherg},
  \citenamefont {Kohlert}, \citenamefont {Sala}, \citenamefont {Pollmann},
  \citenamefont {Hebbe~Madhusudhana}, \citenamefont {Bloch},\ and\
  \citenamefont {Aidelsburger}}]{scherg_nat.commun_12_1_2021}%
  \BibitemOpen
  \bibfield  {author} {\bibinfo {author} {\bibfnamefont {S.}~\bibnamefont
  {Scherg}}, \bibinfo {author} {\bibfnamefont {T.}~\bibnamefont {Kohlert}},
  \bibinfo {author} {\bibfnamefont {P.}~\bibnamefont {Sala}}, \bibinfo {author}
  {\bibfnamefont {F.}~\bibnamefont {Pollmann}}, \bibinfo {author}
  {\bibfnamefont {B.}~\bibnamefont {Hebbe~Madhusudhana}}, \bibinfo {author}
  {\bibfnamefont {I.}~\bibnamefont {Bloch}},\ and\ \bibinfo {author}
  {\bibfnamefont {M.}~\bibnamefont {Aidelsburger}},\ }\bibfield  {title}
  {\bibinfo {title} {Observing non-ergodicity due to kinetic constraints in
  tilted fermi-hubbard chains},\ }\href@noop {} {\bibfield  {journal} {\bibinfo
   {journal} {Nat. Commun.}\ }\textbf {\bibinfo {volume} {12}},\ \bibinfo
  {pages} {1} (\bibinfo {year} {2021})}\BibitemShut {NoStop}%
\bibitem [{\citenamefont {Zhao}\ \emph {et~al.}(2020)\citenamefont {Zhao},
  \citenamefont {Vovrosh}, \citenamefont {Mintert},\ and\ \citenamefont
  {Knolle}}]{zhao_phys.rev.lett_124_160604_2020}%
  \BibitemOpen
  \bibfield  {author} {\bibinfo {author} {\bibfnamefont {H.}~\bibnamefont
  {Zhao}}, \bibinfo {author} {\bibfnamefont {J.}~\bibnamefont {Vovrosh}},
  \bibinfo {author} {\bibfnamefont {F.}~\bibnamefont {Mintert}},\ and\ \bibinfo
  {author} {\bibfnamefont {J.}~\bibnamefont {Knolle}},\ }\bibfield  {title}
  {\bibinfo {title} {Quantum many-body scars in optical lattices},\ }\href
  {https://doi.org/10.1103/PhysRevLett.124.160604} {\bibfield  {journal}
  {\bibinfo  {journal} {Phys. Rev. Lett.}\ }\textbf {\bibinfo {volume} {124}},\
  \bibinfo {pages} {160604} (\bibinfo {year} {2020})}\BibitemShut {NoStop}%
\bibitem [{\citenamefont {Hummel}\ \emph {et~al.}(2023)\citenamefont {Hummel},
  \citenamefont {Richter},\ and\ \citenamefont
  {Schlagheck}}]{Hummel_phys.rev.Lett_130_250402_2023}%
  \BibitemOpen
  \bibfield  {author} {\bibinfo {author} {\bibfnamefont {Q.}~\bibnamefont
  {Hummel}}, \bibinfo {author} {\bibfnamefont {K.}~\bibnamefont {Richter}},\
  and\ \bibinfo {author} {\bibfnamefont {P.}~\bibnamefont {Schlagheck}},\
  }\bibfield  {title} {\bibinfo {title} {Genuine many-body quantum scars along
  unstable modes in bose-hubbard systems},\ }\href
  {https://doi.org/10.1103/PhysRevLett.130.250402} {\bibfield  {journal}
  {\bibinfo  {journal} {Phys. Rev. Lett.}\ }\textbf {\bibinfo {volume} {130}},\
  \bibinfo {pages} {250402} (\bibinfo {year} {2023})}\BibitemShut {NoStop}%
\bibitem [{\citenamefont {Evrard}\ \emph {et~al.}(2024)\citenamefont {Evrard},
  \citenamefont {Pizzi}, \citenamefont {Mistakidis},\ and\ \citenamefont
  {Dag}}]{Evrard_phys.rev.lett_132_020401_2024}%
  \BibitemOpen
  \bibfield  {author} {\bibinfo {author} {\bibfnamefont {B.}~\bibnamefont
  {Evrard}}, \bibinfo {author} {\bibfnamefont {A.}~\bibnamefont {Pizzi}},
  \bibinfo {author} {\bibfnamefont {S.~I.}\ \bibnamefont {Mistakidis}},\ and\
  \bibinfo {author} {\bibfnamefont {C.~B.}\ \bibnamefont {Dag}},\ }\bibfield
  {title} {\bibinfo {title} {Quantum scars and regular eigenstates in a chaotic
  spinor condensate},\ }\href {https://doi.org/10.1103/PhysRevLett.132.020401}
  {\bibfield  {journal} {\bibinfo  {journal} {Phys. Rev. Lett.}\ }\textbf
  {\bibinfo {volume} {132}},\ \bibinfo {pages} {020401} (\bibinfo {year}
  {2024})}\BibitemShut {NoStop}%
\bibitem [{\citenamefont {Alhambra}\ \emph {et~al.}(2020)\citenamefont
  {Alhambra}, \citenamefont {Anshu},\ and\ \citenamefont
  {Wilming}}]{alhambra_phys.rev.b_101_205107_2020}%
  \BibitemOpen
  \bibfield  {author} {\bibinfo {author} {\bibfnamefont {A.~M.}\ \bibnamefont
  {Alhambra}}, \bibinfo {author} {\bibfnamefont {A.}~\bibnamefont {Anshu}},\
  and\ \bibinfo {author} {\bibfnamefont {H.}~\bibnamefont {Wilming}},\
  }\bibfield  {title} {\bibinfo {title} {Revivals imply quantum many-body
  scars},\ }\href {https://doi.org/10.1103/PhysRevB.101.205107} {\bibfield
  {journal} {\bibinfo  {journal} {Phys. Rev. B}\ }\textbf {\bibinfo {volume}
  {101}},\ \bibinfo {pages} {205107} (\bibinfo {year} {2020})}\BibitemShut
  {NoStop}%
\bibitem [{\citenamefont {Pai}\ and\ \citenamefont
  {Pretko}(2019)}]{pai_phys.rev.lett_123_136401_2019}%
  \BibitemOpen
  \bibfield  {author} {\bibinfo {author} {\bibfnamefont {S.}~\bibnamefont
  {Pai}}\ and\ \bibinfo {author} {\bibfnamefont {M.}~\bibnamefont {Pretko}},\
  }\bibfield  {title} {\bibinfo {title} {Dynamical scar states in driven
  fracton systems},\ }\href {https://doi.org/10.1103/PhysRevLett.123.136401}
  {\bibfield  {journal} {\bibinfo  {journal} {Phys. Rev. Lett.}\ }\textbf
  {\bibinfo {volume} {123}},\ \bibinfo {pages} {136401} (\bibinfo {year}
  {2019})}\BibitemShut {NoStop}%
\bibitem [{\citenamefont {Bull}\ \emph {et~al.}(2019)\citenamefont {Bull},
  \citenamefont {Martin},\ and\ \citenamefont {Papi\ifmmode~\acute{c}\else
  \'{c}\fi{}}}]{bull_phys.rev.lett_123_030601_2019}%
  \BibitemOpen
  \bibfield  {author} {\bibinfo {author} {\bibfnamefont {K.}~\bibnamefont
  {Bull}}, \bibinfo {author} {\bibfnamefont {I.}~\bibnamefont {Martin}},\ and\
  \bibinfo {author} {\bibfnamefont {Z.}~\bibnamefont
  {Papi\ifmmode~\acute{c}\else \'{c}\fi{}}},\ }\bibfield  {title} {\bibinfo
  {title} {Systematic construction of scarred many-body dynamics in 1d lattice
  models},\ }\href {https://doi.org/10.1103/PhysRevLett.123.030601} {\bibfield
  {journal} {\bibinfo  {journal} {Phys. Rev. Lett.}\ }\textbf {\bibinfo
  {volume} {123}},\ \bibinfo {pages} {030601} (\bibinfo {year}
  {2019})}\BibitemShut {NoStop}%
\bibitem [{\citenamefont {Ok}\ \emph {et~al.}(2019)\citenamefont {Ok},
  \citenamefont {Choo}, \citenamefont {Mudry}, \citenamefont {Castelnovo},
  \citenamefont {Chamon},\ and\ \citenamefont
  {Neupert}}]{ok_phys.rev.research_1_033144_2019}%
  \BibitemOpen
  \bibfield  {author} {\bibinfo {author} {\bibfnamefont {S.}~\bibnamefont
  {Ok}}, \bibinfo {author} {\bibfnamefont {K.}~\bibnamefont {Choo}}, \bibinfo
  {author} {\bibfnamefont {C.}~\bibnamefont {Mudry}}, \bibinfo {author}
  {\bibfnamefont {C.}~\bibnamefont {Castelnovo}}, \bibinfo {author}
  {\bibfnamefont {C.}~\bibnamefont {Chamon}},\ and\ \bibinfo {author}
  {\bibfnamefont {T.}~\bibnamefont {Neupert}},\ }\bibfield  {title} {\bibinfo
  {title} {Topological many-body scar states in dimensions one, two, and
  three},\ }\href {https://doi.org/10.1103/PhysRevResearch.1.033144} {\bibfield
   {journal} {\bibinfo  {journal} {Phys. Rev. Research}\ }\textbf {\bibinfo
  {volume} {1}},\ \bibinfo {pages} {033144} (\bibinfo {year}
  {2019})}\BibitemShut {NoStop}%
\bibitem [{\citenamefont {Mukherjee}\ \emph {et~al.}(2020)\citenamefont
  {Mukherjee}, \citenamefont {Nandy}, \citenamefont {Sen}, \citenamefont
  {Sen},\ and\ \citenamefont
  {Sengupta}}]{mukherjee_phys.rev.b_101_245107_2020}%
  \BibitemOpen
  \bibfield  {author} {\bibinfo {author} {\bibfnamefont {B.}~\bibnamefont
  {Mukherjee}}, \bibinfo {author} {\bibfnamefont {S.}~\bibnamefont {Nandy}},
  \bibinfo {author} {\bibfnamefont {A.}~\bibnamefont {Sen}}, \bibinfo {author}
  {\bibfnamefont {D.}~\bibnamefont {Sen}},\ and\ \bibinfo {author}
  {\bibfnamefont {K.}~\bibnamefont {Sengupta}},\ }\bibfield  {title} {\bibinfo
  {title} {Collapse and revival of quantum many-body scars via floquet
  engineering},\ }\href {https://doi.org/10.1103/PhysRevB.101.245107}
  {\bibfield  {journal} {\bibinfo  {journal} {Phys. Rev. B}\ }\textbf {\bibinfo
  {volume} {101}},\ \bibinfo {pages} {245107} (\bibinfo {year}
  {2020})}\BibitemShut {NoStop}%
\bibitem [{\citenamefont {Mark}\ \emph {et~al.}(2020)\citenamefont {Mark},
  \citenamefont {Lin},\ and\ \citenamefont
  {Motrunich}}]{mark_phys.rev.b_101_195131_2020}%
  \BibitemOpen
  \bibfield  {author} {\bibinfo {author} {\bibfnamefont {D.~K.}\ \bibnamefont
  {Mark}}, \bibinfo {author} {\bibfnamefont {C.-J.}\ \bibnamefont {Lin}},\ and\
  \bibinfo {author} {\bibfnamefont {O.~I.}\ \bibnamefont {Motrunich}},\
  }\bibfield  {title} {\bibinfo {title} {Unified structure for exact towers of
  scar states in the affleck-kennedy-lieb-tasaki and other models},\ }\href
  {https://doi.org/10.1103/PhysRevB.101.195131} {\bibfield  {journal} {\bibinfo
   {journal} {Phys. Rev. B}\ }\textbf {\bibinfo {volume} {101}},\ \bibinfo
  {pages} {195131} (\bibinfo {year} {2020})}\BibitemShut {NoStop}%
\bibitem [{\citenamefont {Hudomal}\ \emph {et~al.}(2020)\citenamefont
  {Hudomal}, \citenamefont {Vasi{\'c}}, \citenamefont {Regnault},\ and\
  \citenamefont {Papi{\'c}}}]{hudomal_commun.phys_3_1_2020}%
  \BibitemOpen
  \bibfield  {author} {\bibinfo {author} {\bibfnamefont {A.}~\bibnamefont
  {Hudomal}}, \bibinfo {author} {\bibfnamefont {I.}~\bibnamefont {Vasi{\'c}}},
  \bibinfo {author} {\bibfnamefont {N.}~\bibnamefont {Regnault}},\ and\
  \bibinfo {author} {\bibfnamefont {Z.}~\bibnamefont {Papi{\'c}}},\ }\bibfield
  {title} {\bibinfo {title} {Quantum scars of bosons with correlated hopping},\
  }\href@noop {} {\bibfield  {journal} {\bibinfo  {journal} {Commun. Phys.}\
  }\textbf {\bibinfo {volume} {3}},\ \bibinfo {pages} {1} (\bibinfo {year}
  {2020})}\BibitemShut {NoStop}%
\bibitem [{\citenamefont {van Voorden}\ \emph {et~al.}(2020)\citenamefont {van
  Voorden}, \citenamefont {Min\'a\ifmmode~\check{r}\else \v{r}\fi{}},\ and\
  \citenamefont {Schoutens}}]{voorden_phys.rev.b_101_220305_2020}%
  \BibitemOpen
  \bibfield  {author} {\bibinfo {author} {\bibfnamefont {B.}~\bibnamefont {van
  Voorden}}, \bibinfo {author} {\bibfnamefont {J.~c.~v.}\ \bibnamefont
  {Min\'a\ifmmode~\check{r}\else \v{r}\fi{}}},\ and\ \bibinfo {author}
  {\bibfnamefont {K.}~\bibnamefont {Schoutens}},\ }\bibfield  {title} {\bibinfo
  {title} {Quantum many-body scars in transverse field ising ladders and
  beyond},\ }\href {https://doi.org/10.1103/PhysRevB.101.220305} {\bibfield
  {journal} {\bibinfo  {journal} {Phys. Rev. B}\ }\textbf {\bibinfo {volume}
  {101}},\ \bibinfo {pages} {220305} (\bibinfo {year} {2020})}\BibitemShut
  {NoStop}%
\bibitem [{\citenamefont {Lee}\ \emph {et~al.}(2020)\citenamefont {Lee},
  \citenamefont {Melendrez}, \citenamefont {Pal},\ and\ \citenamefont
  {Changlani}}]{lee_phys.rev.b_101_241111_2020}%
  \BibitemOpen
  \bibfield  {author} {\bibinfo {author} {\bibfnamefont {K.}~\bibnamefont
  {Lee}}, \bibinfo {author} {\bibfnamefont {R.}~\bibnamefont {Melendrez}},
  \bibinfo {author} {\bibfnamefont {A.}~\bibnamefont {Pal}},\ and\ \bibinfo
  {author} {\bibfnamefont {H.~J.}\ \bibnamefont {Changlani}},\ }\bibfield
  {title} {\bibinfo {title} {Exact three-colored quantum scars from geometric
  frustration},\ }\href {https://doi.org/10.1103/PhysRevB.101.241111}
  {\bibfield  {journal} {\bibinfo  {journal} {Phys. Rev. B}\ }\textbf {\bibinfo
  {volume} {101}},\ \bibinfo {pages} {241111} (\bibinfo {year}
  {2020})}\BibitemShut {NoStop}%
\bibitem [{\citenamefont {Shiraishi}\ and\ \citenamefont
  {Mori}(2017)}]{shiraishi_phys.rev.lett_119_030601_2017}%
  \BibitemOpen
  \bibfield  {author} {\bibinfo {author} {\bibfnamefont {N.}~\bibnamefont
  {Shiraishi}}\ and\ \bibinfo {author} {\bibfnamefont {T.}~\bibnamefont
  {Mori}},\ }\bibfield  {title} {\bibinfo {title} {Systematic construction of
  counterexamples to the eigenstate thermalization hypothesis},\ }\href
  {https://doi.org/10.1103/PhysRevLett.119.030601} {\bibfield  {journal}
  {\bibinfo  {journal} {Phys. Rev. Lett.}\ }\textbf {\bibinfo {volume} {119}},\
  \bibinfo {pages} {030601} (\bibinfo {year} {2017})}\BibitemShut {NoStop}%
\bibitem [{\citenamefont {Moudgalya}\ \emph {et~al.}(2018)\citenamefont
  {Moudgalya}, \citenamefont {Regnault},\ and\ \citenamefont
  {Bernevig}}]{moudgalya_phys.rev.b_98_235156_2018}%
  \BibitemOpen
  \bibfield  {author} {\bibinfo {author} {\bibfnamefont {S.}~\bibnamefont
  {Moudgalya}}, \bibinfo {author} {\bibfnamefont {N.}~\bibnamefont
  {Regnault}},\ and\ \bibinfo {author} {\bibfnamefont {B.~A.}\ \bibnamefont
  {Bernevig}},\ }\bibfield  {title} {\bibinfo {title} {Entanglement of exact
  excited states of affleck-kennedy-lieb-tasaki models: Exact results,
  many-body scars, and violation of the strong eigenstate thermalization
  hypothesis},\ }\href {https://doi.org/10.1103/PhysRevB.98.235156} {\bibfield
  {journal} {\bibinfo  {journal} {Phys. Rev. B}\ }\textbf {\bibinfo {volume}
  {98}},\ \bibinfo {pages} {235156} (\bibinfo {year} {2018})}\BibitemShut
  {NoStop}%
\bibitem [{\citenamefont {Lin}\ and\ \citenamefont
  {Motrunich}(2019)}]{lin_phys.rev.lett_122_173401_2019}%
  \BibitemOpen
  \bibfield  {author} {\bibinfo {author} {\bibfnamefont {C.-J.}\ \bibnamefont
  {Lin}}\ and\ \bibinfo {author} {\bibfnamefont {O.~I.}\ \bibnamefont
  {Motrunich}},\ }\bibfield  {title} {\bibinfo {title} {Exact quantum many-body
  scar states in the rydberg-blockaded atom chain},\ }\href
  {https://doi.org/10.1103/PhysRevLett.122.173401} {\bibfield  {journal}
  {\bibinfo  {journal} {Phys. Rev. Lett.}\ }\textbf {\bibinfo {volume} {122}},\
  \bibinfo {pages} {173401} (\bibinfo {year} {2019})}\BibitemShut {NoStop}%
\bibitem [{\citenamefont {Chattopadhyay}\ \emph {et~al.}(2020)\citenamefont
  {Chattopadhyay}, \citenamefont {Pichler}, \citenamefont {Lukin},\ and\
  \citenamefont {Ho}}]{chattopadhyay_phys.rev.b_101_174308_2020}%
  \BibitemOpen
  \bibfield  {author} {\bibinfo {author} {\bibfnamefont {S.}~\bibnamefont
  {Chattopadhyay}}, \bibinfo {author} {\bibfnamefont {H.}~\bibnamefont
  {Pichler}}, \bibinfo {author} {\bibfnamefont {M.~D.}\ \bibnamefont {Lukin}},\
  and\ \bibinfo {author} {\bibfnamefont {W.~W.}\ \bibnamefont {Ho}},\
  }\bibfield  {title} {\bibinfo {title} {Quantum many-body scars from virtual
  entangled pairs},\ }\href {https://doi.org/10.1103/PhysRevB.101.174308}
  {\bibfield  {journal} {\bibinfo  {journal} {Phys. Rev. B}\ }\textbf {\bibinfo
  {volume} {101}},\ \bibinfo {pages} {174308} (\bibinfo {year}
  {2020})}\BibitemShut {NoStop}%
\bibitem [{\citenamefont {Mizuta}\ \emph {et~al.}(2020)\citenamefont {Mizuta},
  \citenamefont {Takasan},\ and\ \citenamefont
  {Kawakami}}]{mizuta_phys.rev.research_2_033284_2020}%
  \BibitemOpen
  \bibfield  {author} {\bibinfo {author} {\bibfnamefont {K.}~\bibnamefont
  {Mizuta}}, \bibinfo {author} {\bibfnamefont {K.}~\bibnamefont {Takasan}},\
  and\ \bibinfo {author} {\bibfnamefont {N.}~\bibnamefont {Kawakami}},\
  }\bibfield  {title} {\bibinfo {title} {Exact floquet quantum many-body scars
  under rydberg blockade},\ }\href
  {https://doi.org/10.1103/PhysRevResearch.2.033284} {\bibfield  {journal}
  {\bibinfo  {journal} {Phys. Rev. Research}\ }\textbf {\bibinfo {volume}
  {2}},\ \bibinfo {pages} {033284} (\bibinfo {year} {2020})}\BibitemShut
  {NoStop}%
\bibitem [{\citenamefont {Moudgalya}\ \emph
  {et~al.}(2020{\natexlab{a}})\citenamefont {Moudgalya}, \citenamefont
  {Regnault},\ and\ \citenamefont
  {Bernevig}}]{moudgalya_phys.revb_102_085140_2020}%
  \BibitemOpen
  \bibfield  {author} {\bibinfo {author} {\bibfnamefont {S.}~\bibnamefont
  {Moudgalya}}, \bibinfo {author} {\bibfnamefont {N.}~\bibnamefont
  {Regnault}},\ and\ \bibinfo {author} {\bibfnamefont {B.~A.}\ \bibnamefont
  {Bernevig}},\ }\bibfield  {title} {\bibinfo {title}
  {$\ensuremath{\eta}$-pairing in hubbard models: From spectrum generating
  algebras to quantum many-body scars},\ }\href
  {https://doi.org/10.1103/PhysRevB.102.085140} {\bibfield  {journal} {\bibinfo
   {journal} {Phys. Rev. B}\ }\textbf {\bibinfo {volume} {102}},\ \bibinfo
  {pages} {085140} (\bibinfo {year} {2020}{\natexlab{a}})}\BibitemShut
  {NoStop}%
\bibitem [{\citenamefont {Moudgalya}\ \emph
  {et~al.}(2020{\natexlab{b}})\citenamefont {Moudgalya}, \citenamefont
  {O'Brien}, \citenamefont {Bernevig}, \citenamefont {Fendley},\ and\
  \citenamefont {Regnault}}]{moudgalya_phys.rev.b_102_085120_2020}%
  \BibitemOpen
  \bibfield  {author} {\bibinfo {author} {\bibfnamefont {S.}~\bibnamefont
  {Moudgalya}}, \bibinfo {author} {\bibfnamefont {E.}~\bibnamefont {O'Brien}},
  \bibinfo {author} {\bibfnamefont {B.~A.}\ \bibnamefont {Bernevig}}, \bibinfo
  {author} {\bibfnamefont {P.}~\bibnamefont {Fendley}},\ and\ \bibinfo {author}
  {\bibfnamefont {N.}~\bibnamefont {Regnault}},\ }\bibfield  {title} {\bibinfo
  {title} {Large classes of quantum scarred hamiltonians from matrix product
  states},\ }\href {https://doi.org/10.1103/PhysRevB.102.085120} {\bibfield
  {journal} {\bibinfo  {journal} {Phys. Rev. B}\ }\textbf {\bibinfo {volume}
  {102}},\ \bibinfo {pages} {085120} (\bibinfo {year}
  {2020}{\natexlab{b}})}\BibitemShut {NoStop}%
\bibitem [{\citenamefont {Iadecola}\ and\ \citenamefont
  {Schecter}(2020)}]{iadecola_phys.rev.b_101_024306_2020}%
  \BibitemOpen
  \bibfield  {author} {\bibinfo {author} {\bibfnamefont {T.}~\bibnamefont
  {Iadecola}}\ and\ \bibinfo {author} {\bibfnamefont {M.}~\bibnamefont
  {Schecter}},\ }\bibfield  {title} {\bibinfo {title} {Quantum many-body scar
  states with emergent kinetic constraints and finite-entanglement revivals},\
  }\href {https://doi.org/10.1103/PhysRevB.101.024306} {\bibfield  {journal}
  {\bibinfo  {journal} {Phys. Rev. B}\ }\textbf {\bibinfo {volume} {101}},\
  \bibinfo {pages} {024306} (\bibinfo {year} {2020})}\BibitemShut {NoStop}%
\bibitem [{\citenamefont {Choi}\ \emph {et~al.}(2019)\citenamefont {Choi},
  \citenamefont {Turner}, \citenamefont {Pichler}, \citenamefont {Ho},
  \citenamefont {Michailidis}, \citenamefont {Papi\ifmmode~\acute{c}\else
  \'{c}\fi{}}, \citenamefont {Serbyn}, \citenamefont {Lukin},\ and\
  \citenamefont {Abanin}}]{choi_phys.rev.lett_122_220603_2019}%
  \BibitemOpen
  \bibfield  {author} {\bibinfo {author} {\bibfnamefont {S.}~\bibnamefont
  {Choi}}, \bibinfo {author} {\bibfnamefont {C.~J.}\ \bibnamefont {Turner}},
  \bibinfo {author} {\bibfnamefont {H.}~\bibnamefont {Pichler}}, \bibinfo
  {author} {\bibfnamefont {W.~W.}\ \bibnamefont {Ho}}, \bibinfo {author}
  {\bibfnamefont {A.~A.}\ \bibnamefont {Michailidis}}, \bibinfo {author}
  {\bibfnamefont {Z.}~\bibnamefont {Papi\ifmmode~\acute{c}\else \'{c}\fi{}}},
  \bibinfo {author} {\bibfnamefont {M.}~\bibnamefont {Serbyn}}, \bibinfo
  {author} {\bibfnamefont {M.~D.}\ \bibnamefont {Lukin}},\ and\ \bibinfo
  {author} {\bibfnamefont {D.~A.}\ \bibnamefont {Abanin}},\ }\bibfield  {title}
  {\bibinfo {title} {Emergent su(2) dynamics and perfect quantum many-body
  scars},\ }\href {https://doi.org/10.1103/PhysRevLett.122.220603} {\bibfield
  {journal} {\bibinfo  {journal} {Phys. Rev. Lett.}\ }\textbf {\bibinfo
  {volume} {122}},\ \bibinfo {pages} {220603} (\bibinfo {year}
  {2019})}\BibitemShut {NoStop}%
\bibitem [{\citenamefont {Bull}\ \emph {et~al.}(2020)\citenamefont {Bull},
  \citenamefont {Desaules},\ and\ \citenamefont {Papi\ifmmode~\acute{c}\else
  \'{c}\fi{}}}]{bull_phys.rev.b_101_165139_2020}%
  \BibitemOpen
  \bibfield  {author} {\bibinfo {author} {\bibfnamefont {K.}~\bibnamefont
  {Bull}}, \bibinfo {author} {\bibfnamefont {J.-Y.}\ \bibnamefont {Desaules}},\
  and\ \bibinfo {author} {\bibfnamefont {Z.}~\bibnamefont
  {Papi\ifmmode~\acute{c}\else \'{c}\fi{}}},\ }\bibfield  {title} {\bibinfo
  {title} {Quantum scars as embeddings of weakly broken lie algebra
  representations},\ }\href {https://doi.org/10.1103/PhysRevB.101.165139}
  {\bibfield  {journal} {\bibinfo  {journal} {Phys. Rev. B}\ }\textbf {\bibinfo
  {volume} {101}},\ \bibinfo {pages} {165139} (\bibinfo {year}
  {2020})}\BibitemShut {NoStop}%
\bibitem [{\citenamefont {O'Dea}\ \emph {et~al.}(2020)\citenamefont {O'Dea},
  \citenamefont {Burnell}, \citenamefont {Chandran},\ and\ \citenamefont
  {Khemani}}]{odea_phys.rev.research_2_043305_2020}%
  \BibitemOpen
  \bibfield  {author} {\bibinfo {author} {\bibfnamefont {N.}~\bibnamefont
  {O'Dea}}, \bibinfo {author} {\bibfnamefont {F.}~\bibnamefont {Burnell}},
  \bibinfo {author} {\bibfnamefont {A.}~\bibnamefont {Chandran}},\ and\
  \bibinfo {author} {\bibfnamefont {V.}~\bibnamefont {Khemani}},\ }\bibfield
  {title} {\bibinfo {title} {From tunnels to towers: Quantum scars from lie
  algebras and $q$-deformed lie algebras},\ }\href
  {https://doi.org/10.1103/PhysRevResearch.2.043305} {\bibfield  {journal}
  {\bibinfo  {journal} {Phys. Rev. Research}\ }\textbf {\bibinfo {volume}
  {2}},\ \bibinfo {pages} {043305} (\bibinfo {year} {2020})}\BibitemShut
  {NoStop}%
\bibitem [{\citenamefont {Khemani}\ \emph {et~al.}(2019)\citenamefont
  {Khemani}, \citenamefont {Laumann},\ and\ \citenamefont
  {Chandran}}]{khemani_phys.rev.b_99_161101_2019}%
  \BibitemOpen
  \bibfield  {author} {\bibinfo {author} {\bibfnamefont {V.}~\bibnamefont
  {Khemani}}, \bibinfo {author} {\bibfnamefont {C.~R.}\ \bibnamefont
  {Laumann}},\ and\ \bibinfo {author} {\bibfnamefont {A.}~\bibnamefont
  {Chandran}},\ }\bibfield  {title} {\bibinfo {title} {Signatures of
  integrability in the dynamics of rydberg-blockaded chains},\ }\href
  {https://doi.org/10.1103/PhysRevB.99.161101} {\bibfield  {journal} {\bibinfo
  {journal} {Phys. Rev. B}\ }\textbf {\bibinfo {volume} {99}},\ \bibinfo
  {pages} {161101} (\bibinfo {year} {2019})}\BibitemShut {NoStop}%
\bibitem [{\citenamefont {Yao}\ \emph {et~al.}(2022)\citenamefont {Yao},
  \citenamefont {Pan}, \citenamefont {Liu},\ and\ \citenamefont
  {Zhai}}]{yao_phys.rev.b_105_125123_2022}%
  \BibitemOpen
  \bibfield  {author} {\bibinfo {author} {\bibfnamefont {Z.}~\bibnamefont
  {Yao}}, \bibinfo {author} {\bibfnamefont {L.}~\bibnamefont {Pan}}, \bibinfo
  {author} {\bibfnamefont {S.}~\bibnamefont {Liu}},\ and\ \bibinfo {author}
  {\bibfnamefont {H.}~\bibnamefont {Zhai}},\ }\bibfield  {title} {\bibinfo
  {title} {Quantum many-body scars and quantum criticality},\ }\href
  {https://doi.org/10.1103/PhysRevB.105.125123} {\bibfield  {journal} {\bibinfo
   {journal} {Phys. Rev. B}\ }\textbf {\bibinfo {volume} {105}},\ \bibinfo
  {pages} {125123} (\bibinfo {year} {2022})}\BibitemShut {NoStop}%
\bibitem [{\citenamefont {Turner}\ \emph {et~al.}(2021)\citenamefont {Turner},
  \citenamefont {Desaules}, \citenamefont {Bull},\ and\ \citenamefont
  {Papi\ifmmode~\acute{c}\else \'{c}\fi{}}}]{turner_phys.rev.x_11_021021_2021}%
  \BibitemOpen
  \bibfield  {author} {\bibinfo {author} {\bibfnamefont {C.~J.}\ \bibnamefont
  {Turner}}, \bibinfo {author} {\bibfnamefont {J.-Y.}\ \bibnamefont
  {Desaules}}, \bibinfo {author} {\bibfnamefont {K.}~\bibnamefont {Bull}},\
  and\ \bibinfo {author} {\bibfnamefont {Z.}~\bibnamefont
  {Papi\ifmmode~\acute{c}\else \'{c}\fi{}}},\ }\bibfield  {title} {\bibinfo
  {title} {Correspondence principle for many-body scars in ultracold rydberg
  atoms},\ }\href {https://doi.org/10.1103/PhysRevX.11.021021} {\bibfield
  {journal} {\bibinfo  {journal} {Phys. Rev. X}\ }\textbf {\bibinfo {volume}
  {11}},\ \bibinfo {pages} {021021} (\bibinfo {year} {2021})}\BibitemShut
  {NoStop}%
\bibitem [{\citenamefont {Michailidis}\ \emph
  {et~al.}(2020{\natexlab{a}})\citenamefont {Michailidis}, \citenamefont
  {Turner}, \citenamefont {Papi\ifmmode~\acute{c}\else \'{c}\fi{}},
  \citenamefont {Abanin},\ and\ \citenamefont
  {Serbyn}}]{michailidis_phys.rev.x_10_011055_2020}%
  \BibitemOpen
  \bibfield  {author} {\bibinfo {author} {\bibfnamefont {A.~A.}\ \bibnamefont
  {Michailidis}}, \bibinfo {author} {\bibfnamefont {C.~J.}\ \bibnamefont
  {Turner}}, \bibinfo {author} {\bibfnamefont {Z.}~\bibnamefont
  {Papi\ifmmode~\acute{c}\else \'{c}\fi{}}}, \bibinfo {author} {\bibfnamefont
  {D.~A.}\ \bibnamefont {Abanin}},\ and\ \bibinfo {author} {\bibfnamefont
  {M.}~\bibnamefont {Serbyn}},\ }\bibfield  {title} {\bibinfo {title} {Slow
  quantum thermalization and many-body revivals from mixed phase space},\
  }\href {https://doi.org/10.1103/PhysRevX.10.011055} {\bibfield  {journal}
  {\bibinfo  {journal} {Phys. Rev. X}\ }\textbf {\bibinfo {volume} {10}},\
  \bibinfo {pages} {011055} (\bibinfo {year} {2020}{\natexlab{a}})}\BibitemShut
  {NoStop}%
\bibitem [{\citenamefont {Desaules}\ \emph {et~al.}(2022)\citenamefont
  {Desaules}, \citenamefont {Pietracaprina}, \citenamefont
  {Papi\ifmmode~\acute{c}\else \'{c}\fi{}}, \citenamefont {Goold},\ and\
  \citenamefont {Pappalardi}}]{desaules_phys.rev.lett_129_020601_2022}%
  \BibitemOpen
  \bibfield  {author} {\bibinfo {author} {\bibfnamefont {J.-Y.}\ \bibnamefont
  {Desaules}}, \bibinfo {author} {\bibfnamefont {F.}~\bibnamefont
  {Pietracaprina}}, \bibinfo {author} {\bibfnamefont {Z.}~\bibnamefont
  {Papi\ifmmode~\acute{c}\else \'{c}\fi{}}}, \bibinfo {author} {\bibfnamefont
  {J.}~\bibnamefont {Goold}},\ and\ \bibinfo {author} {\bibfnamefont
  {S.}~\bibnamefont {Pappalardi}},\ }\bibfield  {title} {\bibinfo {title}
  {Extensive multipartite entanglement from su(2) quantum many-body scars},\
  }\href {https://doi.org/10.1103/PhysRevLett.129.020601} {\bibfield  {journal}
  {\bibinfo  {journal} {Phys. Rev. Lett.}\ }\textbf {\bibinfo {volume} {129}},\
  \bibinfo {pages} {020601} (\bibinfo {year} {2022})}\BibitemShut {NoStop}%
\bibitem [{\citenamefont {Mondragon-Shem}\ \emph {et~al.}(2021)\citenamefont
  {Mondragon-Shem}, \citenamefont {Vavilov},\ and\ \citenamefont
  {Martin}}]{mondrago_PRXQuantum_2_030349_2021}%
  \BibitemOpen
  \bibfield  {author} {\bibinfo {author} {\bibfnamefont {I.}~\bibnamefont
  {Mondragon-Shem}}, \bibinfo {author} {\bibfnamefont {M.~G.}\ \bibnamefont
  {Vavilov}},\ and\ \bibinfo {author} {\bibfnamefont {I.}~\bibnamefont
  {Martin}},\ }\bibfield  {title} {\bibinfo {title} {Fate of quantum many-body
  scars in the presence of disorder},\ }\href
  {https://doi.org/10.1103/PRXQuantum.2.030349} {\bibfield  {journal} {\bibinfo
   {journal} {PRX Quantum}\ }\textbf {\bibinfo {volume} {2}},\ \bibinfo {pages}
  {030349} (\bibinfo {year} {2021})}\BibitemShut {NoStop}%
\bibitem [{\citenamefont {Michailidis}\ \emph
  {et~al.}(2020{\natexlab{b}})\citenamefont {Michailidis}, \citenamefont
  {Turner}, \citenamefont {Papi\ifmmode~\acute{c}\else \'{c}\fi{}},
  \citenamefont {Abanin},\ and\ \citenamefont
  {Serbyn}}]{michalidis_phys.rev.research_2_022065_2020}%
  \BibitemOpen
  \bibfield  {author} {\bibinfo {author} {\bibfnamefont {A.~A.}\ \bibnamefont
  {Michailidis}}, \bibinfo {author} {\bibfnamefont {C.~J.}\ \bibnamefont
  {Turner}}, \bibinfo {author} {\bibfnamefont {Z.}~\bibnamefont
  {Papi\ifmmode~\acute{c}\else \'{c}\fi{}}}, \bibinfo {author} {\bibfnamefont
  {D.~A.}\ \bibnamefont {Abanin}},\ and\ \bibinfo {author} {\bibfnamefont
  {M.}~\bibnamefont {Serbyn}},\ }\bibfield  {title} {\bibinfo {title}
  {Stabilizing two-dimensional quantum scars by deformation and
  synchronization},\ }\href {https://doi.org/10.1103/PhysRevResearch.2.022065}
  {\bibfield  {journal} {\bibinfo  {journal} {Phys. Rev. Research}\ }\textbf
  {\bibinfo {volume} {2}},\ \bibinfo {pages} {022065} (\bibinfo {year}
  {2020}{\natexlab{b}})}\BibitemShut {NoStop}%
\bibitem [{\citenamefont {Turner}\ \emph
  {et~al.}(2018{\natexlab{b}})\citenamefont {Turner}, \citenamefont
  {Michailidis}, \citenamefont {Abanin}, \citenamefont {Serbyn},\ and\
  \citenamefont {Papi\ifmmode~\acute{c}\else
  \'{c}\fi{}}}]{turner_phys.rev.b_98_155134_2018}%
  \BibitemOpen
  \bibfield  {author} {\bibinfo {author} {\bibfnamefont {C.~J.}\ \bibnamefont
  {Turner}}, \bibinfo {author} {\bibfnamefont {A.~A.}\ \bibnamefont
  {Michailidis}}, \bibinfo {author} {\bibfnamefont {D.~A.}\ \bibnamefont
  {Abanin}}, \bibinfo {author} {\bibfnamefont {M.}~\bibnamefont {Serbyn}},\
  and\ \bibinfo {author} {\bibfnamefont {Z.}~\bibnamefont
  {Papi\ifmmode~\acute{c}\else \'{c}\fi{}}},\ }\bibfield  {title} {\bibinfo
  {title} {Quantum scarred eigenstates in a rydberg atom chain: Entanglement,
  breakdown of thermalization, and stability to perturbations},\ }\href
  {https://doi.org/10.1103/PhysRevB.98.155134} {\bibfield  {journal} {\bibinfo
  {journal} {Phys. Rev. B}\ }\textbf {\bibinfo {volume} {98}},\ \bibinfo
  {pages} {155134} (\bibinfo {year} {2018}{\natexlab{b}})}\BibitemShut
  {NoStop}%
\bibitem [{\citenamefont {Luukko}\ \emph {et~al.}(2016)\citenamefont {Luukko},
  \citenamefont {Drury}, \citenamefont {Klales}, \citenamefont {Kaplan},
  \citenamefont {Heller},\ and\ \citenamefont
  {R{\"a}s{\"a}nen}}]{Luukko_sci.rep_6_37656_2016}%
  \BibitemOpen
  \bibfield  {author} {\bibinfo {author} {\bibfnamefont {P.~J.~J.}\
  \bibnamefont {Luukko}}, \bibinfo {author} {\bibfnamefont {B.}~\bibnamefont
  {Drury}}, \bibinfo {author} {\bibfnamefont {A.}~\bibnamefont {Klales}},
  \bibinfo {author} {\bibfnamefont {L.}~\bibnamefont {Kaplan}}, \bibinfo
  {author} {\bibfnamefont {E.~J.}\ \bibnamefont {Heller}},\ and\ \bibinfo
  {author} {\bibfnamefont {E.}~\bibnamefont {R{\"a}s{\"a}nen}},\ }\bibfield
  {title} {\bibinfo {title} {Strong quantum scarring by local impurities},\
  }\href {https://doi.org/10.1038/srep37656} {\bibfield  {journal} {\bibinfo
  {journal} {Sci. Rep.}\ }\textbf {\bibinfo {volume} {6}},\ \bibinfo {pages}
  {37656} (\bibinfo {year} {2016})}\BibitemShut {NoStop}%
\bibitem [{\citenamefont {Keski-Rahkonen}\ \emph {et~al.}(2017)\citenamefont
  {Keski-Rahkonen}, \citenamefont {Luukko}, \citenamefont {Kaplan},
  \citenamefont {Heller},\ and\ \citenamefont
  {R{\"a}s{\"a}nen}}]{keski-rahkonen_phys.rev.b_97_094204_2017}%
  \BibitemOpen
  \bibfield  {author} {\bibinfo {author} {\bibfnamefont {J.}~\bibnamefont
  {Keski-Rahkonen}}, \bibinfo {author} {\bibfnamefont {P.~J.~J.}\ \bibnamefont
  {Luukko}}, \bibinfo {author} {\bibfnamefont {L.}~\bibnamefont {Kaplan}},
  \bibinfo {author} {\bibfnamefont {E.~J.}\ \bibnamefont {Heller}},\ and\
  \bibinfo {author} {\bibfnamefont {E.}~\bibnamefont {R{\"a}s{\"a}nen}},\
  }\bibfield  {title} {\bibinfo {title} {Controllable quantum scars in
  semiconductor quantum dots},\ }\href
  {https://doi.org/10.1103/PhysRevB.96.094204} {\bibfield  {journal} {\bibinfo
  {journal} {Phys. Rev. B}\ }\textbf {\bibinfo {volume} {96}},\ \bibinfo
  {pages} {094204} (\bibinfo {year} {2017})}\BibitemShut {NoStop}%
\bibitem [{\citenamefont {Keski-Rahkonen}\ \emph
  {et~al.}(2019{\natexlab{a}})\citenamefont {Keski-Rahkonen}, \citenamefont
  {Luukko}, \citenamefont {{\AA}berg},\ and\ \citenamefont
  {R{\"a}s{\"a}nen}}]{keski-rahkonen_j.phys.conden.matter_31_105301_2019}%
  \BibitemOpen
  \bibfield  {author} {\bibinfo {author} {\bibfnamefont {J.}~\bibnamefont
  {Keski-Rahkonen}}, \bibinfo {author} {\bibfnamefont {P.~J.~J.}\ \bibnamefont
  {Luukko}}, \bibinfo {author} {\bibfnamefont {S.}~\bibnamefont {{\AA}berg}},\
  and\ \bibinfo {author} {\bibfnamefont {E.}~\bibnamefont {R{\"a}s{\"a}nen}},\
  }\bibfield  {title} {\bibinfo {title} {Effects of scarring on quantum chaos
  in disordered quantum wells},\ }\href
  {https://doi.org/10.1088/1361-648x/aaf9fb} {\bibfield  {journal} {\bibinfo
  {journal} {J. Phys.: Condens. Matter}\ }\textbf {\bibinfo {volume} {31}},\
  \bibinfo {pages} {105301} (\bibinfo {year} {2019}{\natexlab{a}})}\BibitemShut
  {NoStop}%
\bibitem [{\citenamefont {Keski-Rahkonen}\ \emph
  {et~al.}(2019{\natexlab{b}})\citenamefont {Keski-Rahkonen}, \citenamefont
  {Ruhanen}, \citenamefont {Heller},\ and\ \citenamefont
  {R{\"a}s{\"a}nen}}]{keski-rahkonen_phys.rev.lett_123_214101_2019}%
  \BibitemOpen
  \bibfield  {author} {\bibinfo {author} {\bibfnamefont {J.}~\bibnamefont
  {Keski-Rahkonen}}, \bibinfo {author} {\bibfnamefont {A.}~\bibnamefont
  {Ruhanen}}, \bibinfo {author} {\bibfnamefont {E.~J.}\ \bibnamefont
  {Heller}},\ and\ \bibinfo {author} {\bibfnamefont {E.}~\bibnamefont
  {R{\"a}s{\"a}nen}},\ }\bibfield  {title} {\bibinfo {title} {Quantum lissajous
  scars},\ }\href {https://doi.org/10.1103/PhysRevLett.123.214101} {\bibfield
  {journal} {\bibinfo  {journal} {Phys. Rev. Lett.}\ }\textbf {\bibinfo
  {volume} {123}},\ \bibinfo {pages} {214101} (\bibinfo {year}
  {2019}{\natexlab{b}})}\BibitemShut {NoStop}%
\bibitem [{\citenamefont {Luukko}\ and\ \citenamefont
  {Rost}(2017)}]{luukko_phys.rev.lett_119_203001_2017}%
  \BibitemOpen
  \bibfield  {author} {\bibinfo {author} {\bibfnamefont {P.~J.~J.}\
  \bibnamefont {Luukko}}\ and\ \bibinfo {author} {\bibfnamefont {J.-M.}\
  \bibnamefont {Rost}},\ }\bibfield  {title} {\bibinfo {title} {Polyatomic
  trilobite rydberg molecules in a dense random gas},\ }\href
  {https://doi.org/10.1103/PhysRevLett.119.203001} {\bibfield  {journal}
  {\bibinfo  {journal} {Phys. Rev. Lett.}\ }\textbf {\bibinfo {volume} {119}},\
  \bibinfo {pages} {203001} (\bibinfo {year} {2017})}\BibitemShut {NoStop}%
\bibitem [{\citenamefont {Kroetz}\ \emph {et~al.}(2016)\citenamefont {Kroetz},
  \citenamefont {Oliveira}, \citenamefont {Portela},\ and\ \citenamefont
  {Viana}}]{kroetz_phys.rev.E_94_022218_2016}%
  \BibitemOpen
  \bibfield  {author} {\bibinfo {author} {\bibfnamefont {T.}~\bibnamefont
  {Kroetz}}, \bibinfo {author} {\bibfnamefont {H.~A.}\ \bibnamefont
  {Oliveira}}, \bibinfo {author} {\bibfnamefont {J.~S.~E.}\ \bibnamefont
  {Portela}},\ and\ \bibinfo {author} {\bibfnamefont {R.~L.}\ \bibnamefont
  {Viana}},\ }\bibfield  {title} {\bibinfo {title} {Dynamical properties of the
  soft-wall elliptical billiard},\ }\href
  {https://doi.org/10.1103/PhysRevE.94.022218} {\bibfield  {journal} {\bibinfo
  {journal} {Phys. Rev. E}\ }\textbf {\bibinfo {volume} {94}},\ \bibinfo
  {pages} {022218} (\bibinfo {year} {2016})}\BibitemShut {NoStop}%
\bibitem [{\citenamefont {Ahn}\ \emph {et~al.}(1999)\citenamefont {Ahn},
  \citenamefont {Richter},\ and\ \citenamefont
  {Lee}}]{kang_phys.rev.lett_83_4144_1999}%
  \BibitemOpen
  \bibfield  {author} {\bibinfo {author} {\bibfnamefont {K.-H.}\ \bibnamefont
  {Ahn}}, \bibinfo {author} {\bibfnamefont {K.}~\bibnamefont {Richter}},\ and\
  \bibinfo {author} {\bibfnamefont {I.-H.}\ \bibnamefont {Lee}},\ }\bibfield
  {title} {\bibinfo {title} {Addition spectra of chaotic quantum dots:
  Interplay between interactions and geometry},\ }\href
  {https://doi.org/10.1103/PhysRevLett.83.4144} {\bibfield  {journal} {\bibinfo
   {journal} {Phys. Rev. Lett.}\ }\textbf {\bibinfo {volume} {83}},\ \bibinfo
  {pages} {4144} (\bibinfo {year} {1999})}\BibitemShut {NoStop}%
\bibitem [{\citenamefont {Marcus}\ \emph {et~al.}(1993)\citenamefont {Marcus},
  \citenamefont {Westervelt}, \citenamefont {Hopkins},\ and\ \citenamefont
  {Gossard}}]{marcus_chaos_3_643_1993}%
  \BibitemOpen
  \bibfield  {author} {\bibinfo {author} {\bibfnamefont {C.~M.}\ \bibnamefont
  {Marcus}}, \bibinfo {author} {\bibfnamefont {R.~M.}\ \bibnamefont
  {Westervelt}}, \bibinfo {author} {\bibfnamefont {P.~F.}\ \bibnamefont
  {Hopkins}},\ and\ \bibinfo {author} {\bibfnamefont {A.~C.}\ \bibnamefont
  {Gossard}},\ }\bibfield  {title} {\bibinfo {title} {Conductance fluctuations
  and quantum chaotic scattering in semiconductor microstructures},\ }\href
  {https://doi.org/10.1063/1.165927} {\bibfield  {journal} {\bibinfo  {journal}
  {Chaos}\ }\textbf {\bibinfo {volume} {3}},\ \bibinfo {pages} {643} (\bibinfo
  {year} {1993})}\BibitemShut {NoStop}%
\bibitem [{\citenamefont
  {Ketzmerick}(1996)}]{Ketzmerick_phys.rev.b_54_10841_1996}%
  \BibitemOpen
  \bibfield  {author} {\bibinfo {author} {\bibfnamefont {R.}~\bibnamefont
  {Ketzmerick}},\ }\bibfield  {title} {\bibinfo {title} {Fractal conductance
  fluctuations in generic chaotic cavities},\ }\href
  {https://doi.org/10.1103/PhysRevB.54.10841} {\bibfield  {journal} {\bibinfo
  {journal} {Phys. Rev. B}\ }\textbf {\bibinfo {volume} {54}},\ \bibinfo
  {pages} {10841} (\bibinfo {year} {1996})}\BibitemShut {NoStop}%
\bibitem [{\citenamefont {Sachrajda}\ \emph {et~al.}(1998)\citenamefont
  {Sachrajda}, \citenamefont {Ketzmerick}, \citenamefont {Gould}, \citenamefont
  {Feng}, \citenamefont {Kelly}, \citenamefont {Delage},\ and\ \citenamefont
  {Wasilewski}}]{Sachrajda_phys.rev.lett_80_1948_1998}%
  \BibitemOpen
  \bibfield  {author} {\bibinfo {author} {\bibfnamefont {A.~S.}\ \bibnamefont
  {Sachrajda}}, \bibinfo {author} {\bibfnamefont {R.}~\bibnamefont
  {Ketzmerick}}, \bibinfo {author} {\bibfnamefont {C.}~\bibnamefont {Gould}},
  \bibinfo {author} {\bibfnamefont {Y.}~\bibnamefont {Feng}}, \bibinfo {author}
  {\bibfnamefont {P.~J.}\ \bibnamefont {Kelly}}, \bibinfo {author}
  {\bibfnamefont {A.}~\bibnamefont {Delage}},\ and\ \bibinfo {author}
  {\bibfnamefont {Z.}~\bibnamefont {Wasilewski}},\ }\bibfield  {title}
  {\bibinfo {title} {Fractal conductance fluctuations in a soft-wall stadium
  and a {S}inai billiard},\ }\href
  {https://doi.org/10.1103/PhysRevLett.80.1948} {\bibfield  {journal} {\bibinfo
   {journal} {Phys. Rev. Lett.}\ }\textbf {\bibinfo {volume} {80}},\ \bibinfo
  {pages} {1948} (\bibinfo {year} {1998})}\BibitemShut {NoStop}%
\bibitem [{\citenamefont {Nakamura}\ and\ \citenamefont
  {Thomas}(1988)}]{nakamura_phys.rev.lett_61_247_1988}%
  \BibitemOpen
  \bibfield  {author} {\bibinfo {author} {\bibfnamefont {K.}~\bibnamefont
  {Nakamura}}\ and\ \bibinfo {author} {\bibfnamefont {H.}~\bibnamefont
  {Thomas}},\ }\bibfield  {title} {\bibinfo {title} {Quantum billiard in a
  magnetic field: Chaos and diamagnetism},\ }\href
  {https://doi.org/10.1103/PhysRevLett.61.247} {\bibfield  {journal} {\bibinfo
  {journal} {Phys. Rev. Lett.}\ }\textbf {\bibinfo {volume} {61}},\ \bibinfo
  {pages} {247} (\bibinfo {year} {1988})}\BibitemShut {NoStop}%
\bibitem [{\citenamefont {Magn{\'u}sd{\'o}ttir}\ and\ \citenamefont
  {Gudmundsson}(2000)}]{magnusdottir_Phys.Rev.B.61.10229_2000}%
  \BibitemOpen
  \bibfield  {author} {\bibinfo {author} {\bibfnamefont {I.}~\bibnamefont
  {Magn{\'u}sd{\'o}ttir}}\ and\ \bibinfo {author} {\bibfnamefont
  {V.}~\bibnamefont {Gudmundsson}},\ }\bibfield  {title} {\bibinfo {title}
  {Magnetization of noncircular quantum dots},\ }\href
  {https://doi.org/10.1103/PhysRevB.61.10229} {\bibfield  {journal} {\bibinfo
  {journal} {Phys. Rev. B}\ }\textbf {\bibinfo {volume} {61}},\ \bibinfo
  {pages} {10229} (\bibinfo {year} {2000})}\BibitemShut {NoStop}%
\bibitem [{\citenamefont {Ji}\ and\ \citenamefont
  {Berggren}(1995)}]{zhen_phys.rev.B_52_1745_1995}%
  \BibitemOpen
  \bibfield  {author} {\bibinfo {author} {\bibfnamefont {Z.-L.}\ \bibnamefont
  {Ji}}\ and\ \bibinfo {author} {\bibfnamefont {K.-F.}\ \bibnamefont
  {Berggren}},\ }\bibfield  {title} {\bibinfo {title} {Transition from chaotic
  to regular behavior of electrons in a stadium-shaped quantum dot in a
  perpendicular magnetic field},\ }\href
  {https://doi.org/10.1103/PhysRevB.52.1745} {\bibfield  {journal} {\bibinfo
  {journal} {Phys. Rev. B}\ }\textbf {\bibinfo {volume} {52}},\ \bibinfo
  {pages} {1745} (\bibinfo {year} {1995})}\BibitemShut {NoStop}%
\bibitem [{\citenamefont {R{\"a}s{\"a}nen}\ \emph {et~al.}(2004)\citenamefont
  {R{\"a}s{\"a}nen}, \citenamefont {K{\"o}nemann}, \citenamefont {Haug},
  \citenamefont {Puska},\ and\ \citenamefont
  {Nieminen}}]{rasanen_phys.rev.B_70_115308_2004}%
  \BibitemOpen
  \bibfield  {author} {\bibinfo {author} {\bibfnamefont {E.}~\bibnamefont
  {R{\"a}s{\"a}nen}}, \bibinfo {author} {\bibfnamefont {J.}~\bibnamefont
  {K{\"o}nemann}}, \bibinfo {author} {\bibfnamefont {R.~J.}\ \bibnamefont
  {Haug}}, \bibinfo {author} {\bibfnamefont {M.~J.}\ \bibnamefont {Puska}},\
  and\ \bibinfo {author} {\bibfnamefont {R.~M.}\ \bibnamefont {Nieminen}},\
  }\bibfield  {title} {\bibinfo {title} {Impurity effects in quantum dots:
  Toward quantitative modeling},\ }\href
  {https://doi.org/10.1103/PhysRevB.70.115308} {\bibfield  {journal} {\bibinfo
  {journal} {Phys. Rev. B}\ }\textbf {\bibinfo {volume} {70}},\ \bibinfo
  {pages} {115308} (\bibinfo {year} {2004})}\BibitemShut {NoStop}%
\bibitem [{\citenamefont {G{\"u}{\c c}l{\"u}}\ \emph
  {et~al.}(2003)\citenamefont {G{\"u}{\c c}l{\"u}}, \citenamefont {Wang},\ and\
  \citenamefont {Guo}}]{guclu_phys.rev.B_68_035304_2003}%
  \BibitemOpen
  \bibfield  {author} {\bibinfo {author} {\bibfnamefont {A.~D.}\ \bibnamefont
  {G{\"u}{\c c}l{\"u}}}, \bibinfo {author} {\bibfnamefont {J.-S.}\ \bibnamefont
  {Wang}},\ and\ \bibinfo {author} {\bibfnamefont {H.}~\bibnamefont {Guo}},\
  }\bibfield  {title} {\bibinfo {title} {Disordered quantum dots: A diffusion
  quantum monte carlo study},\ }\href
  {https://doi.org/10.1103/PhysRevB.68.035304} {\bibfield  {journal} {\bibinfo
  {journal} {Phys. Rev. B}\ }\textbf {\bibinfo {volume} {68}},\ \bibinfo
  {pages} {035304} (\bibinfo {year} {2003})}\BibitemShut {NoStop}%
\bibitem [{\citenamefont {Halonen}\ \emph {et~al.}(1996)\citenamefont
  {Halonen}, \citenamefont {Hyv\"onen}, \citenamefont {Pietil\"ainen},\ and\
  \citenamefont {Chakraborty}}]{halonen_phys.rev.B_53_6971_1996}%
  \BibitemOpen
  \bibfield  {author} {\bibinfo {author} {\bibfnamefont {V.}~\bibnamefont
  {Halonen}}, \bibinfo {author} {\bibfnamefont {P.}~\bibnamefont {Hyv\"onen}},
  \bibinfo {author} {\bibfnamefont {P.}~\bibnamefont {Pietil\"ainen}},\ and\
  \bibinfo {author} {\bibfnamefont {T.}~\bibnamefont {Chakraborty}},\
  }\bibfield  {title} {\bibinfo {title} {Effects of scattering centers on the
  energy spectrum of a quantum dot},\ }\href
  {https://doi.org/10.1103/PhysRevB.53.6971} {\bibfield  {journal} {\bibinfo
  {journal} {Phys. Rev. B}\ }\textbf {\bibinfo {volume} {53}},\ \bibinfo
  {pages} {6971} (\bibinfo {year} {1996})}\BibitemShut {NoStop}%
\bibitem [{\citenamefont {Pilatowsky-Cameo}\ \emph {et~al.}(2021)\citenamefont
  {Pilatowsky-Cameo}, \citenamefont {Villase{\~n}or}, \citenamefont
  {Bastarrachea-Magnani}, \citenamefont {Lerma-Hern{\'a}ndez}, \citenamefont
  {Santos},\ and\ \citenamefont
  {Hirsch}}]{Pilatowsky_New.J.Phys_23_033045_2021}%
  \BibitemOpen
  \bibfield  {author} {\bibinfo {author} {\bibfnamefont {S.}~\bibnamefont
  {Pilatowsky-Cameo}}, \bibinfo {author} {\bibfnamefont {D.}~\bibnamefont
  {Villase{\~n}or}}, \bibinfo {author} {\bibfnamefont {M.~A.}\ \bibnamefont
  {Bastarrachea-Magnani}}, \bibinfo {author} {\bibfnamefont {S.}~\bibnamefont
  {Lerma-Hern{\'a}ndez}}, \bibinfo {author} {\bibfnamefont {L.~F.}\
  \bibnamefont {Santos}},\ and\ \bibinfo {author} {\bibfnamefont {J.~G.}\
  \bibnamefont {Hirsch}},\ }\bibfield  {title} {\bibinfo {title} {Quantum
  scarring in a spin-boson system: fundamental families of periodic orbits},\
  }\href@noop {} {\bibfield  {journal} {\bibinfo  {journal} {New J. Phys.}\
  }\textbf {\bibinfo {volume} {23}},\ \bibinfo {pages} {033045} (\bibinfo
  {year} {2021})}\BibitemShut {NoStop}%
\bibitem [{\citenamefont {Lee}\ and\ \citenamefont
  {Creagh}(2003)}]{Lee_ann.phys_307_392_2003}%
  \BibitemOpen
  \bibfield  {author} {\bibinfo {author} {\bibfnamefont {S.-Y.}\ \bibnamefont
  {Lee}}\ and\ \bibinfo {author} {\bibfnamefont {S.~C.}\ \bibnamefont
  {Creagh}},\ }\bibfield  {title} {\bibinfo {title} {Wavefunction statistics
  using scar states},\ }\href@noop {} {\bibfield  {journal} {\bibinfo
  {journal} {Ann. Phys. (N.Y.)}\ }\textbf {\bibinfo {volume} {307}},\ \bibinfo
  {pages} {392} (\bibinfo {year} {2003})}\BibitemShut {NoStop}%
\bibitem [{\citenamefont {Bies}\ \emph {et~al.}(2001)\citenamefont {Bies},
  \citenamefont {Kaplan},\ and\ \citenamefont
  {Heller}}]{Bies_phys.rev.e_64_016204_2001}%
  \BibitemOpen
  \bibfield  {author} {\bibinfo {author} {\bibfnamefont {W.~E.}\ \bibnamefont
  {Bies}}, \bibinfo {author} {\bibfnamefont {L.}~\bibnamefont {Kaplan}},\ and\
  \bibinfo {author} {\bibfnamefont {E.~J.}\ \bibnamefont {Heller}},\ }\bibfield
   {title} {\bibinfo {title} {Scarring effects on tunneling in chaotic
  double-well potentials},\ }\href {https://doi.org/10.1103/PhysRevE.64.016204}
  {\bibfield  {journal} {\bibinfo  {journal} {Phys. Rev. E}\ }\textbf {\bibinfo
  {volume} {64}},\ \bibinfo {pages} {016204} (\bibinfo {year}
  {2001})}\BibitemShut {NoStop}%
\bibitem [{\citenamefont {Wang}\ \emph {et~al.}(2001)\citenamefont {Wang},
  \citenamefont {Lai},\ and\ \citenamefont
  {Gu}}]{Wang_phys.rev.e_63_056208_2001}%
  \BibitemOpen
  \bibfield  {author} {\bibinfo {author} {\bibfnamefont {J.}~\bibnamefont
  {Wang}}, \bibinfo {author} {\bibfnamefont {C.-H.}\ \bibnamefont {Lai}},\ and\
  \bibinfo {author} {\bibfnamefont {Y.}~\bibnamefont {Gu}},\ }\bibfield
  {title} {\bibinfo {title} {Ergodicity and scars of the quantum cat map in the
  semiclassical regime},\ }\href {https://doi.org/10.1103/PhysRevE.63.056208}
  {\bibfield  {journal} {\bibinfo  {journal} {Phys. Rev. E}\ }\textbf {\bibinfo
  {volume} {63}},\ \bibinfo {pages} {056208} (\bibinfo {year}
  {2001})}\BibitemShut {NoStop}%
\bibitem [{\citenamefont {Deutsch}(1991)}]{Deutsch_Phys.Rev.A_43_2046_1991}%
  \BibitemOpen
  \bibfield  {author} {\bibinfo {author} {\bibfnamefont {J.~M.}\ \bibnamefont
  {Deutsch}},\ }\bibfield  {title} {\bibinfo {title} {Quantum statistical
  mechanics in a closed system},\ }\href
  {https://doi.org/10.1103/PhysRevA.43.2046} {\bibfield  {journal} {\bibinfo
  {journal} {Phys. Rev. A}\ }\textbf {\bibinfo {volume} {43}},\ \bibinfo
  {pages} {2046} (\bibinfo {year} {1991})}\BibitemShut {NoStop}%
\bibitem [{\citenamefont {Srednicki}(1994)}]{Srdnicki_Phys.Rev.E_50_888_1994}%
  \BibitemOpen
  \bibfield  {author} {\bibinfo {author} {\bibfnamefont {M.}~\bibnamefont
  {Srednicki}},\ }\bibfield  {title} {\bibinfo {title} {Chaos and quantum
  thermalization},\ }\href {https://doi.org/10.1103/PhysRevE.50.888} {\bibfield
   {journal} {\bibinfo  {journal} {Phys. Rev. E}\ }\textbf {\bibinfo {volume}
  {50}},\ \bibinfo {pages} {888} (\bibinfo {year} {1994})}\BibitemShut
  {NoStop}%
\bibitem [{\citenamefont {Rigol}\ \emph {et~al.}(2008)\citenamefont {Rigol},
  \citenamefont {Dunjko},\ and\ \citenamefont
  {Olshanii}}]{Rigol_Nature_452_854_2008}%
  \BibitemOpen
  \bibfield  {author} {\bibinfo {author} {\bibfnamefont {M.}~\bibnamefont
  {Rigol}}, \bibinfo {author} {\bibfnamefont {V.}~\bibnamefont {Dunjko}},\ and\
  \bibinfo {author} {\bibfnamefont {M.}~\bibnamefont {Olshanii}},\ }\bibfield
  {title} {\bibinfo {title} {Thermalization and its mechanism for generic
  isolated quantum systems},\ }\href@noop {} {\bibfield  {journal} {\bibinfo
  {journal} {Nature}\ }\textbf {\bibinfo {volume} {452}},\ \bibinfo {pages}
  {854} (\bibinfo {year} {2008})}\BibitemShut {NoStop}%
\bibitem [{\citenamefont {Polkovnikov}\ \emph {et~al.}(2011)\citenamefont
  {Polkovnikov}, \citenamefont {Sengupta}, \citenamefont {Silva},\ and\
  \citenamefont {Vengalattore}}]{Polkovnikov_Rev.Mod.Phys_83_863_2011}%
  \BibitemOpen
  \bibfield  {author} {\bibinfo {author} {\bibfnamefont {A.}~\bibnamefont
  {Polkovnikov}}, \bibinfo {author} {\bibfnamefont {K.}~\bibnamefont
  {Sengupta}}, \bibinfo {author} {\bibfnamefont {A.}~\bibnamefont {Silva}},\
  and\ \bibinfo {author} {\bibfnamefont {M.}~\bibnamefont {Vengalattore}},\
  }\bibfield  {title} {\bibinfo {title} {Colloquium: Nonequilibrium dynamics of
  closed interacting quantum systems},\ }\href
  {https://doi.org/10.1103/RevModPhys.83.863} {\bibfield  {journal} {\bibinfo
  {journal} {Rev. Mod. Phys.}\ }\textbf {\bibinfo {volume} {83}},\ \bibinfo
  {pages} {863} (\bibinfo {year} {2011})}\BibitemShut {NoStop}%
\bibitem [{\citenamefont
  {Deutsch}(2018)}]{Deutsch_rep.prog.phys_81_082001_2018}%
  \BibitemOpen
  \bibfield  {author} {\bibinfo {author} {\bibfnamefont {J.~M.}\ \bibnamefont
  {Deutsch}},\ }\bibfield  {title} {\bibinfo {title} {Eigenstate thermalization
  hypothesis},\ }\href@noop {} {\bibfield  {journal} {\bibinfo  {journal} {Rep.
  Prog. Phys.}\ }\textbf {\bibinfo {volume} {81}},\ \bibinfo {pages} {082001}
  (\bibinfo {year} {2018})}\BibitemShut {NoStop}%
\bibitem [{\citenamefont {Hirose}\ and\ \citenamefont
  {Wingreen}(2002)}]{hirose_phys.rev.B_65_193305_2002}%
  \BibitemOpen
  \bibfield  {author} {\bibinfo {author} {\bibfnamefont {K.}~\bibnamefont
  {Hirose}}\ and\ \bibinfo {author} {\bibfnamefont {N.~S.}\ \bibnamefont
  {Wingreen}},\ }\bibfield  {title} {\bibinfo {title} {Ground-state energy and
  spin in disordered quantum dots},\ }\href
  {https://doi.org/10.1103/PhysRevB.65.193305} {\bibfield  {journal} {\bibinfo
  {journal} {Phys. Rev. B}\ }\textbf {\bibinfo {volume} {65}},\ \bibinfo
  {pages} {193305} (\bibinfo {year} {2002})}\BibitemShut {NoStop}%
\bibitem [{\citenamefont {Hirose}\ \emph {et~al.}(2001)\citenamefont {Hirose},
  \citenamefont {Zhou},\ and\ \citenamefont
  {Wingreen}}]{hirose_phys.Rev.B_63_075301_2001}%
  \BibitemOpen
  \bibfield  {author} {\bibinfo {author} {\bibfnamefont {K.}~\bibnamefont
  {Hirose}}, \bibinfo {author} {\bibfnamefont {F.}~\bibnamefont {Zhou}},\ and\
  \bibinfo {author} {\bibfnamefont {N.~S.}\ \bibnamefont {Wingreen}},\
  }\bibfield  {title} {\bibinfo {title} {Density-functional theory of
  spin-polarized disordered quantum dots},\ }\href
  {https://doi.org/10.1103/PhysRevB.63.075301} {\bibfield  {journal} {\bibinfo
  {journal} {Phys. Rev. B}\ }\textbf {\bibinfo {volume} {63}},\ \bibinfo
  {pages} {075301} (\bibinfo {year} {2001})}\BibitemShut {NoStop}%
\bibitem [{\citenamefont {Blasi}\ \emph {et~al.}(2013)\citenamefont {Blasi},
  \citenamefont {Borunda}, \citenamefont {R{\"a}s{\"a}nen},\ and\ \citenamefont
  {Heller}}]{blasi_phys.rev.B_87_241303_2013}%
  \BibitemOpen
  \bibfield  {author} {\bibinfo {author} {\bibfnamefont {T.}~\bibnamefont
  {Blasi}}, \bibinfo {author} {\bibfnamefont {M.~F.}\ \bibnamefont {Borunda}},
  \bibinfo {author} {\bibfnamefont {E.}~\bibnamefont {R{\"a}s{\"a}nen}},\ and\
  \bibinfo {author} {\bibfnamefont {E.~J.}\ \bibnamefont {Heller}},\ }\bibfield
   {title} {\bibinfo {title} {Optimal local control of coherent dynamics in
  custom-made nanostructures},\ }\href
  {https://doi.org/10.1103/PhysRevB.87.241303} {\bibfield  {journal} {\bibinfo
  {journal} {Phys. Rev. B}\ }\textbf {\bibinfo {volume} {87}},\ \bibinfo
  {pages} {241303} (\bibinfo {year} {2013})}\BibitemShut {NoStop}%
\bibitem [{\citenamefont {Boyd}\ \emph {et~al.}(2011)\citenamefont {Boyd},
  \citenamefont {Storm}, \citenamefont {Samuelson},\ and\ \citenamefont
  {Westervelt}}]{boyd_nanotechnology_22_185201_2011}%
  \BibitemOpen
  \bibfield  {author} {\bibinfo {author} {\bibfnamefont {E.~E.}\ \bibnamefont
  {Boyd}}, \bibinfo {author} {\bibfnamefont {K.}~\bibnamefont {Storm}},
  \bibinfo {author} {\bibfnamefont {L.}~\bibnamefont {Samuelson}},\ and\
  \bibinfo {author} {\bibfnamefont {R.~M.}\ \bibnamefont {Westervelt}},\
  }\bibfield  {title} {\bibinfo {title} {Scanning gate imaging of quantum dots
  in 1d ultra-thin {InAs}/{InP} nanowires},\ }\href
  {https://doi.org/10.1088/0957-4484/22/18/185201} {\bibfield  {journal}
  {\bibinfo  {journal} {Nanotechnology}\ }\textbf {\bibinfo {volume} {22}},\
  \bibinfo {pages} {185201} (\bibinfo {year} {2011})}\BibitemShut {NoStop}%
\bibitem [{\citenamefont {Bleszynski}\ \emph {et~al.}(2007)\citenamefont
  {Bleszynski}, \citenamefont {Zwanenburg}, \citenamefont {Westervelt},
  \citenamefont {Roest}, \citenamefont {Bakkers},\ and\ \citenamefont
  {Kouwenhoven}}]{bleszynski_nano.lett_7_2559_2007}%
  \BibitemOpen
  \bibfield  {author} {\bibinfo {author} {\bibfnamefont {A.~C.}\ \bibnamefont
  {Bleszynski}}, \bibinfo {author} {\bibfnamefont {F.~A.}\ \bibnamefont
  {Zwanenburg}}, \bibinfo {author} {\bibfnamefont {R.~M.}\ \bibnamefont
  {Westervelt}}, \bibinfo {author} {\bibfnamefont {A.~L.}\ \bibnamefont
  {Roest}}, \bibinfo {author} {\bibfnamefont {E.~P. A.~M.}\ \bibnamefont
  {Bakkers}},\ and\ \bibinfo {author} {\bibfnamefont {L.~P.}\ \bibnamefont
  {Kouwenhoven}},\ }\bibfield  {title} {\bibinfo {title} {Scanned probe imaging
  of quantum dots inside inas nanowires},\ }\href
  {https://doi.org/10.1021/nl0621037} {\bibfield  {journal} {\bibinfo
  {journal} {Nano Lett.}\ }\textbf {\bibinfo {volume} {7}},\ \bibinfo {pages}
  {2559} (\bibinfo {year} {2007})}\BibitemShut {NoStop}%
\bibitem [{\citenamefont {Reimann}\ and\ \citenamefont
  {Manninen}(2002)}]{reimann_rev.mod.phys_74_1283_2002}%
  \BibitemOpen
  \bibfield  {author} {\bibinfo {author} {\bibfnamefont {S.~M.}\ \bibnamefont
  {Reimann}}\ and\ \bibinfo {author} {\bibfnamefont {M.}~\bibnamefont
  {Manninen}},\ }\bibfield  {title} {\bibinfo {title} {Electronic structure of
  quantum dots},\ }\href {https://doi.org/10.1103/RevModPhys.74.1283}
  {\bibfield  {journal} {\bibinfo  {journal} {Rev. Mod. Phys.}\ }\textbf
  {\bibinfo {volume} {74}},\ \bibinfo {pages} {1283} (\bibinfo {year}
  {2002})}\BibitemShut {NoStop}%
\bibitem [{\citenamefont {Kouwenhoven}\ \emph {et~al.}(2001)\citenamefont
  {Kouwenhoven}, \citenamefont {Austing},\ and\ \citenamefont
  {Tarucha}}]{kouwenhoven_rep.prog.phys_64_701_2001}%
  \BibitemOpen
  \bibfield  {author} {\bibinfo {author} {\bibfnamefont {L.~P.}\ \bibnamefont
  {Kouwenhoven}}, \bibinfo {author} {\bibfnamefont {D.~G.}\ \bibnamefont
  {Austing}},\ and\ \bibinfo {author} {\bibfnamefont {S.}~\bibnamefont
  {Tarucha}},\ }\bibfield  {title} {\bibinfo {title} {Few-electron quantum
  dots},\ }\href {https://doi.org/10.1088/0034-4885/64/6/201} {\bibfield
  {journal} {\bibinfo  {journal} {Rep. Prog. Phys.}\ }\textbf {\bibinfo
  {volume} {64}},\ \bibinfo {pages} {701} (\bibinfo {year} {2001})}\BibitemShut
  {NoStop}%
\bibitem [{\citenamefont {Bruce}\ and\ \citenamefont
  {Maksym}(2000)}]{bruce_phys.rev.B_61_4718_2000}%
  \BibitemOpen
  \bibfield  {author} {\bibinfo {author} {\bibfnamefont {N.~A.}\ \bibnamefont
  {Bruce}}\ and\ \bibinfo {author} {\bibfnamefont {P.~A.}\ \bibnamefont
  {Maksym}},\ }\bibfield  {title} {\bibinfo {title} {Quantum states of
  interacting electrons in a real quantum dot},\ }\href
  {https://doi.org/10.1103/PhysRevB.61.4718} {\bibfield  {journal} {\bibinfo
  {journal} {Phys. Rev. B}\ }\textbf {\bibinfo {volume} {61}},\ \bibinfo
  {pages} {4718} (\bibinfo {year} {2000})}\BibitemShut {NoStop}%
\bibitem [{\citenamefont {Stopa}(1996)}]{stopa_phys.rev.B_54_13767_1996}%
  \BibitemOpen
  \bibfield  {author} {\bibinfo {author} {\bibfnamefont {M.}~\bibnamefont
  {Stopa}},\ }\bibfield  {title} {\bibinfo {title} {Quantum dot self-consistent
  electronic structure and the coulomb blockade},\ }\href
  {https://doi.org/10.1103/PhysRevB.54.13767} {\bibfield  {journal} {\bibinfo
  {journal} {Phys. Rev. B}\ }\textbf {\bibinfo {volume} {54}},\ \bibinfo
  {pages} {13767} (\bibinfo {year} {1996})}\BibitemShut {NoStop}%
\bibitem [{\citenamefont {Rogge}\ \emph {et~al.}(2010)\citenamefont {Rogge},
  \citenamefont {R\"as\"anen},\ and\ \citenamefont
  {Haug}}]{rogge_phys.rev.lett_105_046802_2010}%
  \BibitemOpen
  \bibfield  {author} {\bibinfo {author} {\bibfnamefont {M.~C.}\ \bibnamefont
  {Rogge}}, \bibinfo {author} {\bibfnamefont {E.}~\bibnamefont {R\"as\"anen}},\
  and\ \bibinfo {author} {\bibfnamefont {R.~J.}\ \bibnamefont {Haug}},\
  }\bibfield  {title} {\bibinfo {title} {Interaction-induced spin polarization
  in quantum dots},\ }\href {https://doi.org/10.1103/PhysRevLett.105.046802}
  {\bibfield  {journal} {\bibinfo  {journal} {Phys. Rev. Lett.}\ }\textbf
  {\bibinfo {volume} {105}},\ \bibinfo {pages} {046802} (\bibinfo {year}
  {2010})}\BibitemShut {NoStop}%
\bibitem [{\citenamefont {R\"as\"anen}\ \emph {et~al.}(2008)\citenamefont
  {R\"as\"anen}, \citenamefont {Saarikoski}, \citenamefont {Harju},
  \citenamefont {Ciorga},\ and\ \citenamefont
  {Sachrajda}}]{rasanen_phys.rev.B_77_041302_2008}%
  \BibitemOpen
  \bibfield  {author} {\bibinfo {author} {\bibfnamefont {E.}~\bibnamefont
  {R\"as\"anen}}, \bibinfo {author} {\bibfnamefont {H.}~\bibnamefont
  {Saarikoski}}, \bibinfo {author} {\bibfnamefont {A.}~\bibnamefont {Harju}},
  \bibinfo {author} {\bibfnamefont {M.}~\bibnamefont {Ciorga}},\ and\ \bibinfo
  {author} {\bibfnamefont {A.~S.}\ \bibnamefont {Sachrajda}},\ }\bibfield
  {title} {\bibinfo {title} {Spin droplets in confined quantum hall systems},\
  }\href {https://doi.org/10.1103/PhysRevB.77.041302} {\bibfield  {journal}
  {\bibinfo  {journal} {Phys. Rev. B}\ }\textbf {\bibinfo {volume} {77}},\
  \bibinfo {pages} {041302} (\bibinfo {year} {2008})}\BibitemShut {NoStop}%
\bibitem [{\citenamefont {Mendoza}\ and\ \citenamefont
  {Schulz}(2003{\natexlab{a}})}]{mendoze_phys.rev.B_68_205302_2003}%
  \BibitemOpen
  \bibfield  {author} {\bibinfo {author} {\bibfnamefont {M.}~\bibnamefont
  {Mendoza}}\ and\ \bibinfo {author} {\bibfnamefont {P.~A.}\ \bibnamefont
  {Schulz}},\ }\bibfield  {title} {\bibinfo {title} {Wave-function mapping
  conditions in open quantum dot structures},\ }\href
  {https://doi.org/10.1103/PhysRevB.68.205302} {\bibfield  {journal} {\bibinfo
  {journal} {Phys. Rev. B}\ }\textbf {\bibinfo {volume} {68}},\ \bibinfo
  {pages} {205302} (\bibinfo {year} {2003}{\natexlab{a}})}\BibitemShut
  {NoStop}%
\bibitem [{\citenamefont {Zozoulenko}\ \emph {et~al.}(1998)\citenamefont
  {Zozoulenko}, \citenamefont {Sachrajda}, \citenamefont {Zawadzki},
  \citenamefont {Berggren}, \citenamefont {Feng},\ and\ \citenamefont
  {Wasilewski}}]{zozoulenko_phys.rev.B_58_10597_1998}%
  \BibitemOpen
  \bibfield  {author} {\bibinfo {author} {\bibfnamefont {I.~V.}\ \bibnamefont
  {Zozoulenko}}, \bibinfo {author} {\bibfnamefont {A.~S.}\ \bibnamefont
  {Sachrajda}}, \bibinfo {author} {\bibfnamefont {P.}~\bibnamefont {Zawadzki}},
  \bibinfo {author} {\bibfnamefont {K.-F.}\ \bibnamefont {Berggren}}, \bibinfo
  {author} {\bibfnamefont {Y.}~\bibnamefont {Feng}},\ and\ \bibinfo {author}
  {\bibfnamefont {Z.}~\bibnamefont {Wasilewski}},\ }\bibfield  {title}
  {\bibinfo {title} {Conductance fluctuations in a rectangular dot at constant
  magnetic fields},\ }\href {https://doi.org/10.1103/PhysRevB.58.10597}
  {\bibfield  {journal} {\bibinfo  {journal} {Phys. Rev. B}\ }\textbf {\bibinfo
  {volume} {58}},\ \bibinfo {pages} {10597} (\bibinfo {year}
  {1998})}\BibitemShut {NoStop}%
\bibitem [{\citenamefont {Bird}(2013)}]{bird_2013_QD}%
  \BibitemOpen
  \bibfield  {author} {\bibinfo {author} {\bibfnamefont {J.}~\bibnamefont
  {Bird}},\ }\href@noop {} {\emph {\bibinfo {title} {Electron Transport in
  Quantum Dots}}}\ (\bibinfo  {publisher} {Springer US},\ \bibinfo {year}
  {2013})\BibitemShut {NoStop}%
\bibitem [{\citenamefont {Jullien}\ \emph {et~al.}(2014)\citenamefont
  {Jullien}, \citenamefont {Roulleau}, \citenamefont {Roche}, \citenamefont
  {Cavanna}, \citenamefont {Jin},\ and\ \citenamefont
  {Glattli}}]{jullien_nature_514_603_2014}%
  \BibitemOpen
  \bibfield  {author} {\bibinfo {author} {\bibfnamefont {T.}~\bibnamefont
  {Jullien}}, \bibinfo {author} {\bibfnamefont {P.}~\bibnamefont {Roulleau}},
  \bibinfo {author} {\bibfnamefont {B.}~\bibnamefont {Roche}}, \bibinfo
  {author} {\bibfnamefont {A.}~\bibnamefont {Cavanna}}, \bibinfo {author}
  {\bibfnamefont {Y.}~\bibnamefont {Jin}},\ and\ \bibinfo {author}
  {\bibfnamefont {D.~C.}\ \bibnamefont {Glattli}},\ }\bibfield  {title}
  {\bibinfo {title} {Quantum tomography of an electron},\ }\href
  {https://doi.org/10.1038/nature13821} {\bibfield  {journal} {\bibinfo
  {journal} {Nature (London)}\ }\textbf {\bibinfo {volume} {514}},\ \bibinfo
  {pages} {603} (\bibinfo {year} {2014})}\BibitemShut {NoStop}%
\bibitem [{\citenamefont {Reynolds}\ and\ \citenamefont
  {Shouppe}(2010)}]{reynolds2010closed}%
  \BibitemOpen
  \bibfield  {author} {\bibinfo {author} {\bibfnamefont {M.}~\bibnamefont
  {Reynolds}}\ and\ \bibinfo {author} {\bibfnamefont {M.}~\bibnamefont
  {Shouppe}},\ }\bibfield  {title} {\bibinfo {title} {Closed, spirograph-like
  orbits in power law central potentials},\ }\href@noop {} {\bibfield
  {journal} {\bibinfo  {journal} {arXiv preprint arXiv:1008.0559}\ } (\bibinfo
  {year} {2010})}\BibitemShut {NoStop}%
\bibitem [{\citenamefont {Kotkin}\ and\ \citenamefont
  {Serbo}(2020)}]{kotkin2020exploring}%
  \BibitemOpen
  \bibfield  {author} {\bibinfo {author} {\bibfnamefont {G.~L.}\ \bibnamefont
  {Kotkin}}\ and\ \bibinfo {author} {\bibfnamefont {V.~G.}\ \bibnamefont
  {Serbo}},\ }\href@noop {} {\emph {\bibinfo {title} {Exploring Classical
  Mechanics: A Collection of 350+ Solved Problems for Students, Lecturers, and
  Researchers-Second Revised and Enlarged English Edition}}}\ (\bibinfo
  {publisher} {Oxford University Press},\ \bibinfo {year} {2020})\BibitemShut
  {NoStop}%
\bibitem [{\citenamefont
  {Kolmogorov}(1953)}]{Kolmogorov_dokl.akad.nauk_93_763_1953}%
  \BibitemOpen
  \bibfield  {author} {\bibinfo {author} {\bibfnamefont {A.~N.}\ \bibnamefont
  {Kolmogorov}},\ }\bibfield  {title} {\bibinfo {title} {On dynamical systems
  with an integral invariant on the torus},\ }\href@noop {} {\bibfield
  {journal} {\bibinfo  {journal} {Dokl. Akad. Nauk SSSR}\ }\textbf {\bibinfo
  {volume} {93}},\ \bibinfo {pages} {763} (\bibinfo {year} {1953})}\BibitemShut
  {NoStop}%
\bibitem [{\citenamefont {Arnold}(1963)}]{Arnold_uspehi.mat.nauk_18_13_1963}%
  \BibitemOpen
  \bibfield  {author} {\bibinfo {author} {\bibfnamefont {V.~I.}\ \bibnamefont
  {Arnold}},\ }\bibfield  {title} {\bibinfo {title} {Proof of a theorem of a.
  n. kolmogorov on the preservation of conditionally periodic motions under a
  small perturbation of the hamiltonian},\ }\href@noop {} {\bibfield  {journal}
  {\bibinfo  {journal} {Uspehi Mat. Nauk}\ }\textbf {\bibinfo {volume} {18}},\
  \bibinfo {pages} {13} (\bibinfo {year} {1963})}\BibitemShut {NoStop}%
\bibitem [{\citenamefont {Moser}(1962)}]{Moser_nachr.akad.wiss_II_1_1962}%
  \BibitemOpen
  \bibfield  {author} {\bibinfo {author} {\bibfnamefont {J.}~\bibnamefont
  {Moser}},\ }\bibfield  {title} {\bibinfo {title} {On invariant curves of
  area-preserving mappings of an annulus},\ }\href@noop {} {\bibfield
  {journal} {\bibinfo  {journal} {Nachr. Akad. Wiss. G{\"o}ttingen, II}\ ,\
  \bibinfo {pages} {1}} (\bibinfo {year} {1962})}\BibitemShut {NoStop}%
\bibitem [{\citenamefont {Luukko}\ and\ \citenamefont
  {R{\"a}s{\"a}nen}(2013)}]{luukko_comput.phys.commun_184_769_2013}%
  \BibitemOpen
  \bibfield  {author} {\bibinfo {author} {\bibfnamefont {P.}~\bibnamefont
  {Luukko}}\ and\ \bibinfo {author} {\bibfnamefont {E.}~\bibnamefont
  {R{\"a}s{\"a}nen}},\ }\bibfield  {title} {\bibinfo {title} {Imaginary time
  propagation code for large-scale two-dimensional eigenvalue problems in
  magnetic fields},\ }\href
  {https://doi.org/https://doi.org/10.1016/j.cpc.2012.09.029} {\bibfield
  {journal} {\bibinfo  {journal} {Comput. Phys. Commun.}\ }\textbf {\bibinfo
  {volume} {184}},\ \bibinfo {pages} {769} (\bibinfo {year}
  {2013})}\BibitemShut {NoStop}%
\bibitem [{\citenamefont {Aichinger}\ \emph {et~al.}(2005)\citenamefont
  {Aichinger}, \citenamefont {Chin},\ and\ \citenamefont
  {Krotscheck}}]{aichinger_comput.phys.commun_171_197_2005}%
  \BibitemOpen
  \bibfield  {author} {\bibinfo {author} {\bibfnamefont {M.}~\bibnamefont
  {Aichinger}}, \bibinfo {author} {\bibfnamefont {S.~A.}\ \bibnamefont
  {Chin}},\ and\ \bibinfo {author} {\bibfnamefont {E.}~\bibnamefont
  {Krotscheck}},\ }\bibfield  {title} {\bibinfo {title} {Fourth-order
  algorithms for solving local schr{\"o}dinger equations in a strong magnetic
  field},\ }\href {https://doi.org/https://doi.org/10.1016/j.cpc.2005.05.006}
  {\bibfield  {journal} {\bibinfo  {journal} {Comput. Phys. Commun.}\ }\textbf
  {\bibinfo {volume} {171}},\ \bibinfo {pages} {197} (\bibinfo {year}
  {2005})}\BibitemShut {NoStop}%
\bibitem [{\citenamefont {Casati}\ \emph {et~al.}(1979)\citenamefont {Casati},
  \citenamefont {Chirikov}, \citenamefont {Izraelev},\ and\ \citenamefont
  {Ford}}]{casati_dynamical_localization}%
  \BibitemOpen
  \bibfield  {author} {\bibinfo {author} {\bibfnamefont {G.}~\bibnamefont
  {Casati}}, \bibinfo {author} {\bibfnamefont {B.~V.}\ \bibnamefont
  {Chirikov}}, \bibinfo {author} {\bibfnamefont {F.~M.}\ \bibnamefont
  {Izraelev}},\ and\ \bibinfo {author} {\bibfnamefont {J.}~\bibnamefont
  {Ford}},\ }\bibfield  {title} {\bibinfo {title} {Stochastic behavior of a
  quantum pendulum under a periodic perturbation},\ }in\ \href@noop {} {\emph
  {\bibinfo {booktitle} {Stochastic Behavior in Classical and Quantum
  Hamiltonian Systems}}},\ \bibinfo {editor} {edited by\ \bibinfo {editor}
  {\bibfnamefont {G.}~\bibnamefont {Casati}}\ and\ \bibinfo {editor}
  {\bibfnamefont {J.}~\bibnamefont {Ford}}}\ (\bibinfo  {publisher} {Springer
  Berlin Heidelberg},\ \bibinfo {address} {Berlin, Heidelberg},\ \bibinfo
  {year} {1979})\ pp.\ \bibinfo {pages} {334--352}\BibitemShut {NoStop}%
\bibitem [{\citenamefont {Grempel}\ \emph {et~al.}(1984)\citenamefont
  {Grempel}, \citenamefont {Prange},\ and\ \citenamefont
  {Fishman}}]{germpel_phys.rev.A_29_1639}%
  \BibitemOpen
  \bibfield  {author} {\bibinfo {author} {\bibfnamefont {D.~R.}\ \bibnamefont
  {Grempel}}, \bibinfo {author} {\bibfnamefont {R.~E.}\ \bibnamefont
  {Prange}},\ and\ \bibinfo {author} {\bibfnamefont {S.}~\bibnamefont
  {Fishman}},\ }\bibfield  {title} {\bibinfo {title} {Quantum dynamics of a
  nonintegrable system},\ }\href {https://doi.org/10.1103/PhysRevA.29.1639}
  {\bibfield  {journal} {\bibinfo  {journal} {Phys. Rev. A}\ }\textbf {\bibinfo
  {volume} {29}},\ \bibinfo {pages} {1639} (\bibinfo {year}
  {1984})}\BibitemShut {NoStop}%
\bibitem [{\citenamefont
  {Shepelyansky}(1987)}]{shepelyansky_physica.d_28_103_1987}%
  \BibitemOpen
  \bibfield  {author} {\bibinfo {author} {\bibfnamefont {D.~L.}\ \bibnamefont
  {Shepelyansky}},\ }\bibfield  {title} {\bibinfo {title} {Localization of
  diffusive excitation in multi-level systems},\ }\href
  {https://doi.org/https://doi.org/10.1016/0167-2789(87)90123-0} {\bibfield
  {journal} {\bibinfo  {journal} {Physica D}\ }\textbf {\bibinfo {volume}
  {28}},\ \bibinfo {pages} {103} (\bibinfo {year} {1987})}\BibitemShut
  {NoStop}%
\bibitem [{\citenamefont {Izrailev}(1990)}]{izrailev_phys.rep_196__299_1990}%
  \BibitemOpen
  \bibfield  {author} {\bibinfo {author} {\bibfnamefont {F.~M.}\ \bibnamefont
  {Izrailev}},\ }\bibfield  {title} {\bibinfo {title} {Simple models of quantum
  chaos: Spectrum and eigenfunctions},\ }\href
  {https://doi.org/https://doi.org/10.1016/0370-1573(90)90067-C} {\bibfield
  {journal} {\bibinfo  {journal} {Phys. Rep.}\ }\textbf {\bibinfo {volume}
  {196}},\ \bibinfo {pages} {299} (\bibinfo {year} {1990})}\BibitemShut
  {NoStop}%
\bibitem [{\citenamefont {Liu}\ \emph {et~al.}(2006)\citenamefont {Liu},
  \citenamefont {Lu}, \citenamefont {Chen},\ and\ \citenamefont
  {Huang}}]{liu_phys.rev.E_74_046214_2006}%
  \BibitemOpen
  \bibfield  {author} {\bibinfo {author} {\bibfnamefont {C.~C.}\ \bibnamefont
  {Liu}}, \bibinfo {author} {\bibfnamefont {T.~H.}\ \bibnamefont {Lu}},
  \bibinfo {author} {\bibfnamefont {Y.~F.}\ \bibnamefont {Chen}},\ and\
  \bibinfo {author} {\bibfnamefont {K.~F.}\ \bibnamefont {Huang}},\ }\bibfield
  {title} {\bibinfo {title} {Wave functions with localizations on classical
  periodic orbits in weakly perturbed quantum billiards},\ }\href
  {https://doi.org/10.1103/PhysRevE.74.046214} {\bibfield  {journal} {\bibinfo
  {journal} {Phys. Rev. E}\ }\textbf {\bibinfo {volume} {74}},\ \bibinfo
  {pages} {046214} (\bibinfo {year} {2006})}\BibitemShut {NoStop}%
\bibitem [{\citenamefont {Chen}\ \emph {et~al.}(2002)\citenamefont {Chen},
  \citenamefont {Huang},\ and\ \citenamefont
  {Lan}}]{chen_phys.rev.E_66_046215_2002}%
  \BibitemOpen
  \bibfield  {author} {\bibinfo {author} {\bibfnamefont {Y.~F.}\ \bibnamefont
  {Chen}}, \bibinfo {author} {\bibfnamefont {K.~F.}\ \bibnamefont {Huang}},\
  and\ \bibinfo {author} {\bibfnamefont {Y.~P.}\ \bibnamefont {Lan}},\
  }\bibfield  {title} {\bibinfo {title} {Localization of wave patterns on
  classical periodic orbits in a square billiard},\ }\href
  {https://doi.org/10.1103/PhysRevE.66.046215} {\bibfield  {journal} {\bibinfo
  {journal} {Phys. Rev. E}\ }\textbf {\bibinfo {volume} {66}},\ \bibinfo
  {pages} {046215} (\bibinfo {year} {2002})}\BibitemShut {NoStop}%
\bibitem [{\citenamefont {Li}\ \emph {et~al.}(2002)\citenamefont {Li},
  \citenamefont {Reichl},\ and\ \citenamefont
  {Wu}}]{li_phys.rev.E_65_056220_2002}%
  \BibitemOpen
  \bibfield  {author} {\bibinfo {author} {\bibfnamefont {W.}~\bibnamefont
  {Li}}, \bibinfo {author} {\bibfnamefont {L.~E.}\ \bibnamefont {Reichl}},\
  and\ \bibinfo {author} {\bibfnamefont {B.}~\bibnamefont {Wu}},\ }\bibfield
  {title} {\bibinfo {title} {Quantum chaos in a ripple billiard},\ }\href
  {https://doi.org/10.1103/PhysRevE.65.056220} {\bibfield  {journal} {\bibinfo
  {journal} {Phys. Rev. E}\ }\textbf {\bibinfo {volume} {65}},\ \bibinfo
  {pages} {056220} (\bibinfo {year} {2002})}\BibitemShut {NoStop}%
\bibitem [{\citenamefont {Kumar}\ and\ \citenamefont
  {Dutta-Roy}(2008)}]{kumar_j.phys.A_41_075306_2008}%
  \BibitemOpen
  \bibfield  {author} {\bibinfo {author} {\bibfnamefont {M.~S.}\ \bibnamefont
  {Kumar}}\ and\ \bibinfo {author} {\bibfnamefont {B.}~\bibnamefont
  {Dutta-Roy}},\ }\bibfield  {title} {\bibinfo {title} {Commensurate
  anisotropic oscillator $su(2)$ coherent states and the classical limit},\
  }\href {https://doi.org/10.1088/1751-8113/41/7/075306} {\bibfield  {journal}
  {\bibinfo  {journal} {J. Phys. A}\ }\textbf {\bibinfo {volume} {41}},\
  \bibinfo {pages} {075306} (\bibinfo {year} {2008})}\BibitemShut {NoStop}%
\bibitem [{\citenamefont {Chen}(2011)}]{chen._phys.rev.A_83_032124_2011}%
  \BibitemOpen
  \bibfield  {author} {\bibinfo {author} {\bibfnamefont {Y.~F.}\ \bibnamefont
  {Chen}},\ }\bibfield  {title} {\bibinfo {title} {Geometry of classical
  periodic orbits and quantum coherent states in coupled oscillators with su(2)
  transformations},\ }\href {https://doi.org/10.1103/PhysRevA.83.032124}
  {\bibfield  {journal} {\bibinfo  {journal} {Phys. Rev. A}\ }\textbf {\bibinfo
  {volume} {83}},\ \bibinfo {pages} {032124} (\bibinfo {year}
  {2011})}\BibitemShut {NoStop}%
\bibitem [{\citenamefont {Chen}\ \emph {et~al.}(2005)\citenamefont {Chen},
  \citenamefont {Lu}, \citenamefont {Su},\ and\ \citenamefont
  {Huang}}]{chen_phys.rev.E_72_056210_2005}%
  \BibitemOpen
  \bibfield  {author} {\bibinfo {author} {\bibfnamefont {Y.~F.}\ \bibnamefont
  {Chen}}, \bibinfo {author} {\bibfnamefont {T.~H.}\ \bibnamefont {Lu}},
  \bibinfo {author} {\bibfnamefont {K.~W.}\ \bibnamefont {Su}},\ and\ \bibinfo
  {author} {\bibfnamefont {K.~F.}\ \bibnamefont {Huang}},\ }\bibfield  {title}
  {\bibinfo {title} {Quantum signatures of nonlinear resonances in mesoscopic
  systems: Efficient extension of localized wave functions},\ }\href
  {https://doi.org/10.1103/PhysRevE.72.056210} {\bibfield  {journal} {\bibinfo
  {journal} {Phys. Rev. E}\ }\textbf {\bibinfo {volume} {72}},\ \bibinfo
  {pages} {056210} (\bibinfo {year} {2005})}\BibitemShut {NoStop}%
\bibitem [{\citenamefont {Chen}\ and\ \citenamefont
  {Huang}(2003)}]{chen_j.phys.A_36_7751_2003}%
  \BibitemOpen
  \bibfield  {author} {\bibinfo {author} {\bibfnamefont {Y.~F.}\ \bibnamefont
  {Chen}}\ and\ \bibinfo {author} {\bibfnamefont {K.~F.}\ \bibnamefont
  {Huang}},\ }\bibfield  {title} {\bibinfo {title} {Vortex structure of quantum
  eigenstates and classical periodic orbits in two-dimensional harmonic
  oscillators},\ }\href {https://doi.org/10.1088/0305-4470/36/28/305}
  {\bibfield  {journal} {\bibinfo  {journal} {J. Phys. A}\ }\textbf {\bibinfo
  {volume} {36}},\ \bibinfo {pages} {7751} (\bibinfo {year}
  {2003})}\BibitemShut {NoStop}%
\bibitem [{\citenamefont {Lee}\ \emph {et~al.}(2008)\citenamefont {Lee},
  \citenamefont {Rim}, \citenamefont {Ryu}, \citenamefont {Kwon}, \citenamefont
  {Choi},\ and\ \citenamefont {Kim}}]{lee_j.phys.A_41_275102_2008}%
  \BibitemOpen
  \bibfield  {author} {\bibinfo {author} {\bibfnamefont {S.-Y.}\ \bibnamefont
  {Lee}}, \bibinfo {author} {\bibfnamefont {S.}~\bibnamefont {Rim}}, \bibinfo
  {author} {\bibfnamefont {J.-W.}\ \bibnamefont {Ryu}}, \bibinfo {author}
  {\bibfnamefont {T.-Y.}\ \bibnamefont {Kwon}}, \bibinfo {author}
  {\bibfnamefont {M.}~\bibnamefont {Choi}},\ and\ \bibinfo {author}
  {\bibfnamefont {C.-M.}\ \bibnamefont {Kim}},\ }\bibfield  {title} {\bibinfo
  {title} {Ray and wave dynamical properties of a spiral-shaped dielectric
  microcavity},\ }\href {https://doi.org/10.1088/1751-8113/41/27/275102}
  {\bibfield  {journal} {\bibinfo  {journal} {J. Phys. A}\ }\textbf {\bibinfo
  {volume} {41}},\ \bibinfo {pages} {275102} (\bibinfo {year}
  {2008})}\BibitemShut {NoStop}%
\bibitem [{\citenamefont {Mendoza}\ and\ \citenamefont
  {Schulz}(2003{\natexlab{b}})}]{mendoza_phys.rev.b_68_205302_2003}%
  \BibitemOpen
  \bibfield  {author} {\bibinfo {author} {\bibfnamefont {M.}~\bibnamefont
  {Mendoza}}\ and\ \bibinfo {author} {\bibfnamefont {P.}~\bibnamefont
  {Schulz}},\ }\bibfield  {title} {\bibinfo {title} {Wave-function mapping
  conditions in open quantum dot structures},\ }\href@noop {} {\bibfield
  {journal} {\bibinfo  {journal} {Phys Rev. B}\ }\textbf {\bibinfo {volume}
  {68}},\ \bibinfo {pages} {205302} (\bibinfo {year}
  {2003}{\natexlab{b}})}\BibitemShut {NoStop}%
\bibitem [{\citenamefont {Ge}\ \emph {et~al.}(2020)\citenamefont {Ge},
  \citenamefont {Joucken}, \citenamefont {Quezada}, \citenamefont {Da~Costa},
  \citenamefont {Davenport}, \citenamefont {Giraldo}, \citenamefont
  {Taniguchi}, \citenamefont {Watanabe}, \citenamefont {Kobayashi},
  \citenamefont {Low} \emph {et~al.}}]{Ge_nano.lett_20_8682_2020}%
  \BibitemOpen
  \bibfield  {author} {\bibinfo {author} {\bibfnamefont {Z.}~\bibnamefont
  {Ge}}, \bibinfo {author} {\bibfnamefont {F.}~\bibnamefont {Joucken}},
  \bibinfo {author} {\bibfnamefont {E.}~\bibnamefont {Quezada}}, \bibinfo
  {author} {\bibfnamefont {D.~R.}\ \bibnamefont {Da~Costa}}, \bibinfo {author}
  {\bibfnamefont {J.}~\bibnamefont {Davenport}}, \bibinfo {author}
  {\bibfnamefont {B.}~\bibnamefont {Giraldo}}, \bibinfo {author} {\bibfnamefont
  {T.}~\bibnamefont {Taniguchi}}, \bibinfo {author} {\bibfnamefont
  {K.}~\bibnamefont {Watanabe}}, \bibinfo {author} {\bibfnamefont {N.~P.}\
  \bibnamefont {Kobayashi}}, \bibinfo {author} {\bibfnamefont {T.}~\bibnamefont
  {Low}}, \emph {et~al.},\ }\bibfield  {title} {\bibinfo {title} {Visualization
  and manipulation of bilayer graphene quantum dots with broken rotational
  symmetry and nontrivial topology},\ }\href@noop {} {\bibfield  {journal}
  {\bibinfo  {journal} {Nano Lett.}\ }\textbf {\bibinfo {volume} {20}},\
  \bibinfo {pages} {8682} (\bibinfo {year} {2020})}\BibitemShut {NoStop}%
\bibitem [{\citenamefont {Ge}\ \emph {et~al.}(2021)\citenamefont {Ge},
  \citenamefont {Wong}, \citenamefont {Lee}, \citenamefont {Joucken},
  \citenamefont {Quezada-Lopez}, \citenamefont {Kahn}, \citenamefont {Tsai},
  \citenamefont {Taniguchi}, \citenamefont {Watanabe}, \citenamefont {Wang}
  \emph {et~al.}}]{Ge_nano.lett_21_8993_2021}%
  \BibitemOpen
  \bibfield  {author} {\bibinfo {author} {\bibfnamefont {Z.}~\bibnamefont
  {Ge}}, \bibinfo {author} {\bibfnamefont {D.}~\bibnamefont {Wong}}, \bibinfo
  {author} {\bibfnamefont {J.}~\bibnamefont {Lee}}, \bibinfo {author}
  {\bibfnamefont {F.}~\bibnamefont {Joucken}}, \bibinfo {author} {\bibfnamefont
  {E.~A.}\ \bibnamefont {Quezada-Lopez}}, \bibinfo {author} {\bibfnamefont
  {S.}~\bibnamefont {Kahn}}, \bibinfo {author} {\bibfnamefont {H.-Z.}\
  \bibnamefont {Tsai}}, \bibinfo {author} {\bibfnamefont {T.}~\bibnamefont
  {Taniguchi}}, \bibinfo {author} {\bibfnamefont {K.}~\bibnamefont {Watanabe}},
  \bibinfo {author} {\bibfnamefont {F.}~\bibnamefont {Wang}}, \emph {et~al.},\
  }\bibfield  {title} {\bibinfo {title} {Imaging quantum interference in
  stadium-shaped monolayer and bilayer graphene quantum dots},\ }\href@noop {}
  {\bibfield  {journal} {\bibinfo  {journal} {Nano Lett.}\ }\textbf {\bibinfo
  {volume} {21}},\ \bibinfo {pages} {8993} (\bibinfo {year}
  {2021})}\BibitemShut {NoStop}%
\end{thebibliography}%

\end{document}